\pdfoutput=1

\documentclass[preprint,aps,nofootinbib,superscriptaddress,floatfix,preprintnumbers,longbibliography]{revtex4-1}

\usepackage[hyperfootnotes=false]{hyperref}
\usepackage{url}

\usepackage{color}

\usepackage{bm}
\usepackage{amsmath}
\usepackage{amssymb}
\usepackage{graphicx}
\usepackage{epstopdf}

\def \beq{\begin{equation}}
\def \eeq{\end{equation}}
\def \bea{\begin{align}}
\def \eea{\end{align}}

\begin{document}

\title{The Tiny (g-2) Muon Wobble from Small-$\boldsymbol{\mu}$ Supersymmetry }

\newcommand{\SU}{\affiliation{Stanford~Institute~for~Theoretical~Physics, Department~of~Physics, Stanford~University, Stanford,~CA~94305, USA}}

\newcommand{\FNL}{\affiliation{Fermi~National~Accelerator~Laboratory, P.~O.~Box~500, Batavia,~IL~60510, USA}}

\newcommand{\UoC}{\affiliation{Enrico~Fermi~Institute and Kavli~Institute~for~Cosmological~Physics, Department~of~Physics, University~of~Chicago, Chicago,~IL~60637, USA}}

\newcommand{\ANL}{\affiliation{HEP~Division, Argonne~National~Laboratory, 9700~Cass~Ave., Argonne,~IL~60439, USA}}

\newcommand{\WS}{\affiliation{Department~of~Physics~\&~Astronomy, Wayne~State~University, Detroit,~MI~48201, USA}}

\author{Sebastian~Baum}
\email{sbaum@stanford.edu}
\SU

\author{Marcela~Carena}
\email{carena@fnal.gov}
\FNL
\UoC

\author{Nausheen~R.~Shah}
\email{nausheen.shah@wayne.edu}
\WS

\author{Carlos~E.~M.~Wagner}
\email{cwagner@uchicago.edu}
\ANL
\UoC

\preprint{EFI-21-3}
\preprint{FERMILAB-PUB-21-184-T}
\preprint{WSU-HEP-2101}

\begin{abstract}
A new measurement of the muon anomalous magnetic moment, $g_\mu-2$, has been reported by the Fermilab Muon g-2 collaboration and shows a $4.2\,\sigma$ departure from the most precise and reliable calculation of this quantity in the Standard Model. Assuming that this discrepancy is due to new physics, we concentrate on a simple supersymmetric model that also provides a dark matter explanation in a previously unexplored region of supersymmetric parameter space. Such interesting region can realize a Bino-like dark matter candidate compatible with all current direct detection constraints for small to moderate values of the Higgsino mass parameter $|\mu|$. This in turn would imply the existence of light additional Higgs bosons and Higgsino particles within reach of the high-luminosity LHC and future colliders. We provide benchmark scenarios that will be tested in the next generation of direct dark matter experiments and at the LHC.
\end{abstract}

\maketitle

\section{Introduction}

The Standard Model (SM) of particle physics has built its reputation on decades of measurements at experiments around the world that testify to its validity. With the discovery of the Higgs boson almost a decade ago~\cite{Aad:2012tfa,Chatrchyan:2012ufa} all SM particles have been observed and the mechanism that gives mass to the SM particles, with the possible exception of the neutrinos, has been established. Nonetheless, we know that physics beyond the SM~(BSM) is required to explain the nature of dark matter~(DM) and the source of the observed matter-antimatter asymmetry. Furthermore, an understanding of some features of the SM such as the hierarchy of the fermion masses or the stability of the electroweak vacuum is lacking. 

The direct discovery of new particles pointing towards new forces or new symmetries in nature will be the most striking and conclusive evidence of BSM physics. However, it may well be the case that BSM particles lie beyond our present experimental reach in mass and/or interaction strength, and that clues for new physics may first come from results for precision observables that depart from their SM expectations. With that in mind, since the discovery of the Higgs boson, we are straining our resources and capabilities to measure the properties of the Higgs boson to higher and higher accuracy, and flavor and electroweak physics experiments at the LHC and elsewhere are pursuing a complementary broad program of precision measurements. Breakthroughs in our understanding of what lies beyond the SM could occur at any time. 

Recently, new results of measurements involving muons have been reported. The LHCb experiment has reported new values of the decay rate of $B$-mesons to a kaon and a pair of muons compared to the decay into a kaon and electrons~\cite{Aaij:2021vac}, providing evidence at the $3\,\sigma$-level of the violation of lepton universality. This so-called $R_K$ anomaly joins the ranks of previously reported anomalies involving heavy-flavor quarks such as the bottom quark forward-backward asymmetry at LEP~\cite{ALEPH:2005ab,ALEPH:2010aa}, and measurements of meson decays at the LHC and $B$-factories such as $R_{K^*}$~\cite{Aaij:2017vbb,Aaij:2020nrf,LHCb:2020gog} and $R_{D^{(*)}}$~\cite{Lees:2012xj,Aaij:2015yra,Huschle:2015rga,Hirose:2016wfn,Aaij:2017uff,Abdesselam:2019dgh}. The Fermilab Muon (g-2) experiment has just reported a new measurement of the anomalous magnetic moment of the muon, $a_\mu \equiv \left( g_{\mu} - 2 \right)/2$. The SM prediction of $a_\mu$ is known with the remarkable relative precision of $4 \times 10^{-8}$, $a_\mu^{\rm SM} = 116~591~810(43) \times 10^{-11}$~\cite{Aoyama:2020ynm,Aoyama:2012wk,Aoyama:2019ryr,Czarnecki:2002nt,Gnendiger:2013pva,Davier:2017zfy,Keshavarzi:2018mgv,Colangelo:2018mtw,Hoferichter:2019mqg,Davier:2019can,Keshavarzi:2019abf,Kurz:2014wya,Melnikov:2003xd,Masjuan:2017tvw,Colangelo:2017fiz,Hoferichter:2018kwz,Gerardin:2019vio,Bijnens:2019ghy,Colangelo:2019uex,Blum:2019ugy,Colangelo:2014qya}. From the new Fermilab Muon (g-2) experiment, the measured value is $a_\mu^{\rm exp,~FNAL} = 116~592~040(54) \times 10^{-11}$~\cite{Abi:2021gix}, which combined with the previous E821 result $a_\mu^{\rm exp,~E821} = 116~592~089(63) \times 10^{-11}$~\cite{Bennett:2006fi}, yields a value $a_\mu^{\rm exp} = 116~592~061(41) \times 10^{-11} $.

An important point when considering the tension between experimental results and the SM predictions are the current limitations on theoretical tools in computing the hadronic vacuum polarization (HVP) contribution to $a_\mu^{\rm SM}$, which is governed by the strong interaction and is particularly challenging to calculate from first principles. The most accurate result of the HVP contribution is based on a data-driven result, extracting its value from precise and reliable low-energy ($e^+ e^- \to {\rm hadrons}$) cross section measurements via dispersion theory. Assuming no contribution from new physics to the low energy processes and conservatively accounting for experimental errors, this yields a value $a_\mu^{\rm HVP} = 685.4(4.0) \times10^{-10}$~\cite{Aoyama:2020ynm,Davier:2017zfy,Keshavarzi:2018mgv,Colangelo:2018mtw,Hoferichter:2019mqg,Davier:2019can,Keshavarzi:2019abf,Kurz:2014wya}, implying an uncertainty of 0.6\,\% in this contribution.\footnote{The HVP contribution has recently been computed in lattice QCD, yielding a higher value of $a_\mu^{\rm HVP}=708.7(5.3) \times 10^{-10}$~\cite{Borsanyi:2020mff}. Given the high complexity of this calculation, independent lattice calculations with commiserate precision are needed before confronting this result with the well tested data-driven one. We stress that if a larger value of the HVP contribution were confirmed, which would (partially) explain the $(g_{\mu}-2)$ anomaly, new physics contributions will be needed to bring theory and measurements of ($e^+ e^- \to {\rm hadrons}$) in agreement~\cite{Lehner:2020crt,Crivellin:2020zul,Keshavarzi:2020bfy,deRafael:2020uif}.}
The SM prediction for the anomalous magnetic moment of the muon and the measured value then differ by $4.2\,\sigma$,
\begin{equation} \label{eq:Damu}
   \Delta a_\mu \equiv (a_\mu^{\rm exp} - a_\mu^{\rm SM}) = \left(251 \pm 59\right) \times 10^{-11} \,.
\end{equation}

It is imperative to ask what these anomalies may imply for new physics. The most relevant questions that come to mind are: Can the $a_\mu$ and $R_{K^{(*)}}$ anomalies be explained by the same BSM physics? Can they give guidance about the nature of DM? Are they related to cosmological discrepancies? How constrained are the possible solutions by other experimental searches? What are future experimental prospects for the possible solutions? 

In Sec.~\ref{sec:ModelReview} we provide a brief overview of the many models which have been previously proposed in the literature to explain the $(g_{\mu}-2)$ anomaly and consider their impact on other possible anomalies and on unresolved questions of the SM. Then, in Sec.~\ref{sec:MSSM}, we discuss a supersymmetric solution in the most simplistic supersymmetric model at hand, the Minimal Supersymmetric Standard Model (MSSM). We focus on a region of the parameter space of the MSSM where the $(g_{\mu}-2)$ anomaly can be realized simultaneously with a viable DM candidate. We show that in the region of moderate $\left|\mu\right|$ and moderate-to-large values of $\tan\beta$, a Bino-like DM candidate can be realized in the proximity of {\it blind spots}~(that require $ \mu \times M_1 <0$) for spin-independent direct detection (SIDD) experiments~\cite{Huang:2014xua}. In this way, our MSSM scenario explores a different region of parameter space than the one considered in the study of Refs.~\cite{Chakraborti:2020vjp,Chakraborti:2021kkr}, which considers regions of large $\mu$ as a way to accommodate current SIDD bounds. We summarize and conclude in Sec.~\ref{sec:Conclusions}. In Appendix~\ref{app:LHCconstraints}, we give details about the LHC constraints on these scenarios.

\section{$\boldsymbol{(g_{\mu}-2)}$ connections to cosmic puzzles and the LHC} \label{sec:ModelReview}
In order to bridge the gap between the SM prediction and the measured value for the anomalous magnetic moment of the muon, a BSM contribution of order $\Delta a_\mu = (20$--$30) \times 10^{-10}$ is needed. Taking the $a_\mu$ anomaly as a guidance for new physics, it is natural to ask how it can be connected to other anomalies, specially those in the muon sector, or to solving puzzles of our universe's early history. There are two broad classes of solutions to the ($g_{\mu} - 2$) anomaly that may be considered in the light of the above:
\begin{itemize}
   \item New relatively light particles with small couplings to muons, typically featuring particles with $\mathcal{O}(100)\,$MeV masses and $\mathcal{O}(10^{-3})$ couplings to muons. Examples of such models we will discuss here are new (light) scalars and new (light) ($Z'$) vector bosons. These new light particles may have left important clues in the cosmos.

   \item New heavy fermions or scalars (possibly accompanied by additional new particles), as well as leptoquark particles, with larger couplings to muons. Similar solutions appear also in supersymmetric extensions of the SM that we shall discuss separately in some detail in Sec.~\ref{sec:MSSM}. In addition, new gauge symmetries, spontaneously broken at low energies, can induce $Z'$ vector bosons with masses comparable to the electroweak scale and $\mathcal{O}(1)$ couplings to muons. These types of new particles can be sought for at the LHC and other terrestrial experiments.
\end{itemize}

The most recent LHCb measurement~\cite{Aaij:2021vac}, $R_K = {\rm BR}(B \rightarrow K \mu^+ \mu^-) / {\rm BR}(B \rightarrow K e^+ e^-) = 0.846^{+0.044}_{-0.041}$ in the kinematic regime of $1.1\,{\rm GeV}^2 \leq q^2 \leq 6.0\,{\rm GeV}^2 $ implies a violation of lepton universality and differs from the SM expectation at the $3.1\,\sigma$ level. Since $R_K$ also involves muons, it naturally appears related to the $(g_\mu-2)$ anomaly. However, as we shall discuss, it is hard to simultaneously fit both $R_K$ and $(g_\mu-2)$. 

{\bf Scalar solutions:} This is perhaps the simplest scenario for the explanation of the observed $\Delta a_\mu$. A scalar particle, with mass $\lesssim 200\,$MeV and couplings to muons of similar size as the corresponding SM-Higgs coupling, can lead to a satisfactory explanation of $\Delta a_\mu$~\cite{Zhou:2001ew,Barger:2010aj,Chen:2015vqy,Batell:2016ove,Davoudiasl:2018fbb,Liu:2018xkx}. One can construct models with such a scalar particle and suppressed couplings to other leptons or quarks in a straightforward way~\cite{Liu:2018xkx}. Alternatively, one can construct models with appropriate values of the couplings of the new scalar to quarks to lead to an explanation of some flavor anomalies, for example the KOTO anomaly~\cite{Ahn:2020opg}, but the constraints tend to be more severe and the model-building becomes more involved~\cite{Liu:2020qgx}. It is important to stress that it proves impossible to fully explain the $R_K$ anomaly with scalars without violating $B_s \rightarrow \mu^+ \mu^-$ measurements~\cite{Aaij:2017vad}; see, for example, Ref.~\cite{Altmannshofer:2021qrr}.

A pseudoscalar particle may also lead to an explanation of $\Delta a_\mu$, provided it couples not only to muons, but also to photons. The typical example are axion-like particles~\cite{Marciano:2016yhf,Bauer:2019gfk}, although obtaining the proper $\Delta a_\mu$ requires a delicate interplay between the muon and photon couplings.\footnote{A similar mechanism applies for $(g_e-2)$ in the case of the QCD axion; see, for instance, Ref.~\cite{Liu:2021wap}} Alternatively, a positive contribution to $a_\mu$ can arise from a two loop Barr-Zee diagram mediated by the pseudoscalar couplings to heavier quarks and leptons~\cite{Gunion:2008dg,Wang:2018hnw}. 

{\bf Fermionic solutions:} Another interesting solution occurs in the case of vector-like leptons, which may induce a contribution to $a_\mu$ via gauge boson and Higgs mediated interactions~\cite{Kannike:2011ng,Dermisek:2013gta}. Note that the mixing between the SM leptons and the new heavy leptons must be carefully controlled to prevent dangerous flavor-changing neutral currents in the lepton sector. A recent analysis shows that consistency with the measured values of $\Delta a_\mu$ may be obtained for vector-like leptons with masses of the order of a few TeV~\cite{Dermisek:2021ajd}.\footnote{See Ref.~\cite{Megias:2017dzd} for an attempt to adress both $a_\mu$ and $R_{K^*}$ in a vector-like lepton model with extra dimensions.} 

{\bf Leptoquark solutions:} This is one of the most interesting solutions to $\Delta a_\mu$, since it can also lead to an explanation of the $R_K$ anomaly; see, for example, Refs.~\cite{Bauer:2015knc,Crivellin:2020tsz,Crivellin:2020mjs,Hiller:2021pul}. A directly related and particularly attractive realization arises in R-parity violating supersymmetry, which enables the same type of interactions as a leptoquark theory; see, for example, Ref.~\cite{Altmannshofer:2020axr}. This solution requires the scalar partner of the right-handed bottom quark to have masses of a few TeV, which may be tested at future LHC runs. Similar to the vector-like lepton scenarios, a careful choice of the leptoquark couplings is necessary to avoid flavor-changing neutral currents. This tuning is perhaps the least attractive feature of such scenarios, although it may be the result of symmetries~\cite{Hiller:2021pul}.

{\bf Gauge boson solutions:} New gauge bosons coupled to muons are an attractive solution to the $a_\mu$ anomaly, since they can be incorporated in an anomaly-free framework that can also lead to an explanation of the $R_{K^{(*)}}$ anomalies. Of particular interest is the gauged $(L_\mu - L_\tau)$ scenario~\cite{Heeck:2011wj}, since it avoids the coupling to electrons.\footnote{Models with $(L_\mu + L_\tau)$ give an intriguing connection to a novel mechanism of electroweak baryogenesis with CP-violation triggered in a dark sector that allows for a suitable DM candidate~\cite{Carena:2018cjh,Carena:2019xrr}. Unfortunately, solutions to $(g_{\mu}-2)$ in this appealing scenario are ruled out by $(B\rightarrow K\mu^+\mu^-)$ constraints due to contributions from the anomalous $WWZ'$ coupling.} The $R_{K^{(*)}}$ anomalies may be explained by the addition of vector-like quarks that mix with the second and third generation SM quarks~\cite{Altmannshofer:2014cfa,Altmannshofer:2016oaq,Altmannshofer:2016jzy}, connecting the ($L_\mu - L_\tau$) gauge boson to baryons. A common explanation of both $R_{K^{(*)}}$ and $a_\mu$ is, however, strongly constrained by neutrino trident bounds on $Z'$ bosons coupled to muons~\cite{Altmannshofer:2014pba,Geiregat:1990gz,Mishra:1991bv}.\footnote{There are also bounds from Coherent $\nu$-Nucleus Scattering (CE$\nu$NS), although these are not yet competitive with the bounds from neutrino trident processes~\cite{Abdullah:2018ykz,Amaral:2020tga}.} In addition, bounds from BaBar~\cite{TheBABAR:2016rlg} and CMS~\cite{Sirunyan:2018nnz} from [$e^+ e^-/pp \to \mu^+ \mu^- + (Z' \to \mu^+ \mu^-)$] rule out the values of the new gauge coupling which could explain the observed value of $a_\mu$ for $m_{Z'} \geq 2 m_\mu \simeq 210\,$MeV. Due to these experimental constraints, explaining the $\Delta a_\mu$ anomaly with a light new gauge boson requires $m_{Z'} \lesssim 200\,$MeV. Explanations of the flavor anomalies require larger gauge boson masses, preventing simultaneous explanations of $R_{K^{(*)}}$ and $a_\mu$.

It is interesting to note that explanations of the $(g_\mu-2)$ anomaly via gauged $(L_\mu - L_\tau)$ may have a relation to some of the cosmological puzzles, in particular the tensions of the late and early time determinations of the Hubble constant, $H_0$~\cite{Escudero:2019gzq,Amaral:2020tga}. In the $m_{Z'} \sim 10\,$MeV region, the effective number of degrees of freedom can be enhanced by $\Delta N_{\rm eff} \approx 0.2$, alleviating the $H_0$-tension. Note that constraints from solar neutrino scattering in Borexino~\cite{Harnik:2012ni,Bilmis:2015lja,Amaral:2020tga} and $\Delta N_{\rm eff}$ bounds~\cite{Escudero:2019gzq} rule out the couplings preferred by the $a_\mu$ anomaly for $m_{Z'} \lesssim 5\,$MeV.

Before considering minimal supersymmetric scenarios for the $(g_{\mu}-2)$ anomaly in some detail, let us summarize the discussion above as follows: 1) All the above solutions, with a broad range of masses and couplings of the new particles, can readily explain the $(g_{\mu}-2)$ anomaly, but it is difficult to simultaneously accommodate the $R_{K^{(*)}}$ anomalies. This difficulty mainly arises from experimental constraints.\footnote{See also Refs.~\cite{Yin:2020afe,Capdevilla:2021rwo} for prospects of probing models adressing the ($g_\mu - 2$) anomaly at high energy muon colliders.} In the rare examples of models where both solutions can be accommodated simultaneously, it is only possible at the cost of significant tuning of the parameters. 2) In most scenarios, a DM candidate can be included in the model (with different levels of complexity). However, there does not appear to be a compelling connection offering a unique guidance for model building. On the other hand, in low-energy SUSY models with R-parity conservation, an explanation of the $(g_{\mu}-2)$ anomaly is naturally connected to the presence of a DM candidate and other new particles within the reach of the (HL-)LHC and future colliders. We explore this possibility in its simplest realization in the next section.

\section{Tiny ($\boldsymbol{g_{\mu}-2}$) Muon Wobble with Small $\boldsymbol{|\mu|}$ in the MSSM} \label{sec:MSSM}
Supersymmetric extensions of the SM remain among the most compelling BSM scenarios~\cite{Nilles:1983ge,Haber:1984rc,Martin:1997ns}, not least because the stability of the Higgs mass parameter under quantum corrections can be ensured. In minimal supersymmetric extensions of the SM, the SM-like Higgs is naturally light~\cite{Casas:1994us,Carena:1995bx,Carena:1995wu,Haber:1996fp,Degrassi:2002fi,Bagnaschi:2014rsa,Draper:2013oza,Lee:2015uza,Vega:2015fna,Bahl:2017aev,Slavich:2020zjv} and the corrections to electroweak precision as well as flavor observables tend to be small, leading to good agreement with observations. Supersymmetric extensions can also lead to gauge coupling unification and provide a natural DM candidate, namely the lightest neutralino. 

In this section, we propose simultaneous ($g_\mu-2$) and DM solutions in the Minimal Supersymmetric Standard Model (MSSM)~\cite{Nilles:1983ge,Haber:1984rc,Martin:1997ns} which have not been explored before. Related recent (but prior to the publication of the Fermilab Muon (g-2) result) studies can, for example, be found in Refs.~\cite{Okada:2016wlm,Cox:2018qyi,Carena:2018nlf,Endo:2019bcj,Badziak:2019gaf,Abdughani:2019wai,Chakraborti:2020vjp,Chakraborti:2021kkr}. One crucial difference between our study and the very recent work in Refs.~\cite{Chakraborti:2020vjp,Chakraborti:2021kkr} is that the spin-independent direct detection (SIDD) cross section is suppressed not by decoupling the Higgsino and heavy Higgs contributions, but by a partial cancellation between the amplitudes mediated by the two neutral CP-even Higgs boson mass eigenstates. This cancellation requires opposite signs of the Higgsino and the Bino mass parameters, $(\mu \times M_1) < 0$~\cite{Huang:2014xua}. Demonstrating that one can explain the $a_\mu$ anomaly in this region of parameter space is non-trivial, as this combination of the Higgsino and Bino mass parameters renders the contribution of the neutralino-smuon loop to $a_\mu$ negative, while the experimentally observed value is {\it larger} than the SM prediction. Explaining the experimental measurement is only possible if the chargino-sneutrino contribution to $a_\mu$ is positive and has larger absolute magnitude than the neutralino-smuon contribution, and if the values of the individual contributions are such that the observed anomaly, $\Delta a_\mu = (20-30) \times 10^{-10}$, can be explained. Moreover, this can only be achieved for moderate (absolute) values of the Higgsino mass parameter $|\mu| \lesssim 500\,$GeV, and values of the heavy Higgs boson masses than are not far away from the current experimental limit coming from direct searches.

\subsection{$\boldsymbol{\Delta a_\mu}$ and Direct Dark Matter Detection Constraints}
The MSSM contributions to $a_\mu$ have been discussed extensively in the literature, see, for example, Refs.~\cite{Barbieri:1982aj,Ellis:1982by,Kosower:1983yw,Moroi:1995yh,Carena:1996qa,Feng:2001tr,Martin:2001st,Badziak:2019gaf}. The most important contributions arise via chargino-sneutrino and neutralino-smuon loops, approximately described by~\cite{Badziak:2019gaf} 
\begin{align} \label{eq:amu_cha}
   a_\mu^{\widetilde{\chi}^\pm - \widetilde{\nu}_\mu} &\simeq \frac{\alpha m_\mu^2 \mu M_2 \tan\beta}{4 \pi \sin^2 \theta_W m_{\widetilde{\nu}_\mu}^2} \left[ \frac{ f_{\chi^\pm}\left( M_2^2 / m_{\widetilde{\nu}_\mu}^2 \right) - f_{\chi^\pm}\left( \mu^2 / m_{\widetilde{\nu}_\mu}^2 \right) }{ M_2^2 - \mu^2 } \right] \;, \\
   a_\mu^{\widetilde{\chi}^0 - \widetilde{\mu}} &\simeq \frac{\alpha m_\mu^2 M_1 \left( \mu\tan\beta - A_\mu \right)}{4 \pi \cos^2 \theta_W \left( m_{\widetilde{\mu}_R}^2 - m_{\widetilde{\mu}_L}^2 \right)} \left[ \frac{ f_{\chi^0}\left( M_1^2 / m_{\widetilde{\mu}_R}^2 \right)}{m_{\widetilde{\mu}_R}^2 } - \frac{f_{\chi^0}\left( M_1^2 / m_{\widetilde{\mu}_L}^2 \right) }{m_{\widetilde{\mu}_L}^2 } \right] \;, \label{eq:amu_neu}
\end{align}
where $M_2$ is the Wino mass parameter and $m_{\widetilde{f}}$ are the scalar particle $\widetilde{f}$ masses, with the loop functions
\begin{align}
   f_{\chi^\pm}(x) &= \frac{x^2 - 4 x + 3 + 2 \ln (x)}{\left( 1-x \right)^3} \;, \\
   f_{\chi^0}(x) &= \frac{x^2 - 1 - 2 x \ln (x)}{\left( 1-x \right)^3} \;;
\end{align}
see Refs.~\cite{Moroi:1995yh,Martin:2001st} for the full (one-loop) expressions. It is interesting to note that these two contributions can be of the same order of magnitude: The chargino-sneutrino contribution is proportional to Higgsino-Wino mixing which can be sizeable, but suppressed by the smallness of the Higgsino-sneutrino-muon coupling which is proportional to the muon Yukawa coupling, $\propto m_\mu \tan\beta/v$, with the SM Higgs vacuum expectation value $v$. The neutralino-smuon contribution, on the other hand, arises via muon-smuon-neutralino vertices which are proportional to the gauge couplings, but is suppressed by the small smuon left-right mixing, $\propto m_\mu (\mu \tan\beta - A_\mu)/(m_{\widetilde{\mu}_R}^2 - m_{\widetilde{\mu}_L}^2)$. Regarding corrections beyond one-loop~\cite{Marchetti:2008hw,Athron:2015rva}, the most relevant contribution is associated with corrections to the muon Yukawa coupling, $\Delta_\mu$. These corrections become relevant at large values of $\mu\tan\beta$ and can be re-summed at all orders of perturbation theory~\cite{Carena:1999py}. While these corrections lead to small modifications of $a_\mu$, they do not change the overall dependence of $\Delta a_\mu$ on the masses of the supersymmetric particles. 

From Eqs.~\eqref{eq:amu_cha}--\eqref{eq:amu_neu} we can observe that the signs of the MSSM contributions to $a_\mu$ depend sensitively on the relative signs of the gaugino masses $M_1$ and $M_2$ and the Higgsino mass parameter $\mu$. As emphasized before, a DM candidate compatible with the current null-results from direct detection experiments can be realized for $|\mu| \lesssim 500$\,GeV if $M_1$ and $\mu$ have opposite signs. For this combinations of signs, the contribution from the neutralino-smuon loop to $a_\mu$ will be negative, $a_\mu^{\widetilde{\chi}^0 - \widetilde{\mu}} < 0$. Since the measured value of $a_\mu$ is {\it larger} than the SM prediction by $\Delta a_\mu \simeq 25 \times 10^{-10}$, we require the chargino-sneutrino contribution to be positive and larger than the neutralino-smuon contribution. This can be realized if $M_2$ has the same sign as $\mu$ and if $|M_2|$ is of similar size as $|\mu|$ and the soft smuon masses. In the regime of moderate or large values of $\tan\beta$, and assuming all weakly interacting sparticles have masses of the same order, $\widetilde{m}$, one obtains approximately 
\begin{equation} \label{eq:approxamu}
   \Delta a_\mu \simeq 1.3 \times 10^{-9} \tan\beta \times \left(\frac{100\,{\rm GeV}}{\widetilde{m}} \right)^2 \;. 
\end{equation}
The factor 1.3 reduces to values closer to 1 if $M_1$ and $M_2$ have opposite signs. This implies that for values of $\tan\beta \simeq 10$, sparticles with masses $\widetilde{m} \sim 200\,$GeV can lead to an explanation of the observed $\Delta a_\mu$ anomaly, while for $\tan\beta = 60$, the characteristic scale of the weakly interacting sparticle masses may be as large as $\widetilde{m} \sim 500$\,GeV. 

The range of $\tan\beta$ and of sparticle masses consistent with the observed $\Delta a_\mu$ has implications on the DM properties. We will concentrate on DM candidates with masses comparable to the weak scale, such that the thermal DM relic density reproduces the observed value. In the MSSM, DM candidates in this mass range can be realized if the lightest supersymmetric particle is an almost-pure Bino, $m_\chi \simeq |M_1|$.

For the moderate-to-large values of $\tan\beta$ required to explain the $(g_\mu-2)$ anomaly, the SIDD amplitude for the scattering of DM with nuclei ($N$) is proportional to 
\begin{equation}
   {\mathcal M}_p^{\rm SI} \propto \frac{v}{\mu^2}\left[2 \frac{( M_1 + \mu \sin 2 \beta)}{m_h^2} - \frac{\mu \cos 2\beta}{m_H^2} \tan\beta \right],
   \label{eq:approxSI}
\end{equation}
where $m_h$ and $m_H$ are the masses of the SM-like and the new heavy neutral Higgs boson, respectively. We see that the SIDD amplitude depends in a crucial way on the sizes and signs of $M_1$ and $\mu$. There are two options to lower the SIDD amplitude: For large values of $|\mu|$, the Higgsino components of the DM candidate become small and the SIDD amplitude is suppressed. Alternatively, the light and heavy CP-even Higgs contributions (first and second terms inside the brackets in Eq.~\eqref{eq:approxSI}) may interfere destructively, leading to a suppression of the SIDD amplitude. The latter option is particularly interesting since it allows $|\mu|$ to remain of the order of the electroweak scale; see, for example Ref.~\cite{Drees:2021pbh} for a recent discussion of naturalness and the connection with direct detection bounds.

Regarding the first term in Eq.~\eqref{eq:approxSI}, if $M_1 \simeq -\mu\sin2\beta$, the contributions of the Higgsino-up and the Higgsino-down admixtures to the $(\chi\chi h)$ interaction cancel. The second term is the contribution to the $(\chi N \to \chi N)$ amplitude arising from the $t$-channel exchange of the non-SM-like heavy Higgs boson $H$. The {\it generalized blind spot} condition for the SIDD cross section of a Bino-like DM candidate is then~\cite{Huang:2014xua}
\begin{equation} \label{eq:gBS}
   \frac{2 \left( M_1 + \mu \sin2\beta \right)}{m_h^2} \approx \frac{\mu \tan\beta \cos 2\beta}{m_H^2} \;.
\end{equation}
If the condition in Eq.~\eqref{eq:gBS} is satisfied, the amplitudes mediated by $h$ and by $H$ exchange interfere destructively, suppressing the SIDD cross section; a property that also holds at the one-loop level~\cite{Han:2018gej}. In general, even if one is not in the proximity of the blind spot solution, if the neutralino is mostly Bino-like, for a given value of $|\mu|$ and $M_1$, the cross section is suppressed (enhanced) if $\mu$ and $M_1$ have opposite (the same) sign.\footnote{Note that $\cos(2\beta) = (1- \tan^2\beta) / (1+\tan^2\beta) \simeq -1$ for moderate-to-large values of $\tan\beta$.} 

The mass of the heavy Higgs boson plays an important role in the blind-spot cancellation. In the presence of light electroweakinos, the current LHC bounds on $m_H$ coming from searches for heavy Higgs bosons decaying into tau-leptons~\cite{Aaboud:2017sjh,Sirunyan:2018zut,Sirunyan:2019xjg,Aad:2020zxo} can be approximated by
\begin{equation}
   m_H \gtrsim 250\,{\rm GeV} \times \sqrt{\tan\beta}\sim 2 ~m_h \sqrt{\tan\beta} ~.
   \label{eq:mHbound}
\end{equation}
For values of $m_H$ close to this bound, the SIDD amplitude is proportional to
\begin{equation}
   {\mathcal M}_p^{\rm SI} \propto \frac{M_1 v}{\mu^2} \left[ 1 + \frac{\mu}{2 M_1} \left( \frac{4}{\tan\beta} + \frac{1}{4} \right) \right] . 
\label{eq:lowestSIDD}
\end{equation} 
To exemplify the relevance of the relative sign and size of $\mu$ and $M_1$, consider ${\mathcal M}_p^{\rm SI}$ for $\tan\beta = 16$. As a reference value for the SIDD amplitude, let us set $\mu \simeq -M_1$. Keeping $M_1$ fixed, but increasing the value of $\left|\mu\right|$ to $\mu \simeq -2 M_1$, the value of ${\mathcal M}_p^{\rm SI}$ becomes a factor of $\approx 1/6$ smaller. Let us compare this to the situation for which $\mu$ and $M_1$ have the same sign. First, we can note that for $\mu = M_1$, the SIDD amplitude is almost a factor 2 larger than for $\mu = -M_1$. Furthermore, in order to obtain a reduction of ${\mathcal M}_p^{\rm SI}$ by a factor of $1/6$, one would have to raise the value of $\left|\mu\right|$ from $\mu \sim M_1$ to $\mu \sim 4 M_1$. This exemplifies that obtaining SIDD cross sections compatible with experimental limits either requires $(\mu M_1) < 0$ (blind spot solution) or, to compensate for a positive sign of this product, one must sufficiently enhance the ratio $\mu/ M_1$ (large-$\mu$ solution).

The spin dependent~(SD) interactions are instead dominated by $Z$-exchange, and can only be suppressed by lowering the Higgsino component of the lightest neutralino. At moderate or large values of $\tan\beta$, the amplitude for SD interactions is proportional to~\cite{Carena:2018nlf}
\begin{equation}
   {\cal{M}}^{\rm SD} \propto \left( \frac{v}{\mu} \right)^2 \cos 2\beta \;.
   \label{eq:lowestSDDD}
\end{equation}
Comparison with the results from direct detection experiments~\cite{Behnke:2016lsk,Fu:2016ega,Aprile:2019dbj,Amole:2019fdf} leads to an approximate bound on $\mu$,
\begin{equation}
   |\mu| \gtrsim 300\,{\rm GeV}\; ,
 \end{equation} 
with a mild dependence on $M_1$. 

\begin{figure}
   \centering
   \includegraphics[scale=0.7]{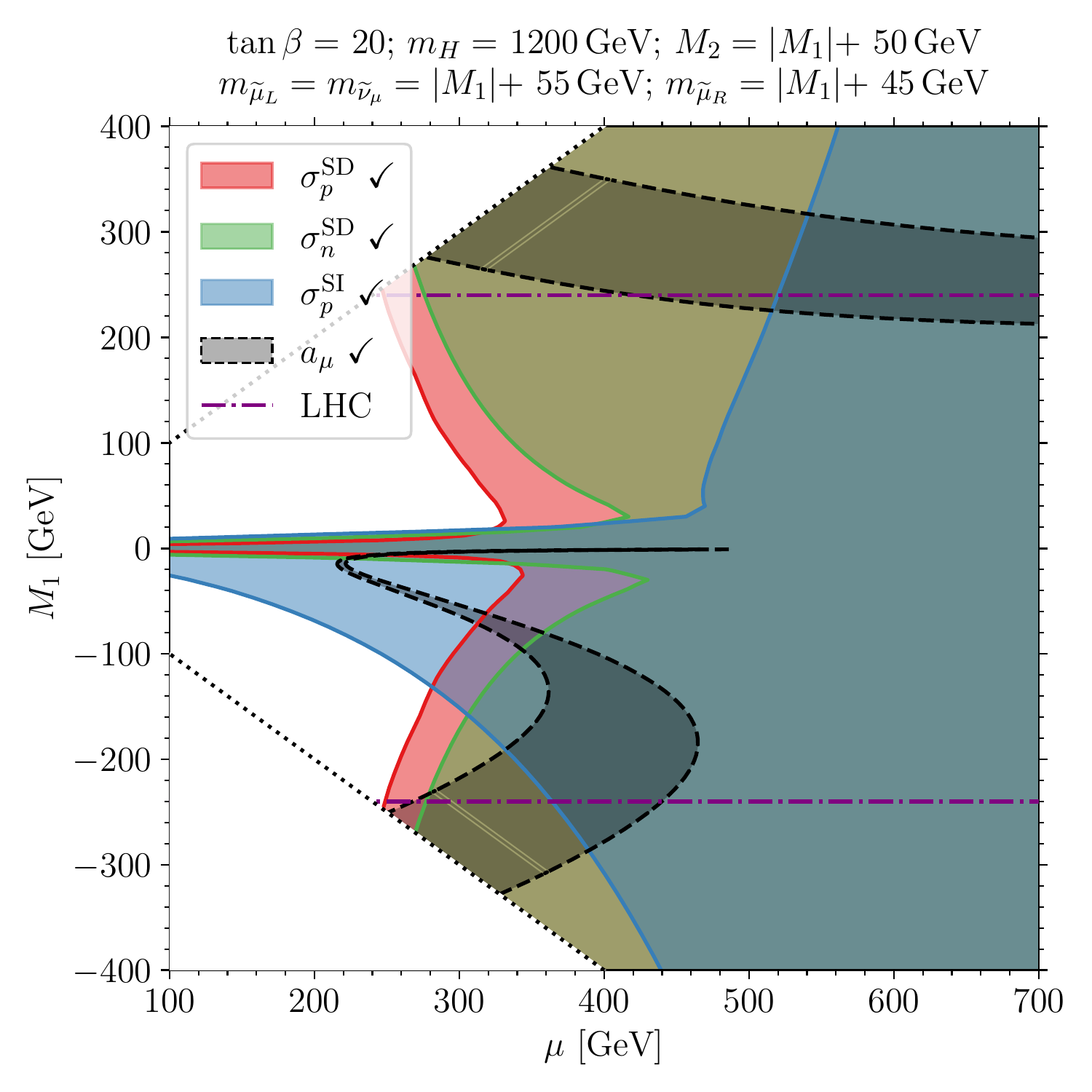}
   \caption{The colored shades show approximate regions in the $\mu$--$M_1$ parameter plane allowed by current DM direct detection constraints on the spin-dependent WIMP-proton, spin-dependent WIMP-neutron, and spin-independent WIMP-nucleon cross section for $\tan\beta = 20$ and values of the slepton, Higgs and Wino mass parameters as indicated in the plot. In the gray areas bounded by the dashed black lines we find a MSSM contribution $\Delta a_\mu = (25.1 \pm 5.9) \times 10^{-10}$, explaining the value observed by the Fermilab and Brookhaven Muon (g-2) experiments. The dash-dotted purple lines indicate constraints arising from tau-leptons $+ {\rm missing~transverse~energy} (+ {\rm jet})$ searches at the LHC, applicable if the mass of the lightest stau is approximately in the middle of the lightest chargino and neutralino masses~\cite{CMS:2019zmn}.
   }
   \label{fig:DDconstraintstb15}
\end{figure}

\begin{figure}
   \centering
   \includegraphics[scale=0.7]{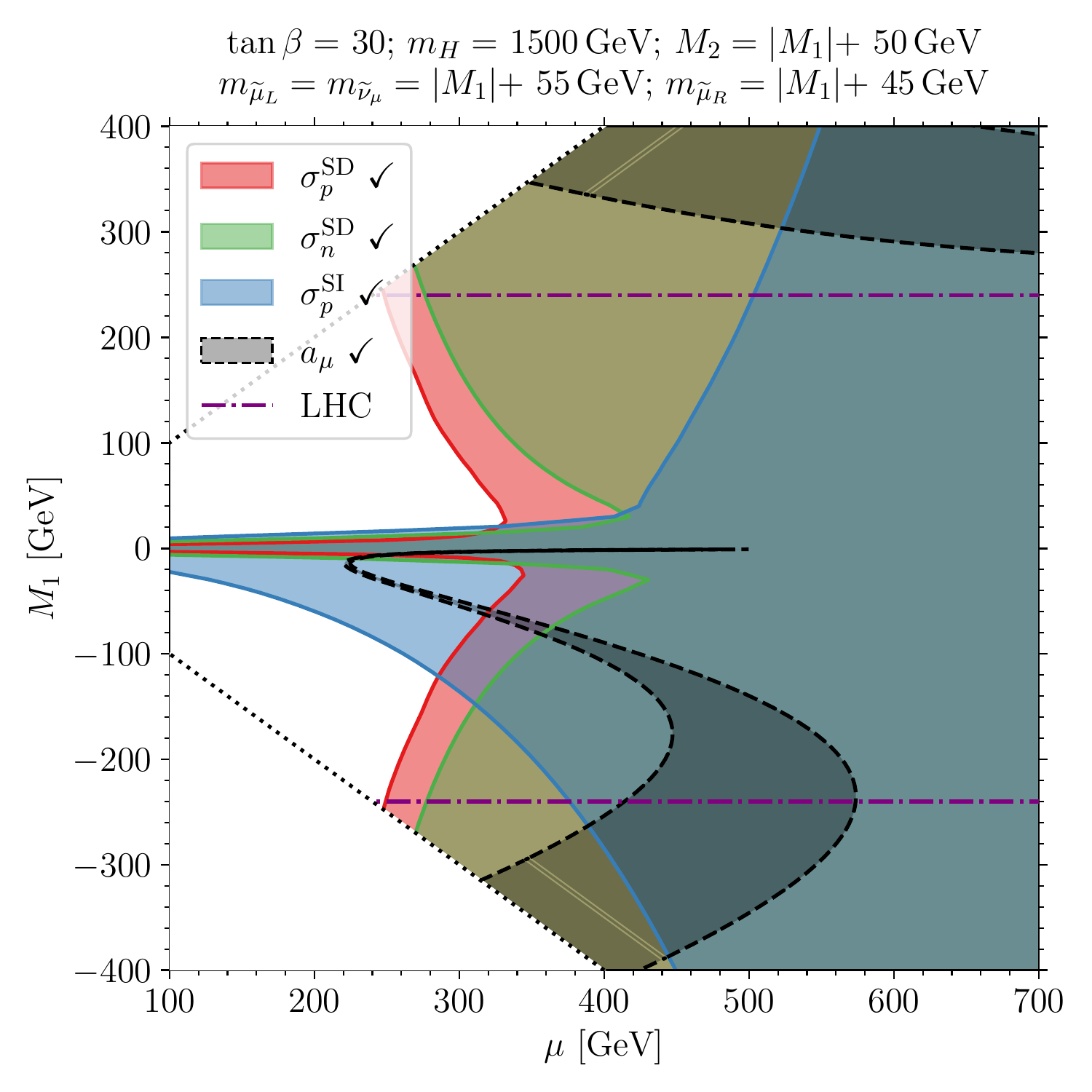}
   \caption{Same as Fig.~\ref{fig:DDconstraintstb15} but for $\tan\beta = 30$ and $m_H = 1500$\,GeV. }
   \label{fig:DDconstraintstb30}
\end{figure}

To summarize this discussion, we show the qualitative behavior of the direct detection cross sections in Figs.~\ref{fig:DDconstraintstb15} and~\ref{fig:DDconstraintstb30} in the $M_1$--$\mu$ plane. We use approximate analytic expressions for the cross sections and set the masses of the heavy Higgs boson and $\tan\beta$ to characteristic values. The values of $M_2$ and the slepton masses have been chosen to lead to a compressed spectrum, alleviating constraints from slepton and chargino searches at the LHC, see, for example, Refs.~\cite{Aad:2014vma,Aad:2019vnb,CMS:2019eln,Aad:2019byo,Aad:2019qnd,Aaboud:2017nhr,Sirunyan:2018ubx,Aaboud:2018jiw,Aaboud:2018sua,Aaboud:2018ngk,ATLAS-CONF-2019-008,ATLAS:2021moa}. The regions shaded in the different colors denote the region allowed by current direct detection constraints on the SD-proton~\cite{Behnke:2016lsk,Amole:2019fdf}, SD-neutron~\cite{Fu:2016ega,Aprile:2019dbj}, and SI~\cite{Angloher:2015ewa,Agnes:2018ves,Aprile:2018dbl,Aprile:2019xxb} scattering cross section. We see that whereas the SD constraints provide an approximately symmetric lower bound on $\mu$, due to the SIDD constraints, the values of $|\mu|$ need to be significantly larger for positive $\mu \times M_1$ than for negative $\mu \times M_1$. We show the region where the MSSM contribution explains the $(g_\mu-2)$ anomaly in Figs.~\ref{fig:DDconstraintstb15} and~\ref{fig:DDconstraintstb30} with the gray shade bounded by the dashed black line. The shape of the region preferred by $\Delta a_\mu$ may be understood from the interplay between the Bino- and Wino-mediated contributions. For large values of $|\mu|$, the Bino contribution tends to be the most relevant one. If one considers positive values of $\mu \times M_1$, the Bino-smuon loop gives a positive contribution to $\Delta a_\mu$ which can account for the $(g_{\mu}-2)$ anomaly for sufficiently large values of $\tan \beta$. However, for smaller values of $|\mu|$ and negative values of $\mu \times M_1$, as required by the blind spot solution, the Bino contribution tends to be subdominant and neither has the sign nor the magnitude to account for the $(g_{\mu}-2)$ anomaly. If anything, depending on the sign of $M_1 \times M_2$, it will partially cancel the Wino contribution to $\Delta a_{\mu}$. For smaller values of $|\mu|$, an explanation for the $(g_{\mu}-2)$ anomaly requires a Wino-mediated contribution enabled by $\mu \times M_2 > 0$ and moderate values of $M_2$. 

For the LHC constraints indicated in Figs.~\ref{fig:DDconstraintstb15} and~\ref{fig:DDconstraintstb30} ($\left| M_1 \right| \gtrsim 240\,$GeV, shown with the purple dash-dotted line) we have assumed that the lightest stau is the next-to-lightest supersymmetric particle, with a mass such that the proper relic density is obtained by co-annihilation of the lightest stau with the lightest neutralino. In such a case, the Wino-like chargino and neutralino have sizable branching ratios into staus, increasing the stau production rate. In order to estimate the LHC limits, we use a recent analysis~\cite{CMS:2019zmn} searching for tau-lepton final states, which assumed that the mass gap between the lightest chargino and neutralino is 50\,GeV and the lightest stau mass lies in the middle of the lightest chargino and neutralino masses, which is close to the situation found under our assumptions. This shows that the LHC is already putting strong constraints on the realization of this scenario. Note that we chose the Wino- ($M_2$) and the first and second generation slepton ($M_L^{1,2}$, $M_R^{1,2}$) mass parameters to be approximately degenerate ($M_2 \approx M_L^{1,2} \approx M_R^{1,2} \approx \left| M_1 \right| + 50\,$GeV) such that current LHC limits for direct slepton searches are avoided for slepton masses above $\sim 200\,$GeV~\cite{Aad:2014vma,Aad:2019vnb,CMS:2019eln,Aad:2019byo,Aad:2019qnd,Aaboud:2017nhr,Sirunyan:2018ubx,Aaboud:2018jiw,Aaboud:2018sua,Aaboud:2018ngk,ATLAS-CONF-2019-008,ATLAS:2021moa}.

Additional constraints from LHC searches with charged leptons in the final states can arise from production of the Higgsino-like neutralino and chargino states. These states decay into gauge and Higgs bosons and lighter charginos and neutralinos. If $\mu$ is large compared to $M_2$, Higgsino production and their decays can lead to relevant signals at the LHC despite Higgsinos having much smaller production cross section that Winos, because for such a choice of parameters, the Higgsino-decays lead to final states with much harder leptons than the leptons arising from Wino or slepton production. We have checked a number of example points from the preferred regions of Figs.~\ref{fig:DDconstraintstb15} and~\ref{fig:DDconstraintstb30} (i.e., where all direct detection constraints are satisfied, where $\left| M_1 \right| \gtrsim 240\,$GeV, avoiding the LHC constrain indicated in the figures, and where the $\Delta a_\mu$ contribution explains the observed value) using \texttt{checkmate2} to check that no additional LHC constraints arise from slepton and electroweakino (including Higgsino) production.

\subsection{Benchmark Points Explaining $\boldsymbol{\Delta a_\mu}$, Dark Matter and Avoiding LHC Constraints}

For a Bino-like DM candidate with mass in the few-hundred GeV range, the observed relic density can be realized via thermal production through different mechanisms, such as co-annihilation with sleptons or charginos~\cite{Ellis:1998kh,Ellis:1999mm,Buckley:2013sca,Han:2013gba,Cabrera:2016wwr,Baker:2018uox}, $t$-channel annihilation via light left-right mixed staus~\cite{Pierce:2013rda} or smuons~\cite{Fukushima:2014yia}, or resonant $s$-channel annihilation~\cite{Han:2013gba,Cabrera:2016wwr}. In Table~\ref{Table:Table1} we present a few benchmark scenarios which simultaneously accommodate the ($g_{\mu}-2$) anomaly and a viable DM candidate. All of them are consistent with the observed relic density, the observed value of $\Delta a_\mu$, and satisfy LHC constraints as well as constraints from direct detection. For all benchmark points, we set the parameters in the squark and gluino sectors such that experimental bounds are satisfied and that the observed mass of the SM-like Higgs boson is reproduced. In general, the supersymmetric partners of the color-charged particles must have masses of the order of a few TeV to satisfy current experimental bounds (see, for instance, Refs.~\cite{Vami:2019slp,Aad:2020aze,Aad:2020aob,Aad:2021zyy}). Note that in some of our benchmark scenarios, the hierarchy between the gluino and the weak gaugino masses is larger than the hierarchy induced by the running of the gaugino masses from (approximately) universal values at the Grand Unification scale. While a Grand Unified Theory (GUT) is theoretically attractive, we do not know if any GUT is realized in nature. The symmetries of the low energy theory do not impose any constraint on the hierarchy between the gluino and the weak gaugino masses. Furthermore, even in GUT models, higher order operators at the GUT scale can lead to departures from universal gaugino masses~\cite{Martin:2009ad}. A somewhat related point is that the small-$\left|\mu\right|$ region we are interested in here tends to require particular choices for the soft supersymmetry breaking masses of the Higgs doublets. As is well known, the large stop masses required to reproduce a 125\,GeV SM-like Higgs boson result in large radiative corrections to $m_{H_u}^2$. Starting from universal soft parameters $m_{H_u}^2 = m_{H_d}^2$ at high energy scales, $m_{H_u}^2$ is driven to large negative values at the electroweak scale, while $m_{H_d}^2$ receives much smaller radiative corrections. The correct electroweak symmetry breaking (and, in particular, the correct mass of the electroweak gauge bosons) is then achieved for $-m_{H_u}^2 \approx \left|\mu\right|^2 \gg m_{H_d}^2$ at the electroweak scale. We leave a dedicated investigation of the required values of the soft parameters to achieve the correct electroweak symmetry breaking pattern for future work, however, let us note that (radiative) electroweak symmetry breaking as well as the correct mass of the electroweak gauge bosons are straightforward to achieve regardless of the value of $|\mu|$ if one allows for different values of the soft supersymmetry breaking parameters $m_{H_u}^2$ and $m_{H_d}^2$ at high energy scales.

The constraints from Higgsino and Wino pair production depend on a careful consideration of the production cross sections and decay branching ratios~\cite{Liu:2020muv,Liu:2020ctf}. Here, we consider a compressed spectrum, for which the electroweakino and slepton constraints are weakened. The results for the spectrum, $\Delta a_\mu$, the relic density, as well as the SI and SD cross sections have been obtained with \texttt{Micromegas~5.2.7.a}~\cite{Belanger:2008sj,Belanger:2018ccd,Belanger:2020gnr}. We use \texttt{SUSY-HIT~1.5}~\cite{Djouadi:2006bz} to compute branching ratios relevant for checking the electroweakino and slepton constraints. One problem in the analysis of the LHC limits is that, in many cases, signals can be obtained from the chain decay of many different electroweak particles, and therefore it is difficult to directly apply the bounds from LHC analyses which are typically presented in terms of simplified models. In order to solve this problem, we use \texttt{checkmate2}~\cite{Dercks:2016npn, Alwall:2014hca, Sjostrand:2014zea, deFavereau:2013fsa}, that uses Monte Carlo event generation to compare all production and decay channels for the neutralinos, charginos and sleptons with the current LHC analyses~\cite{ATLAS:2013qzt, ATLAS:2013lcn, ATLAS:2014hzd, ATLAS:2014ikz, ATLAS:2014gmw, ATLAS:2014bgg, ATLAS:2014zve, ATLAS:2014kpx, ATLAS:2014jxt, ATLAS:2014mmo, ATLAS:2014hqe, ATLAS:2014kci, ATLAS:2015tfi, ATLAS:2015qlt, ATLAS:2015xmt, ATLAS:2015vch, ATLAS:2015zwy, ATLAS:2015dcr, ATLAS-CONF-2012-104, ATLAS-CONF-2013-024, ATLAS-CONF-2013-049, ATLAS-CONF-2013-061, ATLAS-CONF-2013-089, ATLAS-CONF-2015-004, CMS:2013cgf, CMS:2014jvv, CMS:2015kjt, CMS:2015zwg, CMS:2016aro, ATLAS:2016zxj, ATLAS:2016gty, ATLAS:2016wgc, ATLAS:2017nga, ATLAS:2017tmw, ATLAS:2017qwn, ATLAS:2017drc, ATLAS:2017vat, ATLAS:2017mjy, ATLAS:2018nud, ATLAS:2018ojr, ATLAS:2018zdn, ATLAS:2019gdh, ATLAS:2019lff, ATLAS:2019fag, ATLAS:2019gti, ATLAS:2019lng, ATLAS:2020qlk, ATLAS:2020dsf, ATLAS:2021twp, ATLAS:2021yyr, ATLAS:2021fbt, ATLAS-CONF-2015-082, ATLAS-CONF-2016-013, ATLAS-CONF-2016-050, ATLAS-CONF-2016-054, ATLAS-CONF-2016-066, ATLAS-CONF-2016-076, ATLAS-CONF-2016-096, ATLAS-CONF-2017-060, ATLAS-CONF-2018-041, ATLAS-CONF-2019-040, ATLAS-CONF-2019-020, ATLAS-CONF-2020-048, CMS:2017moi}. Although most of the relevant LHC analyses have been included in \texttt{checkmate2}, a few of the most recent analyses are not yet implemented in this code. In these cases, we check the compatibility of our points by using conservative estimates of the particle contributions to the different search signals, as explained in Appendix~\ref{app:LHCconstraints}. 
 
The scenarios presented below correspond to different origins of the observed DM relic density and should serve as a guidance for experimental probes of the supersymmetric explanation of the muon $(g-2)$. 

\begin{table}
\begin{tabular}{|c|c|c|c|c|}
\hline
 & BMSM & BMST & BMW & BMH \\ 
\hline
$M_1$ [GeV] & -352 & -258 & -274 & 63 \\
\hline
$M_2$ [GeV] & 400 & 310 & 310 & 700\\
\hline
$\mu$ [GeV]& 690 & 475 & 500 & 470 \\
\hline
$M_{L}^{1,2}$ [GeV] & 360 & 320 & 350 & 750 \\
\hline
$M_{L}^3$ [GeV] & 500 & 320 & 350 & 750 \\
\hline
$M_R^{1,2}$ [GeV] & 360 & 320 & 350 & 750 \\
\hline
$M_R^3$ [GeV] & 500 & 320 & 350 & 750 \\
\hline
$M_A$ [GeV] & 2000 & 1800 & 1600 & 3000 \\
\hline
$\tan\beta$ & 60 & 40 & 35 & 65 \\
\hline
\end{tabular}
\quad
\begin{tabular}{|c|c|c|c|c|}
\hline
 & BMSM & BMST & BMW & BMH \\
\hline
$m_\chi$ [GeV] & 350.2 & 255.3 & 271.4 & 61.0 (124.9) \\
\hline
$m_{\tilde{\tau}_1}$ [GeV] & 414.4 & 264.2 & 305.3 & 709.5 \\
\hline
$m_{\tilde{\mu}_1}$ [GeV] & 362.7 & 323.0 & 352.8 & 751.3\\
\hline
$m_{\tilde{\nu}_\tau}$ [GeV] & 496.0 & 313.7 & 344.2 & 747.3 \\
\hline
$m_{\tilde{\nu}_\mu}$ [GeV] & 354.4 & 313.7 & 344.2 & 747.3 \\
\hline
$m_{\chi^{\pm}_1}$ [GeV] & 392.3 & 296.2 & 297.9 & 469.6 \\
\hline
$\Delta a_\mu$ [$10^{-9}$] & 2.10 & 2.89 & 2.35 & 1.93 \\
\hline
$\Omega_{\rm DM} h^2$ & 0.121 & 0.116 & 0.124 & 0.121 \\
\hline
$\sigma_p^{\rm SI}$ [$10^{-10}\,$pb] & 0.645 & 1.58 & 1.42 & 0.315 \\
\hline
$\sigma_p^{\rm SD}$ [$10^{-6}\,$pb] & 1.03 & 5.11 & 4.23 & 3.01 \\
\hline
$\sigma_n^{\rm SI}$ [$10^{-10}\,$pb] & 0.632 & 1.57 & 1.41 & 0.330 \\
\hline
$\sigma_n^{\rm SD}$ [$10^{-6}\,$pb] & 0.882 & 4.10 & 3.42 & 2.34 \\
\hline
\end{tabular}
\caption{Values of the MSSM parameters, mass spectrum and quantities relevant for DM and $(g_\mu -2)$ for the case of Bino-like DM co-annihilating with light sleptons (BMSM), co-annihilating with a light stau (BMST), co-annihilating with a Wino (BMW) and resonant $s$-channel annihilation via the SM-like Higgs boson (BMH). For BMH we also provide the mass of the SM-like Higgs boson $m_h$ between brackets.}
\label{Table:Table1}
\end{table}

\begin{itemize}
   \item {\bf BMSM:} A DM production scenario closely related to the relatively low masses of the muon (neutrino) superpartners required to address the $a_\mu$-anomaly is co-annihilation of the lightest neutralino with the light slepton states. The benchmark BMSM gives a representation of this possibility, where we set the masses of the tau-lepton superpartners to be larger than those of the first and second generation sleptons. Since multiple production channels contribute to final states containing leptons at the LHC, current searches strongly constrain the presence of light electroweak interacting particles in this scenario. In order to be compatible with $\Delta a_\mu$, DM phenomenology and LHC searches, BMSM features the largest values of $|\mu|$ and $\tan\beta$ of the benchmark points presented in this article.

   \item {\bf BMST:} A similar solution to BMSM is associated with the co-annihilation of a light stau with the lightest neutralino. For universal soft slepton masses, this happens naturally at large values of $\tan\beta$, where the lightest stau is pushed to masses lower than those of the sneutrinos. BMST gives a representative example of this possibility. 

   \item {\bf BMW:} The lightest neutralino may co-annihilate with the lightest chargino. The benchmark BMW represents such a possibility.

   \item {\bf BMH:} The lightest neutralino can acquire the proper relic density via resonant $s$-channel annihilation via the SM-like Higgs boson. BMH represents such a possibility.
\end{itemize}
Although the mechanisms controlling the relic density are different for the different benchmark points, they share many characteristics. They feature masses of weakly interacting sparticles masses lower than about 500\,GeV and values of $\tan\beta$ of the order of a few 10's, leading to values of $\Delta a_\mu$ in the desired range. Apart from BMH, which we will discuss further below, all benchmark points in Table~\ref{Table:Table1} have negative values of $\mu \times M_1$ and positive values of $\mu \times M_2$. 

The observed relic density for a Bino-like DM candidate may also be obtained via resonant $s$-channel annihilation via the heavy Higgs boson $A$ and $H$. However, for the values of $\tan\beta$ necessary to enhance $\Delta a_\mu$, LHC bounds on the heavy Higgs bosons become very strong, implying a heavy spectrum. Using the bound on $m_H$ provided in Eq.~\eqref{eq:mHbound}, the approximate expression for $\Delta a_\mu$ in Eq.~\eqref{eq:approxamu}, and assuming that all the weakly interacting sparticles have masses close to $m_H/2$, the maximal value for $\Delta a_\mu$ that may be obtained is
\begin{equation}
   \Delta a_\mu \simeq 10^{-9} \tan\beta \frac{4 }{m_H^2} (100\,{\rm GeV})^2 \lesssim 7 \times 10^{-10},
\end{equation}
which is a factor of a few smaller than the observed anomaly. Therefore, we shall not discuss this particular solution further. 

Resonant $s$-channel annihilation via the $Z$-boson presents similar characteristics to resonant annihilation mediated by the SM-like Higgs, $h$. Thus, we present only an example of the latter case here, BMH. For such small values of $M_1 \simeq m_h/2 = 62.5$\,GeV, values of $|\mu| \sim 500$\,GeV may lead to the desired suppression of the SIDD cross section for either sign of $\mu$. This follows, for instance, from Eq.~\eqref{eq:approxSI}, where we also observe that for positive values of $\mu \times M1$, values of $m_H$ significantly larger than the current experimental bounds are preferred. Note that, for BMH, we chose the sleptons and the Winos to be heavy to avoid the bounds from the LHC. Hence, obtaining the proper value of $\Delta a_\mu$ requires relatively large values of $\tan\beta$.

\subsection{Future Prospects}

The benchmark points presented above are compatible with current experimental limits, but will be tested in the near future in several ways. 

First, all four benchmark points will be probed by the next generation of direct detection experiments: The SIDD cross sections of all four benchmark points are within the projected sensitivities of the LZ and XENONnT experiments~\cite{LUX-ZEPLIN:2018poe,XENON:2020kmp}. More generally, for $\mu \times M_1 < 0$, and for fixed values of $M_1$, $\mu$ and $\tan\beta$, the smallest possible value of the SIDD cross section is associated with the smallest allowed value of the heavy Higgs mass, see Eq.~\eqref{eq:approxSI}. For masses $200\,{\rm GeV} \lesssim |M_1| \lesssim 500$\,GeV, a hierarchy $1 \lesssim |\mu/M_1| \lesssim 3$, and $\tan\beta \gtrsim 20$, compatible with collider physics, muon $(g-2)$, and Dark Matter relic density constraints, the smallest possible SIDD cross section is (see Eq.~\eqref{eq:lowestSIDD})
\begin{equation}
   \sigma_p^{\rm SI} > \mathcal{O} \left(10^{-10}\right)\,{\rm pb} \times \left(\frac{M_1}{250\,{\rm GeV}}\right)^2 \times \left(\frac{500\,{\rm GeV}}{\mu} \right)^4 \;. 
\label{eq:SIest}
\end{equation}
The LZ and XENONnT experiments will probe cross sections as small as $\sigma_p^{\rm SI} \sim \mathcal{O}\left(10^{-12}\right)\,$pb for $\left|M_1\right| \sim 40\,$GeV, growing to $\sigma_p^{\rm SI} \sim \mathcal{O}\left(10^{-11}\right)\,$pb for $\left|M_1\right| \sim 500\,$GeV, implying full coverage of this representative region of parameters.

Furthermore, the spin-dependent WIMP-neutron cross sections can be probed by LZ and XENONnT, while the next generation of the PICO experiment will probe the spin-dependent WIMP-proton cross sections~\cite{PICO:2015amc}.
From Eq.~\eqref{eq:lowestSDDD} we can see that the spin-dependent WIMP-nucleon cross sections are 
\begin{equation}
   \sigma_n^{\rm SD} \sim \sigma_p^{\rm SD} > \mathcal{O}\left(10^{-6}\right)\,{\rm pb} \times \left(\frac{500\,{\rm GeV}}{\mu} \right)^4 \;,
\end{equation}
with a mild dependence on $M_1$. The future sensitivities of LZ/XENONnT on $\sigma_n^{\rm SI}$ move from a few times $10^{-7}\,$pb for $\left|M_1\right| \sim 100\,$GeV to $\sim 10^{-6}\,$pb for $\left|M_1\right| \sim 500\,$GeV, while PICO-500 will probe $\sigma_p^{\rm SD} \sim 10^{-6}\,$pb for $\left|M_1\right| \sim 100\,$GeV and $\sigma_p^{\rm SD} \sim 5 \times 10^{-6}\,$pb for $\left|M_1\right| \sim 500\,$GeV. Hence, these experiments will probe the region of parameter space where $\left| \mu \right| \lesssim 500\,$GeV. In particular, LZ, XENONnT and PICO-500 will probe the spin-dependent cross sections of the benchmark points BMST, BMW, BMH, while BMSM has spin-dependent interactions smaller than the projected sensitivities of these experiments.

Second, for all benchmark points with $M_1 \times \mu < 0$ (BMSM, BMST, and BMW), the SIDD cross section is suppressed below current experimental limits due to the destructive interference between the amplitudes mediated by the SM-like and the heavy Higgs bosons discussed above. For this suppression to be effective, the masses of the non-SM-like Higgs bosons must be low enough to within the reach (see, for example, Ref.~\cite{Bahl:2020kwe}) of future runs of the LHC: The high-luminosity LHC will be sensitive to Higgs bosons with masses of about a factor 1.5 larger than current exclusion limits (keeping all other parameters, in particular $\tan\beta$, fixed). From the expression of the SIDD cross section, Eq.~\eqref{eq:approxSI} we see that increasing $m_H \to 1.5 \ m_H$ corresponds to a factor 2-3 increase of the SIDD cross section. Such SIDD cross sections would be in conflict with {\it current} experimental constraints, or conversely, values of the heavy Higgs mass allowed by current direct detection bounds will be efficiently probed by the high-luminosity LHC. For BMSH, on the other hand, the SIDD cross section is suppressed by a large hierarchy between the Higgsino and Bino mass parameters, $|\mu| \gg |M_1|$. Such ``large $\left| \mu \right|$'' solutions to suppressing the SIDD cross sections allow for heavy Higgs masses beyond the projected reach of the high-luminosity LHC. 

Last but not least, our benchmark scenarios are also testable in searches for electroweakly interacting particles at future runs of the LHC, see, for example, Refs.~\cite{ATL-PHYS-PUB-2018-048,CMS-PAS-FTR-18-001}. We note that some of these projections have already been surpassed by innovative searches with current LHC data, like those presented in Ref.~\cite{ATLAS:2021yqv}, further bolstering the prospects of probing our benchmark points and similar scenarios in the upcoming runs of the LHC. The extrapolation of these conclusions to the whole region of parameters analyzed in this article should be the object of an independent dedicated study, that we plan to perform but is beyond the scope of the current article. Let us also emphasize that future lepton colliders play an important role to probe sleptons and charginos, especially for (semi-)compressed spectra, see Refs.~\cite{Freitas:2003yp,Berggren:2013vna,Fujii:2015jha,ATL-PHYS-PUB-2018-031,CEPCStudyGroup:2018ghi,deBlas:2018mhx,CidVidal:2018eel,Strategy:2019vxc,Baer:2019gvu,Habermehl:2020njb,Berggren:2020tle,Baum:2020gjj,Natalia:2021ssb}.

\section{Summary and Conclusions} \label{sec:Conclusions}

A wide range of possible extensions of the Standard Model (SM) can lead to an explanation of the value of $\Delta a_\mu$ measured at the Fermilab and Brookhaven experiments. While arguably the simplest explanation is the addition of a scalar particle, one can also rely on new gauge bosons, vector-like fermions or leptoquark models. The leptoquark (or R-parity violating supersymmetry) solution seems to be interesting since it can accommodate not only the values of $\Delta a_\mu$, but can also lead to an explanation of the flavor anomalies, although at the prize of a delicate choice of the couplings of the leptoquarks.

In this work, we explore a solution based on the (R-parity conserving) Minimal Supersymmetric extension of the SM, in which, although one cannot address the flavor anomalies, one can find solutions leading to a compelling DM explanation. In particular, we discuss the conditions that are required to be consistent with the observed $\Delta a_\mu$, existing direct dark matter (DM) detection constraints, and the bounds from the LHC on new Higgs bosons and supersymmetric particles. We look for solution in which direct DM detection constraints are fulfilled by a partial cancellation of the light and heavy CP-even Higgs mediated contributions which significantly differ from previous studies relying on very heavy Higgs and Higgsino particles. This cancellation requires negative values of $\mu \times M_1$. Since the observed value of $a_\mu$ is {\it larger} than the SM prediction, the Bino contribution to $a_\mu$, which is proportional to $\mu \times M_1$, must be subdominant. This can only be realized for small-to-moderate values of $|\mu|$. We present corresponding benchmark scenarios associated with different DM production mechanisms to achieve the observed relic density, including co-annihilation with sleptons, or resonant $s$-channel annihilation mediated by the SM-like Higgs or $Z$ bosons.

The corresponding spectra have a number of interesting consequences: 1) The relatively small values of the Higgsino mass parameter lead to a more natural model, in terms of the electroweak hierarchy problem. 2) The DM candidates have spin-independent and spin-dependent direct detection cross sections which can be probed in the next generation of direct detection experiments. 3) Explaining the anomalous magnetic moment of the muon for negative values of $\mu \times M_1$ requires light sleptons and electroweakinos, which should be probed at run 3 of the LHC, the HL-LHC, or, ultimately, at future lepton colliders. 4) The suppression of the direct detection cross section is only possible for relatively light non-SM-like Higgs bosons in the MSSM, which can be probed at run 3 of the LHC and the HL-LHC.

\acknowledgements{
SB is supported in part by NSF Grant PHY-2014215, DOE HEP QuantISED award \#100495, and the Gordon and Betty Moore Foundation Grant GBMF7946.
Fermilab is operated by Fermi Research Alliance, LLC under Contract No. DE-AC02-07CH11359 with the U.S. Department of Energy. The work of CW at the University of Chicago has been also supported by the DOE grant DE-SC0013642. Work at ANL is supported in part by the U.S. Department of Energy~(DOE), Div.~of HEP, Contract DE-AC02-06CH11357. This work was supported in part by the DOE under Task TeV of contract DE-FGO2-96-ER40956. NRS is supported by U.S. Department of Energy under
Contract No.DE-SC0021497. This work was performed in part at Aspen Center for Physics, which is supported by National Science Foundation grant PHY-1607611.
}

\begin{appendix}

\section{LHC constraints from chargino and slepton searches}
\label{app:LHCconstraints}

In this appendix, we discuss the constraints from chargino and slepton searches on our benchmark points presented in Table~\ref{Table:Table1}. The most severe chargino constraints tend to stem from production of the lightest chargino ($\widetilde{\chi}_1^\pm$) and the next-to-lightest neutralino ($\widetilde{\chi}_2^0$) at the LHC, $pp \to \widetilde{\chi}_1^\pm \widetilde{\chi}_2^0$. Note that for all of our benchmark points, the lightest neutralino is Bino-like, while $\widetilde{\chi}_2^0$ is Wino-like (for BMSM, BMST and BMW) or Higgsino-like (for BMH) depending on the hierarchy of $\left|\mu\right|$ and $\left|M_2\right|$. Hence, $\widetilde{\chi}_2^0$ and $\widetilde{\chi}_1^\pm$ will typically be mass degenerate. All the benchmark points presented in this article fulfill the current LHC constraints~\cite{ATLAS:2013qzt, ATLAS:2013lcn, ATLAS:2014hzd, ATLAS:2014ikz, ATLAS:2014gmw, ATLAS:2014bgg, ATLAS:2014zve, ATLAS:2014kpx, ATLAS:2014jxt, ATLAS:2014mmo, ATLAS:2014hqe, ATLAS:2014kci, ATLAS:2015tfi, ATLAS:2015qlt, ATLAS:2015xmt, ATLAS:2015vch, ATLAS:2015zwy, ATLAS:2015dcr, ATLAS-CONF-2012-104, ATLAS-CONF-2013-024, ATLAS-CONF-2013-049, ATLAS-CONF-2013-061, ATLAS-CONF-2013-089, ATLAS-CONF-2015-004, CMS:2013cgf, CMS:2014jvv, CMS:2015kjt, CMS:2015zwg, CMS:2016aro, ATLAS:2016zxj, ATLAS:2016gty, ATLAS:2016wgc, ATLAS:2017nga, ATLAS:2017tmw, ATLAS:2017qwn, ATLAS:2017drc, ATLAS:2017vat, ATLAS:2017mjy, ATLAS:2018nud, ATLAS:2018ojr, ATLAS:2018zdn, ATLAS:2019gdh, ATLAS:2019lff, ATLAS:2019fag, ATLAS:2019gti, ATLAS:2019lng, ATLAS:2020qlk, ATLAS:2020dsf, ATLAS:2021twp, ATLAS:2021yyr, ATLAS:2021fbt, ATLAS-CONF-2015-082, ATLAS-CONF-2016-013, ATLAS-CONF-2016-050, ATLAS-CONF-2016-054, ATLAS-CONF-2016-066, ATLAS-CONF-2016-076, ATLAS-CONF-2016-096, ATLAS-CONF-2017-060, ATLAS-CONF-2018-041, ATLAS-CONF-2019-040, ATLAS-CONF-2019-020, ATLAS-CONF-2020-048, CMS:2017moi} implemented in~\texttt{checkmate2}~\cite{Dercks:2016npn, Alwall:2014hca, Sjostrand:2014zea, deFavereau:2013fsa}. We also check compatibility with very recent LHC searches which are not yet implemented in \texttt{checkmate2} by using conservative estimates of the particle contribution to these search channels.

In order to gain a physical intuition of how the benchmark points avoid the LHC constraints, we provide a brief discussion of their properties. 

{\bf BMSM}: The lightest neutralino has a mass $m_{\widetilde{\chi}_1^0} = 350\,$GeV, and the Wino-like next-to-lightest neutralino and lightest chargino have masses $m_{\widetilde{\chi}_2^0} = m_{\widetilde{\chi}_1^\pm} = 392\,$GeV. We have computed the $\widetilde{\chi}_2^0 + \widetilde{\chi}_1^\pm$ production cross section at the 13\,TeV LHC with \texttt{MadGraph5\_v3.1.1}~\cite{Alwall:2014hca}, finding $\sigma(pp \to \widetilde{\chi}_2^0 + \widetilde{\chi}_1^\pm) = 0.08\,$pb . Comparing this to the upper limit from Ref.~\cite{ATLAS:2021moa} $\sigma(pp \to WZ+2\widetilde{\chi}_1^0) \lesssim 0.6\,$pb at these masses (this search is not yet implemented in \texttt{checkmate2}; we have taken the limit from the supplementary material of Ref.~\cite{ATLAS:2021moa} accessible via \texttt{HEPdata} or the CERN Document Server), we see that this benchmark point is not constrained by this search even before taking into account that the $\widetilde{\chi}_2^0$ and $\widetilde{\chi}_1^\pm$ decay branching ratios into gauge bosons are small in this scenario. For this benchmark point, however, the Wino- and Higgsino-like neutralinos and charginos can undergo cascade decays involving the light sleptons, giving rise to potentially detectable signatures in searches for charged leptons and missing energy at the LHC. Due to BMSM's mass spectrum, the production of the Wino-like $\widetilde{\chi}_2^0$ and $\widetilde{\chi}_1^\pm$ gives rise to relatively soft leptons. The most sensitive search corresponding to this final state currently implemented in \texttt{checkmate2} is Ref.~\cite{ATLAS:2019lng}, for which we find a signal strength of $r \sim 0.8$. The signal strength here is defined as the ratio of the number of events predicted for the model point and the observed limits on the number of events in the most constraining signal region of any given LHC search. Production of the Higgsino-like $\widetilde{\chi}_3^0$, $\widetilde{\chi}_4^0$, and $\widetilde{\chi}_2^\pm$, on the other hand, leads to final states with much harder charged leptons and larger missing transverse energy. The most sensitive LHC search currently implemented in \texttt{checkmate2} for such signatures is Ref.~\cite{CMS:2017moi}, for which we find a signal strength of $r \sim 0.8$. Regarding direct slepton searches, this benchmark point features approximately mass degenerate left- and right-handed selectrons and smuons with $m_{\widetilde{\ell}^\pm} = 313\,$GeV. Such compressed spectra with $m_{\widetilde{\ell}^\pm} - m_{\widetilde{\chi}_1^0} = 10\,$GeV are not constrained by current LHC searches for direct slepton production, see, for example, Ref.~\cite{ATLAS-CONF-2019-008}.
 
{\bf BMST}: The lightest neutralino has mass $m_{\widetilde{\chi}_1^0} = 255\,$GeV, and the Wino-like next-to-lightest neutralino and lightest chargino have masses $m_{\widetilde{\chi}_2^0} = m_{\widetilde{\chi}_1^\pm} = 296\,$GeV. Both the next-to-lightest neutralino and the lightest chargino decay into staus for this benchmark point, ${\rm BR}(\widetilde{\chi}_2^0 \to \widetilde{\tau}_1^\pm + \tau^\mp) = {\rm BR}(\widetilde{\chi}_1^\pm \to \widetilde{\tau}_1^\pm + \nu_\tau) = 100\,\%$ from our \texttt{SUSY-HIT} results. The staus in turn decay into tau-leptons, ${\rm BR}(\widetilde{\tau}^\pm \to \tau^\pm + \widetilde{\chi}_1^0) = 100\,\%$, leading to tau-leptons + missing transverse energy final states from $\widetilde{\chi}_2^0 $ and $\widetilde{\chi}_1^\pm$ production at the LHC. Although the corresponding searches are quite challenging, studies with initial state radiation jets in a compressed region with chargino-neutralino mass gap $m_{\widetilde{\chi}_1^\pm} - m_{\widetilde{\chi}_1^0} \approx 50\,$GeV and stau masses in the middle between the chargino and neutralino, $(m_{\widetilde{\chi}_1^\pm} + m_{\widetilde{\chi}_1^0})/2 = m_{\widetilde{\tau}}$, constrain the chargino mass to $m_{\widetilde{\chi}_1^\pm} \gtrsim 290$\,GeV~\cite{CMS:2019zmn}. We have arranged the spectrum of BMST such that this bound is approximately applicable, and accordingly, we chose the masses of the Wino-like next-to-lightest neutralino and the lightest chargino to be larger than 290\,GeV. Regarding the slepton searches, the selectrons and smuons have masses $m_{\widetilde{\ell}^\pm} = 323\,$GeV for BMST. Hence, $m_{\widetilde{\ell}^\pm} - m_{\widetilde{\chi}_1^0} = 68\,$GeV, which is below the mass gaps excluded by current LHC searches~\cite{ATLAS-CONF-2019-008}. We note that out of the searches implemented in \texttt{checkmate2}, BMST has the largest signal strength ($r \sim 0.3$) for the search in Ref.~\cite{ATLAS:2020qlk}.

{\bf BMW}: The lightest neutralino has mass $m_{\widetilde{\chi}_1^0} = 271\,$GeV, and the Wino-like next-to-lightest neutralino and lightest chargino have masses $m_{\widetilde{\chi}_2^0} = m_{\widetilde{\chi}_1^\pm} = 298\,$GeV. For the $\widetilde{\chi}_2^0 + \widetilde{\chi}_1^\pm$ production cross section, we find $\sigma(pp \to \widetilde{\chi}_2^0 + \widetilde{\chi}_1^\pm) = 0.26\,$pb. While this point is not constrained by any of the analyses included in \texttt{checkmate2}, it may be constrained by the recent bounds coming from the multi-lepton final state analyses in Ref.~\cite{ATLAS:2021moa}, which is not yet implemented in \texttt{checkmate2}. Note that the dominant production mechanism of charged lepton final states from the charginos and neutralinos in BMS is via tau-leptons. In Ref.~\cite{ATLAS:2021moa}, however, the limits are obtained assuming decays of the charginos and neutralinos into gauge bosons and the lightest neutralinos and hence the limits are not directly applicable to this case. In order to make a conservative comparison to the upper limit $\sigma(pp \to WZ+2\widetilde{\chi}_1^0) \lesssim 0.9\,$pb~\cite{ATLAS:2021moa} at these masses, we can note that, including the dominant contribution coming from $\tau$ lepton decays, the total leptonic branching ratio from $\widetilde{\chi}_2^0 + \widetilde{\chi}_1^\pm$ production\footnote{Here, we define the leptonic branching ratio as the sum of the branching ratios of $\widetilde{\chi}_2^0$/$\widetilde{\chi}_1^\pm$ involving $Z$/$W^\pm$-bosons multiplied by their leptonic branching ratios and the (stau-mediated) decays into tau-lepton(s) multiplied with the leptonic branching ratio of the taus.} for this benchmark points is 4.5\,\%, while Ref.~\cite{ATLAS:2021moa} assumed ${\rm BR}(\widetilde{\chi}_2^0 \to \widetilde{\chi}_1^0 Z) = {\rm BR}(\widetilde{\chi}_1^\pm \to \widetilde{\chi}_1^0 W^\pm) = 100\,\%$, corresponding to a total leptonic branching ratio of $1.9\,\%$, leading to an estimate of the signal strength of $r = (0.26\,{\rm pb}/0.9\,{\rm pb}) \times (4.5\,\%/1.9\,\%) \sim 0.7$. Note that the true signal strength is most likely significantly lower since leptons coming from tau-lepton decays are softer than those coming from direct lepton production. 

Regarding direct slepton searches, the lightest charged sleptons for this benchmark point are the staus, $m_{\widetilde{\tau}_1} = 305\,$GeV, followed by the selectrons and smuons with $m_{\widetilde{\ell}^\pm} = 353\,$GeV. Such mass gaps, $m_{\widetilde{\ell}^\pm} - m_{\widetilde{\chi}_1^0} = 82\,$GeV, are not constrained by current LHC bounds even under the assumption of ${\rm BR}(\widetilde{\ell}^\pm \to \ell^\pm + \chi_1^0) = 100\,\%$ for all four of the left- and right-handed selectron and smuon states, with stau bounds being even weaker. For this benchmark point, the left-handed charged sleptons decay preferentially into charginos, ${\rm BR}(\widetilde{\ell}_L^\pm \to \nu_\ell + \chi_1^\pm) = 53\,\%$, and have sizeable branching ratios into the next-to-lightest (Wino-like) neutralino, ${\rm BR}(\widetilde{\ell}_L^\pm \to \ell^\pm + \chi_2^0) = 28\,\%$. The reduced decay branching ratios into the lightest neutralino implies softer spectra of visible decay products at the LHC and hence even weaker bounds. Moreover, due to the compressed chargino and neutralino spectrum, no relevant additional constraints emerge from the decay of the sleptons into the Wino-like states. 

We note that out of the searches implemented in \texttt{checkmate2}, BMW has the largest signal strength ($r \sim 0.4$) for the search in Ref.~\cite{ATLAS:2020qlk}.

{\bf BMH}: The lightest neutralino has mass $m_{\widetilde{\chi}_1^0} = 61\,$GeV, and the next-to-lightest neutralino and lightest chargino have masses $m_{\widetilde{\chi}_2^0} = m_{\widetilde{\chi}_1^\pm} = 470\,$GeV. Unlike for all of the other benchmark points, $m_{\widetilde{\chi}_2^0}$ and $m_{\widetilde{\chi}_1^\pm}$ are Higgsino-like, leading to a relatively small $\widetilde{\chi}_2^0 + \widetilde{\chi}_1^\pm$ production cross section of $\sigma(pp \to \widetilde{\chi}_2^0 + \widetilde{\chi}_1^\pm) = 0.013\,$pb at the 13\,TeV LHC. This is significantly below the upper limit from Ref.~\cite{ATLAS:2021moa} at theses masses, $\sigma(pp \to WZ+2\widetilde{\chi}_1^0) \lesssim 0.02\,$pb, even before taking the branching ratios of $\widetilde{\chi}_2^0$ and $\widetilde{\chi}_1^\pm$ into account [${\rm BR}(\widetilde{\chi}_2^0 \to \widetilde{\chi}_1^0 + h) = 62\,\%$ and ${\rm BR}(\widetilde{\chi}_3^0 \to \widetilde{\chi}_1^0 + h) = 34\,\%$ for this point]. The masses of the Wino-like state are $m_{\widetilde{\chi}_4^0} = m_{\widetilde{\chi}_2^\pm} = 745\,$GeV, beyond the current limit on these states~\cite{Aaboud:2018ngk,ATLAS:2018ojr,CMS:2021cox}, even before accounting for the decay patterns of the heavy Winos. The Wino-like states dominantly decay into the intermediate Higgsino-like states, $\widetilde{\chi}_4^0/\widetilde{\chi}_2^\pm \to \widetilde{\chi}_3^0/\widetilde{\chi}_2^0/\widetilde{\chi}_1^\pm + W^\pm/Z/h$. Thus, production of Wino-like states at the LHC will mostly lead to cascade decays with softer visible final states than if the Wino-like states would directly decay into the lightest neutralino, $\widetilde{\chi}_4^0/\widetilde{\chi}_2^\pm \to \widetilde{\chi}_1^0 + W^\pm/Z/h$, complicating experimental searches. Furthermore, let us stress that the bounds on Wino production presented by the experimental collaborations assume that the squarks are decoupled, and accordingly ignore the important $t$-channel squark-mediated contributions to the Wino production cross section which can lower the cross section by order one factors depending on the exact squark masses~\cite{Liu:2020ctf}. These arguments apply to very recent searches for Winos in hadronic final states~\cite{ATLAS:2021yqv,CMS-PAS-SUS-21-002} that would rule out this scenario in the absence of cascade decays and the $t$-channel squark contributions. Nonetheless, these impressive searches clearly show the potential of the experimental collaborations to test the regions of parameters represented by our scenarios in future runs of the LHC. Regarding the slepton searches, the lightest charged sleptons for this benchmark point are the staus, $m_{\widetilde{\tau}_1} = 710\,$GeV, and the selectrons and smuons have masses $m_{\widetilde{\ell}^\pm} = 751\,$GeV, beyond the reach of current LHC searches~\cite{ATLAS-CONF-2019-008}. We note that out of the searches implemented in \texttt{checkmate2}, BMH has the largest signal strength ($r \sim 0.4$) for the search in Ref.~\cite{ATLAS:2020qlk}.

\end{appendix}

\bibliography{theBib}

\begin{thebibliography}{253}%
\makeatletter
\providecommand \@ifxundefined [1]{%
 \@ifx{#1\undefined}
}%
\providecommand \@ifnum [1]{%
 \ifnum #1\expandafter \@firstoftwo
 \else \expandafter \@secondoftwo
 \fi
}%
\providecommand \@ifx [1]{%
 \ifx #1\expandafter \@firstoftwo
 \else \expandafter \@secondoftwo
 \fi
}%
\providecommand \natexlab [1]{#1}%
\providecommand \enquote  [1]{``#1''}%
\providecommand \bibnamefont  [1]{#1}%
\providecommand \bibfnamefont [1]{#1}%
\providecommand \citenamefont [1]{#1}%
\providecommand \href@noop [0]{\@secondoftwo}%
\providecommand \href [0]{\begingroup \@sanitize@url \@href}%
\providecommand \@href[1]{\@@startlink{#1}\@@href}%
\providecommand \@@href[1]{\endgroup#1\@@endlink}%
\providecommand \@sanitize@url [0]{\catcode `\\12\catcode `\$12\catcode
  `\&12\catcode `\#12\catcode `\^12\catcode `\_12\catcode `\%12\relax}%
\providecommand \@@startlink[1]{}%
\providecommand \@@endlink[0]{}%
\providecommand \url  [0]{\begingroup\@sanitize@url \@url }%
\providecommand \@url [1]{\endgroup\@href {#1}{\urlprefix }}%
\providecommand \urlprefix  [0]{URL }%
\providecommand \Eprint [0]{\href }%
\providecommand \doibase [0]{http://dx.doi.org/}%
\providecommand \selectlanguage [0]{\@gobble}%
\providecommand \bibinfo  [0]{\@secondoftwo}%
\providecommand \bibfield  [0]{\@secondoftwo}%
\providecommand \translation [1]{[#1]}%
\providecommand \BibitemOpen [0]{}%
\providecommand \bibitemStop [0]{}%
\providecommand \bibitemNoStop [0]{.\EOS\space}%
\providecommand \EOS [0]{\spacefactor3000\relax}%
\providecommand \BibitemShut  [1]{\csname bibitem#1\endcsname}%
\let\auto@bib@innerbib\@empty
\bibitem [{\citenamefont {Aad}\ \emph {et~al.}(2012)\citenamefont {Aad} \emph
  {et~al.}}]{Aad:2012tfa}%
  \BibitemOpen
  \bibfield  {author} {\bibinfo {author} {\bibfnamefont {Georges}\ \bibnamefont
  {Aad}} \emph {et~al.} (\bibinfo {collaboration} {ATLAS}),\ }\bibfield
  {title} {\enquote {\bibinfo {title} {{Observation of a new particle in the
  search for the Standard Model Higgs boson with the ATLAS detector at the
  LHC}},}\ }\href {\doibase 10.1016/j.physletb.2012.08.020} {\bibfield
  {journal} {\bibinfo  {journal} {Phys. Lett. B}\ }\textbf {\bibinfo {volume}
  {716}},\ \bibinfo {pages} {1--29} (\bibinfo {year} {2012})},\ \Eprint
  {http://arxiv.org/abs/1207.7214} {arXiv:1207.7214 [hep-ex]} \BibitemShut
  {NoStop}%
\bibitem [{\citenamefont {Chatrchyan}\ \emph {et~al.}(2012)\citenamefont
  {Chatrchyan} \emph {et~al.}}]{Chatrchyan:2012ufa}%
  \BibitemOpen
  \bibfield  {author} {\bibinfo {author} {\bibfnamefont {Serguei}\ \bibnamefont
  {Chatrchyan}} \emph {et~al.} (\bibinfo {collaboration} {CMS}),\ }\bibfield
  {title} {\enquote {\bibinfo {title} {{Observation of a New Boson at a Mass of
  125 GeV with the CMS Experiment at the LHC}},}\ }\href {\doibase
  10.1016/j.physletb.2012.08.021} {\bibfield  {journal} {\bibinfo  {journal}
  {Phys. Lett. B}\ }\textbf {\bibinfo {volume} {716}},\ \bibinfo {pages}
  {30--61} (\bibinfo {year} {2012})},\ \Eprint {http://arxiv.org/abs/1207.7235}
  {arXiv:1207.7235 [hep-ex]} \BibitemShut {NoStop}%
\bibitem [{\citenamefont {Aaij}\ \emph
  {et~al.}(2021{\natexlab{a}})\citenamefont {Aaij} \emph
  {et~al.}}]{Aaij:2021vac}%
  \BibitemOpen
  \bibfield  {author} {\bibinfo {author} {\bibfnamefont {Roel}\ \bibnamefont
  {Aaij}} \emph {et~al.} (\bibinfo {collaboration} {LHCb}),\ }\bibfield
  {title} {\enquote {\bibinfo {title} {{Test of lepton universality in
  beauty-quark decays}},}\ }\href@noop {} {\  (\bibinfo {year}
  {2021}{\natexlab{a}})},\ \Eprint {http://arxiv.org/abs/2103.11769}
  {arXiv:2103.11769 [hep-ex]} \BibitemShut {NoStop}%
\bibitem [{\citenamefont {Schael}\ \emph {et~al.}(2006)\citenamefont {Schael}
  \emph {et~al.}}]{ALEPH:2005ab}%
  \BibitemOpen
  \bibfield  {author} {\bibinfo {author} {\bibfnamefont {S.}~\bibnamefont
  {Schael}} \emph {et~al.} (\bibinfo {collaboration} {ALEPH, DELPHI, L3, OPAL,
  SLD, LEP Electroweak Working Group, SLD Electroweak Group, SLD Heavy Flavour
  Group}),\ }\bibfield  {title} {\enquote {\bibinfo {title} {{Precision
  electroweak measurements on the $Z$ resonance}},}\ }\href {\doibase
  10.1016/j.physrep.2005.12.006} {\bibfield  {journal} {\bibinfo  {journal}
  {Phys. Rept.}\ }\textbf {\bibinfo {volume} {427}},\ \bibinfo {pages}
  {257--454} (\bibinfo {year} {2006})},\ \Eprint
  {http://arxiv.org/abs/hep-ex/0509008} {arXiv:hep-ex/0509008} \BibitemShut
  {NoStop}%
\bibitem [{\citenamefont {ALEPH}(2010)}]{ALEPH:2010aa}%
  \BibitemOpen
  \bibfield  {author} {\bibinfo {author} {\bibfnamefont {D0~DELPHI L3 OPAL SLD
  LEP Electroweak Working Group Tevatron Electroweak Working Group SLD
  Electroweak Heavy Flavour~Groups}\ \bibnamefont {ALEPH}, \bibfnamefont
  {CDF}},\ }\bibfield  {title} {\enquote {\bibinfo {title} {{Precision
  Electroweak Measurements and Constraints on the Standard Model}},}\
  }\href@noop {} {\  (\bibinfo {year} {2010})},\ \Eprint
  {http://arxiv.org/abs/1012.2367} {arXiv:1012.2367 [hep-ex]} \BibitemShut
  {NoStop}%
\bibitem [{\citenamefont {Aaij}\ \emph
  {et~al.}(2017{\natexlab{a}})\citenamefont {Aaij} \emph
  {et~al.}}]{Aaij:2017vbb}%
  \BibitemOpen
  \bibfield  {author} {\bibinfo {author} {\bibfnamefont {R.}~\bibnamefont
  {Aaij}} \emph {et~al.} (\bibinfo {collaboration} {LHCb}),\ }\bibfield
  {title} {\enquote {\bibinfo {title} {{Test of lepton universality with $B^{0}
  \rightarrow K^{*0}\ell^{+}\ell^{-}$ decays}},}\ }\href {\doibase
  10.1007/JHEP08(2017)055} {\bibfield  {journal} {\bibinfo  {journal} {JHEP}\
  }\textbf {\bibinfo {volume} {08}},\ \bibinfo {pages} {055} (\bibinfo {year}
  {2017}{\natexlab{a}})},\ \Eprint {http://arxiv.org/abs/1705.05802}
  {arXiv:1705.05802 [hep-ex]} \BibitemShut {NoStop}%
\bibitem [{\citenamefont {Aaij}\ \emph {et~al.}(2020)\citenamefont {Aaij} \emph
  {et~al.}}]{Aaij:2020nrf}%
  \BibitemOpen
  \bibfield  {author} {\bibinfo {author} {\bibfnamefont {Roel}\ \bibnamefont
  {Aaij}} \emph {et~al.} (\bibinfo {collaboration} {LHCb}),\ }\bibfield
  {title} {\enquote {\bibinfo {title} {{Measurement of $CP$-Averaged
  Observables in the $B^{0}\rightarrow K^{*0}\mu^{+}\mu^{-}$ Decay}},}\ }\href
  {\doibase 10.1103/PhysRevLett.125.011802} {\bibfield  {journal} {\bibinfo
  {journal} {Phys. Rev. Lett.}\ }\textbf {\bibinfo {volume} {125}},\ \bibinfo
  {pages} {011802} (\bibinfo {year} {2020})},\ \Eprint
  {http://arxiv.org/abs/2003.04831} {arXiv:2003.04831 [hep-ex]} \BibitemShut
  {NoStop}%
\bibitem [{\citenamefont {Aaij}\ \emph
  {et~al.}(2021{\natexlab{b}})\citenamefont {Aaij} \emph
  {et~al.}}]{LHCb:2020gog}%
  \BibitemOpen
  \bibfield  {author} {\bibinfo {author} {\bibfnamefont {Roel}\ \bibnamefont
  {Aaij}} \emph {et~al.} (\bibinfo {collaboration} {LHCb}),\ }\bibfield
  {title} {\enquote {\bibinfo {title} {{Angular Analysis of the
  $B^{+}\rightarrow K^{\ast+}\mu^{+}\mu^{-}$ Decay}},}\ }\href {\doibase
  10.1103/PhysRevLett.126.161802} {\bibfield  {journal} {\bibinfo  {journal}
  {Phys. Rev. Lett.}\ }\textbf {\bibinfo {volume} {126}},\ \bibinfo {pages}
  {161802} (\bibinfo {year} {2021}{\natexlab{b}})},\ \Eprint
  {http://arxiv.org/abs/2012.13241} {arXiv:2012.13241 [hep-ex]} \BibitemShut
  {NoStop}%
\bibitem [{\citenamefont {Lees}\ \emph {et~al.}(2012)\citenamefont {Lees} \emph
  {et~al.}}]{Lees:2012xj}%
  \BibitemOpen
  \bibfield  {author} {\bibinfo {author} {\bibfnamefont {J.~P.}\ \bibnamefont
  {Lees}} \emph {et~al.} (\bibinfo {collaboration} {BaBar}),\ }\bibfield
  {title} {\enquote {\bibinfo {title} {{Evidence for an excess of $\bar{B} \to
  D^{(*)} \tau^-\bar{\nu}_\tau$ decays}},}\ }\href {\doibase
  10.1103/PhysRevLett.109.101802} {\bibfield  {journal} {\bibinfo  {journal}
  {Phys. Rev. Lett.}\ }\textbf {\bibinfo {volume} {109}},\ \bibinfo {pages}
  {101802} (\bibinfo {year} {2012})},\ \Eprint {http://arxiv.org/abs/1205.5442}
  {arXiv:1205.5442 [hep-ex]} \BibitemShut {NoStop}%
\bibitem [{\citenamefont {Aaij}\ \emph {et~al.}(2015)\citenamefont {Aaij} \emph
  {et~al.}}]{Aaij:2015yra}%
  \BibitemOpen
  \bibfield  {author} {\bibinfo {author} {\bibfnamefont {Roel}\ \bibnamefont
  {Aaij}} \emph {et~al.} (\bibinfo {collaboration} {LHCb}),\ }\bibfield
  {title} {\enquote {\bibinfo {title} {{Measurement of the ratio of branching
  fractions $\mathcal{B}(\bar{B}^0 \to
  D^{*+}\tau^{-}\bar{\nu}_{\tau})/\mathcal{B}(\bar{B}^0 \to
  D^{*+}\mu^{-}\bar{\nu}_{\mu})$}},}\ }\href {\doibase
  10.1103/PhysRevLett.115.111803} {\bibfield  {journal} {\bibinfo  {journal}
  {Phys. Rev. Lett.}\ }\textbf {\bibinfo {volume} {115}},\ \bibinfo {pages}
  {111803} (\bibinfo {year} {2015})},\ \bibinfo {note} {[Erratum:
  Phys.Rev.Lett. 115, 159901 (2015)]},\ \Eprint
  {http://arxiv.org/abs/1506.08614} {arXiv:1506.08614 [hep-ex]} \BibitemShut
  {NoStop}%
\bibitem [{\citenamefont {Huschle}\ \emph {et~al.}(2015)\citenamefont {Huschle}
  \emph {et~al.}}]{Huschle:2015rga}%
  \BibitemOpen
  \bibfield  {author} {\bibinfo {author} {\bibfnamefont {M.}~\bibnamefont
  {Huschle}} \emph {et~al.} (\bibinfo {collaboration} {Belle}),\ }\bibfield
  {title} {\enquote {\bibinfo {title} {{Measurement of the branching ratio of
  $\bar{B} \to D^{(\ast)} \tau^- \bar{\nu}_\tau$ relative to $\bar{B} \to
  D^{(\ast)} \ell^- \bar{\nu}_\ell$ decays with hadronic tagging at Belle}},}\
  }\href {\doibase 10.1103/PhysRevD.92.072014} {\bibfield  {journal} {\bibinfo
  {journal} {Phys. Rev. D}\ }\textbf {\bibinfo {volume} {92}},\ \bibinfo
  {pages} {072014} (\bibinfo {year} {2015})},\ \Eprint
  {http://arxiv.org/abs/1507.03233} {arXiv:1507.03233 [hep-ex]} \BibitemShut
  {NoStop}%
\bibitem [{\citenamefont {Hirose}\ \emph {et~al.}(2017)\citenamefont {Hirose}
  \emph {et~al.}}]{Hirose:2016wfn}%
  \BibitemOpen
  \bibfield  {author} {\bibinfo {author} {\bibfnamefont {S.}~\bibnamefont
  {Hirose}} \emph {et~al.} (\bibinfo {collaboration} {Belle}),\ }\bibfield
  {title} {\enquote {\bibinfo {title} {{Measurement of the $\tau$ lepton
  polarization and $R(D^*)$ in the decay $\bar{B} \to D^* \tau^-
  \bar{\nu}_\tau$}},}\ }\href {\doibase 10.1103/PhysRevLett.118.211801}
  {\bibfield  {journal} {\bibinfo  {journal} {Phys. Rev. Lett.}\ }\textbf
  {\bibinfo {volume} {118}},\ \bibinfo {pages} {211801} (\bibinfo {year}
  {2017})},\ \Eprint {http://arxiv.org/abs/1612.00529} {arXiv:1612.00529
  [hep-ex]} \BibitemShut {NoStop}%
\bibitem [{\citenamefont {Aaij}\ \emph {et~al.}(2018)\citenamefont {Aaij} \emph
  {et~al.}}]{Aaij:2017uff}%
  \BibitemOpen
  \bibfield  {author} {\bibinfo {author} {\bibfnamefont {R.}~\bibnamefont
  {Aaij}} \emph {et~al.} (\bibinfo {collaboration} {LHCb}),\ }\bibfield
  {title} {\enquote {\bibinfo {title} {{Measurement of the ratio of the $B^0
  \to D^{*-} \tau^+ \nu_{\tau}$ and $B^0 \to D^{*-} \mu^+ \nu_{\mu}$ branching
  fractions using three-prong $\tau$-lepton decays}},}\ }\href {\doibase
  10.1103/PhysRevLett.120.171802} {\bibfield  {journal} {\bibinfo  {journal}
  {Phys. Rev. Lett.}\ }\textbf {\bibinfo {volume} {120}},\ \bibinfo {pages}
  {171802} (\bibinfo {year} {2018})},\ \Eprint
  {http://arxiv.org/abs/1708.08856} {arXiv:1708.08856 [hep-ex]} \BibitemShut
  {NoStop}%
\bibitem [{\citenamefont {Abdesselam}\ \emph {et~al.}(2019)\citenamefont
  {Abdesselam} \emph {et~al.}}]{Abdesselam:2019dgh}%
  \BibitemOpen
  \bibfield  {author} {\bibinfo {author} {\bibfnamefont {A.}~\bibnamefont
  {Abdesselam}} \emph {et~al.} (\bibinfo {collaboration} {Belle}),\ }\bibfield
  {title} {\enquote {\bibinfo {title} {{Measurement of $\mathcal{R}(D)$ and
  $\mathcal{R}(D^{\ast})$ with a semileptonic tagging method}},}\ }\href@noop
  {} {\  (\bibinfo {year} {2019})},\ \Eprint {http://arxiv.org/abs/1904.08794}
  {arXiv:1904.08794 [hep-ex]} \BibitemShut {NoStop}%
\bibitem [{\citenamefont {Aoyama}\ \emph {et~al.}(2020)\citenamefont {Aoyama}
  \emph {et~al.}}]{Aoyama:2020ynm}%
  \BibitemOpen
  \bibfield  {author} {\bibinfo {author} {\bibfnamefont {T.}~\bibnamefont
  {Aoyama}} \emph {et~al.},\ }\bibfield  {title} {\enquote {\bibinfo {title}
  {{The anomalous magnetic moment of the muon in the Standard Model}},}\ }\href
  {\doibase 10.1016/j.physrep.2020.07.006} {\bibfield  {journal} {\bibinfo
  {journal} {Phys. Rept.}\ }\textbf {\bibinfo {volume} {887}},\ \bibinfo
  {pages} {1--166} (\bibinfo {year} {2020})},\ \Eprint
  {http://arxiv.org/abs/2006.04822} {arXiv:2006.04822 [hep-ph]} \BibitemShut
  {NoStop}%
\bibitem [{\citenamefont {Aoyama}\ \emph {et~al.}(2012)\citenamefont {Aoyama},
  \citenamefont {Hayakawa}, \citenamefont {Kinoshita},\ and\ \citenamefont
  {Nio}}]{Aoyama:2012wk}%
  \BibitemOpen
  \bibfield  {author} {\bibinfo {author} {\bibfnamefont {Tatsumi}\ \bibnamefont
  {Aoyama}}, \bibinfo {author} {\bibfnamefont {Masashi}\ \bibnamefont
  {Hayakawa}}, \bibinfo {author} {\bibfnamefont {Toichiro}\ \bibnamefont
  {Kinoshita}}, \ and\ \bibinfo {author} {\bibfnamefont {Makiko}\ \bibnamefont
  {Nio}},\ }\bibfield  {title} {\enquote {\bibinfo {title} {{Complete
  Tenth-Order QED Contribution to the Muon g-2}},}\ }\href {\doibase
  10.1103/PhysRevLett.109.111808} {\bibfield  {journal} {\bibinfo  {journal}
  {Phys. Rev. Lett.}\ }\textbf {\bibinfo {volume} {109}},\ \bibinfo {pages}
  {111808} (\bibinfo {year} {2012})},\ \Eprint {http://arxiv.org/abs/1205.5370}
  {arXiv:1205.5370 [hep-ph]} \BibitemShut {NoStop}%
\bibitem [{\citenamefont {Aoyama}\ \emph {et~al.}(2019)\citenamefont {Aoyama},
  \citenamefont {Kinoshita},\ and\ \citenamefont {Nio}}]{Aoyama:2019ryr}%
  \BibitemOpen
  \bibfield  {author} {\bibinfo {author} {\bibfnamefont {Tatsumi}\ \bibnamefont
  {Aoyama}}, \bibinfo {author} {\bibfnamefont {Toichiro}\ \bibnamefont
  {Kinoshita}}, \ and\ \bibinfo {author} {\bibfnamefont {Makiko}\ \bibnamefont
  {Nio}},\ }\bibfield  {title} {\enquote {\bibinfo {title} {{Theory of the
  Anomalous Magnetic Moment of the Electron}},}\ }\href {\doibase
  10.3390/atoms7010028} {\bibfield  {journal} {\bibinfo  {journal} {Atoms}\
  }\textbf {\bibinfo {volume} {7}},\ \bibinfo {pages} {28} (\bibinfo {year}
  {2019})}\BibitemShut {NoStop}%
\bibitem [{\citenamefont {Czarnecki}\ \emph {et~al.}(2003)\citenamefont
  {Czarnecki}, \citenamefont {Marciano},\ and\ \citenamefont
  {Vainshtein}}]{Czarnecki:2002nt}%
  \BibitemOpen
  \bibfield  {author} {\bibinfo {author} {\bibfnamefont {Andrzej}\ \bibnamefont
  {Czarnecki}}, \bibinfo {author} {\bibfnamefont {William~J.}\ \bibnamefont
  {Marciano}}, \ and\ \bibinfo {author} {\bibfnamefont {Arkady}\ \bibnamefont
  {Vainshtein}},\ }\bibfield  {title} {\enquote {\bibinfo {title} {{Refinements
  in electroweak contributions to the muon anomalous magnetic moment}},}\
  }\href {\doibase 10.1103/PhysRevD.67.073006} {\bibfield  {journal} {\bibinfo
  {journal} {Phys. Rev. D}\ }\textbf {\bibinfo {volume} {67}},\ \bibinfo
  {pages} {073006} (\bibinfo {year} {2003})},\ \bibinfo {note} {[Erratum:
  Phys.Rev.D 73, 119901 (2006)]},\ \Eprint
  {http://arxiv.org/abs/hep-ph/0212229} {arXiv:hep-ph/0212229} \BibitemShut
  {NoStop}%
\bibitem [{\citenamefont {Gnendiger}\ \emph {et~al.}(2013)\citenamefont
  {Gnendiger}, \citenamefont {St\"ockinger},\ and\ \citenamefont
  {St\"ockinger-Kim}}]{Gnendiger:2013pva}%
  \BibitemOpen
  \bibfield  {author} {\bibinfo {author} {\bibfnamefont {C.}~\bibnamefont
  {Gnendiger}}, \bibinfo {author} {\bibfnamefont {D.}~\bibnamefont
  {St\"ockinger}}, \ and\ \bibinfo {author} {\bibfnamefont {H.}~\bibnamefont
  {St\"ockinger-Kim}},\ }\bibfield  {title} {\enquote {\bibinfo {title} {{The
  electroweak contributions to $(g-2)_\mu$ after the Higgs boson mass
  measurement}},}\ }\href {\doibase 10.1103/PhysRevD.88.053005} {\bibfield
  {journal} {\bibinfo  {journal} {Phys. Rev. D}\ }\textbf {\bibinfo {volume}
  {88}},\ \bibinfo {pages} {053005} (\bibinfo {year} {2013})},\ \Eprint
  {http://arxiv.org/abs/1306.5546} {arXiv:1306.5546 [hep-ph]} \BibitemShut
  {NoStop}%
\bibitem [{\citenamefont {Davier}\ \emph {et~al.}(2017)\citenamefont {Davier},
  \citenamefont {Hoecker}, \citenamefont {Malaescu},\ and\ \citenamefont
  {Zhang}}]{Davier:2017zfy}%
  \BibitemOpen
  \bibfield  {author} {\bibinfo {author} {\bibfnamefont {Michel}\ \bibnamefont
  {Davier}}, \bibinfo {author} {\bibfnamefont {Andreas}\ \bibnamefont
  {Hoecker}}, \bibinfo {author} {\bibfnamefont {Bogdan}\ \bibnamefont
  {Malaescu}}, \ and\ \bibinfo {author} {\bibfnamefont {Zhiqing}\ \bibnamefont
  {Zhang}},\ }\bibfield  {title} {\enquote {\bibinfo {title} {{Reevaluation of
  the hadronic vacuum polarisation contributions to the Standard Model
  predictions of the muon $g-2$ and ${\alpha (m_Z^2)}$ using newest hadronic
  cross-section data}},}\ }\href {\doibase 10.1140/epjc/s10052-017-5161-6}
  {\bibfield  {journal} {\bibinfo  {journal} {Eur. Phys. J. C}\ }\textbf
  {\bibinfo {volume} {77}},\ \bibinfo {pages} {827} (\bibinfo {year} {2017})},\
  \Eprint {http://arxiv.org/abs/1706.09436} {arXiv:1706.09436 [hep-ph]}
  \BibitemShut {NoStop}%
\bibitem [{\citenamefont {Keshavarzi}\ \emph {et~al.}(2018)\citenamefont
  {Keshavarzi}, \citenamefont {Nomura},\ and\ \citenamefont
  {Teubner}}]{Keshavarzi:2018mgv}%
  \BibitemOpen
  \bibfield  {author} {\bibinfo {author} {\bibfnamefont {Alexander}\
  \bibnamefont {Keshavarzi}}, \bibinfo {author} {\bibfnamefont {Daisuke}\
  \bibnamefont {Nomura}}, \ and\ \bibinfo {author} {\bibfnamefont {Thomas}\
  \bibnamefont {Teubner}},\ }\bibfield  {title} {\enquote {\bibinfo {title}
  {{Muon $g-2$ and $\alpha(M_Z^2)$: a new data-based analysis}},}\ }\href
  {\doibase 10.1103/PhysRevD.97.114025} {\bibfield  {journal} {\bibinfo
  {journal} {Phys. Rev. D}\ }\textbf {\bibinfo {volume} {97}},\ \bibinfo
  {pages} {114025} (\bibinfo {year} {2018})},\ \Eprint
  {http://arxiv.org/abs/1802.02995} {arXiv:1802.02995 [hep-ph]} \BibitemShut
  {NoStop}%
\bibitem [{\citenamefont {Colangelo}\ \emph {et~al.}(2019)\citenamefont
  {Colangelo}, \citenamefont {Hoferichter},\ and\ \citenamefont
  {Stoffer}}]{Colangelo:2018mtw}%
  \BibitemOpen
  \bibfield  {author} {\bibinfo {author} {\bibfnamefont {Gilberto}\
  \bibnamefont {Colangelo}}, \bibinfo {author} {\bibfnamefont {Martin}\
  \bibnamefont {Hoferichter}}, \ and\ \bibinfo {author} {\bibfnamefont {Peter}\
  \bibnamefont {Stoffer}},\ }\bibfield  {title} {\enquote {\bibinfo {title}
  {{Two-pion contribution to hadronic vacuum polarization}},}\ }\href {\doibase
  10.1007/JHEP02(2019)006} {\bibfield  {journal} {\bibinfo  {journal} {JHEP}\
  }\textbf {\bibinfo {volume} {02}},\ \bibinfo {pages} {006} (\bibinfo {year}
  {2019})},\ \Eprint {http://arxiv.org/abs/1810.00007} {arXiv:1810.00007
  [hep-ph]} \BibitemShut {NoStop}%
\bibitem [{\citenamefont {Hoferichter}\ \emph {et~al.}(2019)\citenamefont
  {Hoferichter}, \citenamefont {Hoid},\ and\ \citenamefont
  {Kubis}}]{Hoferichter:2019mqg}%
  \BibitemOpen
  \bibfield  {author} {\bibinfo {author} {\bibfnamefont {Martin}\ \bibnamefont
  {Hoferichter}}, \bibinfo {author} {\bibfnamefont {Bai-Long}\ \bibnamefont
  {Hoid}}, \ and\ \bibinfo {author} {\bibfnamefont {Bastian}\ \bibnamefont
  {Kubis}},\ }\bibfield  {title} {\enquote {\bibinfo {title} {{Three-pion
  contribution to hadronic vacuum polarization}},}\ }\href {\doibase
  10.1007/JHEP08(2019)137} {\bibfield  {journal} {\bibinfo  {journal} {JHEP}\
  }\textbf {\bibinfo {volume} {08}},\ \bibinfo {pages} {137} (\bibinfo {year}
  {2019})},\ \Eprint {http://arxiv.org/abs/1907.01556} {arXiv:1907.01556
  [hep-ph]} \BibitemShut {NoStop}%
\bibitem [{\citenamefont {Davier}\ \emph {et~al.}(2020)\citenamefont {Davier},
  \citenamefont {Hoecker}, \citenamefont {Malaescu},\ and\ \citenamefont
  {Zhang}}]{Davier:2019can}%
  \BibitemOpen
  \bibfield  {author} {\bibinfo {author} {\bibfnamefont {M.}~\bibnamefont
  {Davier}}, \bibinfo {author} {\bibfnamefont {A.}~\bibnamefont {Hoecker}},
  \bibinfo {author} {\bibfnamefont {B.}~\bibnamefont {Malaescu}}, \ and\
  \bibinfo {author} {\bibfnamefont {Z.}~\bibnamefont {Zhang}},\ }\bibfield
  {title} {\enquote {\bibinfo {title} {{A new evaluation of the hadronic vacuum
  polarisation contributions to the muon anomalous magnetic moment and to
  $\mathbf{\boldsymbol\alpha(m_Z^2)}$}},}\ }\href {\doibase
  10.1140/epjc/s10052-020-7792-2} {\bibfield  {journal} {\bibinfo  {journal}
  {Eur. Phys. J. C}\ }\textbf {\bibinfo {volume} {80}},\ \bibinfo {pages} {241}
  (\bibinfo {year} {2020})},\ \bibinfo {note} {[Erratum: Eur.Phys.J.C 80, 410
  (2020)]},\ \Eprint {http://arxiv.org/abs/1908.00921} {arXiv:1908.00921
  [hep-ph]} \BibitemShut {NoStop}%
\bibitem [{\citenamefont {Keshavarzi}\ \emph
  {et~al.}(2020{\natexlab{a}})\citenamefont {Keshavarzi}, \citenamefont
  {Nomura},\ and\ \citenamefont {Teubner}}]{Keshavarzi:2019abf}%
  \BibitemOpen
  \bibfield  {author} {\bibinfo {author} {\bibfnamefont {Alexander}\
  \bibnamefont {Keshavarzi}}, \bibinfo {author} {\bibfnamefont {Daisuke}\
  \bibnamefont {Nomura}}, \ and\ \bibinfo {author} {\bibfnamefont {Thomas}\
  \bibnamefont {Teubner}},\ }\bibfield  {title} {\enquote {\bibinfo {title}
  {{$g-2$ of charged leptons, $\alpha (M^2_Z)$ , and the hyperfine splitting of
  muonium}},}\ }\href {\doibase 10.1103/PhysRevD.101.014029} {\bibfield
  {journal} {\bibinfo  {journal} {Phys. Rev. D}\ }\textbf {\bibinfo {volume}
  {101}},\ \bibinfo {pages} {014029} (\bibinfo {year} {2020}{\natexlab{a}})},\
  \Eprint {http://arxiv.org/abs/1911.00367} {arXiv:1911.00367 [hep-ph]}
  \BibitemShut {NoStop}%
\bibitem [{\citenamefont {Kurz}\ \emph {et~al.}(2014)\citenamefont {Kurz},
  \citenamefont {Liu}, \citenamefont {Marquard},\ and\ \citenamefont
  {Steinhauser}}]{Kurz:2014wya}%
  \BibitemOpen
  \bibfield  {author} {\bibinfo {author} {\bibfnamefont {Alexander}\
  \bibnamefont {Kurz}}, \bibinfo {author} {\bibfnamefont {Tao}\ \bibnamefont
  {Liu}}, \bibinfo {author} {\bibfnamefont {Peter}\ \bibnamefont {Marquard}}, \
  and\ \bibinfo {author} {\bibfnamefont {Matthias}\ \bibnamefont
  {Steinhauser}},\ }\bibfield  {title} {\enquote {\bibinfo {title} {{Hadronic
  contribution to the muon anomalous magnetic moment to next-to-next-to-leading
  order}},}\ }\href {\doibase 10.1016/j.physletb.2014.05.043} {\bibfield
  {journal} {\bibinfo  {journal} {Phys. Lett. B}\ }\textbf {\bibinfo {volume}
  {734}},\ \bibinfo {pages} {144--147} (\bibinfo {year} {2014})},\ \Eprint
  {http://arxiv.org/abs/1403.6400} {arXiv:1403.6400 [hep-ph]} \BibitemShut
  {NoStop}%
\bibitem [{\citenamefont {Melnikov}\ and\ \citenamefont
  {Vainshtein}(2004)}]{Melnikov:2003xd}%
  \BibitemOpen
  \bibfield  {author} {\bibinfo {author} {\bibfnamefont {Kirill}\ \bibnamefont
  {Melnikov}}\ and\ \bibinfo {author} {\bibfnamefont {Arkady}\ \bibnamefont
  {Vainshtein}},\ }\bibfield  {title} {\enquote {\bibinfo {title} {{Hadronic
  light-by-light scattering contribution to the muon anomalous magnetic moment
  revisited}},}\ }\href {\doibase 10.1103/PhysRevD.70.113006} {\bibfield
  {journal} {\bibinfo  {journal} {Phys. Rev. D}\ }\textbf {\bibinfo {volume}
  {70}},\ \bibinfo {pages} {113006} (\bibinfo {year} {2004})},\ \Eprint
  {http://arxiv.org/abs/hep-ph/0312226} {arXiv:hep-ph/0312226} \BibitemShut
  {NoStop}%
\bibitem [{\citenamefont {Masjuan}\ and\ \citenamefont
  {Sanchez-Puertas}(2017)}]{Masjuan:2017tvw}%
  \BibitemOpen
  \bibfield  {author} {\bibinfo {author} {\bibfnamefont {Pere}\ \bibnamefont
  {Masjuan}}\ and\ \bibinfo {author} {\bibfnamefont {Pablo}\ \bibnamefont
  {Sanchez-Puertas}},\ }\bibfield  {title} {\enquote {\bibinfo {title}
  {{Pseudoscalar-pole contribution to the $(g_{\mu}-2)$: a rational
  approach}},}\ }\href {\doibase 10.1103/PhysRevD.95.054026} {\bibfield
  {journal} {\bibinfo  {journal} {Phys. Rev. D}\ }\textbf {\bibinfo {volume}
  {95}},\ \bibinfo {pages} {054026} (\bibinfo {year} {2017})},\ \Eprint
  {http://arxiv.org/abs/1701.05829} {arXiv:1701.05829 [hep-ph]} \BibitemShut
  {NoStop}%
\bibitem [{\citenamefont {Colangelo}\ \emph {et~al.}(2017)\citenamefont
  {Colangelo}, \citenamefont {Hoferichter}, \citenamefont {Procura},\ and\
  \citenamefont {Stoffer}}]{Colangelo:2017fiz}%
  \BibitemOpen
  \bibfield  {author} {\bibinfo {author} {\bibfnamefont {Gilberto}\
  \bibnamefont {Colangelo}}, \bibinfo {author} {\bibfnamefont {Martin}\
  \bibnamefont {Hoferichter}}, \bibinfo {author} {\bibfnamefont {Massimiliano}\
  \bibnamefont {Procura}}, \ and\ \bibinfo {author} {\bibfnamefont {Peter}\
  \bibnamefont {Stoffer}},\ }\bibfield  {title} {\enquote {\bibinfo {title}
  {{Dispersion relation for hadronic light-by-light scattering: two-pion
  contributions}},}\ }\href {\doibase 10.1007/JHEP04(2017)161} {\bibfield
  {journal} {\bibinfo  {journal} {JHEP}\ }\textbf {\bibinfo {volume} {04}},\
  \bibinfo {pages} {161} (\bibinfo {year} {2017})},\ \Eprint
  {http://arxiv.org/abs/1702.07347} {arXiv:1702.07347 [hep-ph]} \BibitemShut
  {NoStop}%
\bibitem [{\citenamefont {Hoferichter}\ \emph {et~al.}(2018)\citenamefont
  {Hoferichter}, \citenamefont {Hoid}, \citenamefont {Kubis}, \citenamefont
  {Leupold},\ and\ \citenamefont {Schneider}}]{Hoferichter:2018kwz}%
  \BibitemOpen
  \bibfield  {author} {\bibinfo {author} {\bibfnamefont {Martin}\ \bibnamefont
  {Hoferichter}}, \bibinfo {author} {\bibfnamefont {Bai-Long}\ \bibnamefont
  {Hoid}}, \bibinfo {author} {\bibfnamefont {Bastian}\ \bibnamefont {Kubis}},
  \bibinfo {author} {\bibfnamefont {Stefan}\ \bibnamefont {Leupold}}, \ and\
  \bibinfo {author} {\bibfnamefont {Sebastian~P.}\ \bibnamefont {Schneider}},\
  }\bibfield  {title} {\enquote {\bibinfo {title} {{Dispersion relation for
  hadronic light-by-light scattering: pion pole}},}\ }\href {\doibase
  10.1007/JHEP10(2018)141} {\bibfield  {journal} {\bibinfo  {journal} {JHEP}\
  }\textbf {\bibinfo {volume} {10}},\ \bibinfo {pages} {141} (\bibinfo {year}
  {2018})},\ \Eprint {http://arxiv.org/abs/1808.04823} {arXiv:1808.04823
  [hep-ph]} \BibitemShut {NoStop}%
\bibitem [{\citenamefont {G\'erardin}\ \emph {et~al.}(2019)\citenamefont
  {G\'erardin}, \citenamefont {Meyer},\ and\ \citenamefont
  {Nyffeler}}]{Gerardin:2019vio}%
  \BibitemOpen
  \bibfield  {author} {\bibinfo {author} {\bibfnamefont {Antoine}\ \bibnamefont
  {G\'erardin}}, \bibinfo {author} {\bibfnamefont {Harvey~B.}\ \bibnamefont
  {Meyer}}, \ and\ \bibinfo {author} {\bibfnamefont {Andreas}\ \bibnamefont
  {Nyffeler}},\ }\bibfield  {title} {\enquote {\bibinfo {title} {{Lattice
  calculation of the pion transition form factor with $N_f=2+1$ Wilson
  quarks}},}\ }\href {\doibase 10.1103/PhysRevD.100.034520} {\bibfield
  {journal} {\bibinfo  {journal} {Phys. Rev. D}\ }\textbf {\bibinfo {volume}
  {100}},\ \bibinfo {pages} {034520} (\bibinfo {year} {2019})},\ \Eprint
  {http://arxiv.org/abs/1903.09471} {arXiv:1903.09471 [hep-lat]} \BibitemShut
  {NoStop}%
\bibitem [{\citenamefont {Bijnens}\ \emph {et~al.}(2019)\citenamefont
  {Bijnens}, \citenamefont {Hermansson-Truedsson},\ and\ \citenamefont
  {Rodr\'\i{}guez-S\'anchez}}]{Bijnens:2019ghy}%
  \BibitemOpen
  \bibfield  {author} {\bibinfo {author} {\bibfnamefont {Johan}\ \bibnamefont
  {Bijnens}}, \bibinfo {author} {\bibfnamefont {Nils}\ \bibnamefont
  {Hermansson-Truedsson}}, \ and\ \bibinfo {author} {\bibfnamefont {Antonio}\
  \bibnamefont {Rodr\'\i{}guez-S\'anchez}},\ }\bibfield  {title} {\enquote
  {\bibinfo {title} {{Short-distance constraints for the HLbL contribution to
  the muon anomalous magnetic moment}},}\ }\href {\doibase
  10.1016/j.physletb.2019.134994} {\bibfield  {journal} {\bibinfo  {journal}
  {Phys. Lett. B}\ }\textbf {\bibinfo {volume} {798}},\ \bibinfo {pages}
  {134994} (\bibinfo {year} {2019})},\ \Eprint
  {http://arxiv.org/abs/1908.03331} {arXiv:1908.03331 [hep-ph]} \BibitemShut
  {NoStop}%
\bibitem [{\citenamefont {Colangelo}\ \emph {et~al.}(2020)\citenamefont
  {Colangelo}, \citenamefont {Hagelstein}, \citenamefont {Hoferichter},
  \citenamefont {Laub},\ and\ \citenamefont {Stoffer}}]{Colangelo:2019uex}%
  \BibitemOpen
  \bibfield  {author} {\bibinfo {author} {\bibfnamefont {Gilberto}\
  \bibnamefont {Colangelo}}, \bibinfo {author} {\bibfnamefont {Franziska}\
  \bibnamefont {Hagelstein}}, \bibinfo {author} {\bibfnamefont {Martin}\
  \bibnamefont {Hoferichter}}, \bibinfo {author} {\bibfnamefont {Laetitia}\
  \bibnamefont {Laub}}, \ and\ \bibinfo {author} {\bibfnamefont {Peter}\
  \bibnamefont {Stoffer}},\ }\bibfield  {title} {\enquote {\bibinfo {title}
  {{Longitudinal short-distance constraints for the hadronic light-by-light
  contribution to $(g-2)_\mu$ with large-$N_c$ Regge models}},}\ }\href
  {\doibase 10.1007/JHEP03(2020)101} {\bibfield  {journal} {\bibinfo  {journal}
  {JHEP}\ }\textbf {\bibinfo {volume} {03}},\ \bibinfo {pages} {101} (\bibinfo
  {year} {2020})},\ \Eprint {http://arxiv.org/abs/1910.13432} {arXiv:1910.13432
  [hep-ph]} \BibitemShut {NoStop}%
\bibitem [{\citenamefont {Blum}\ \emph {et~al.}(2020)\citenamefont {Blum},
  \citenamefont {Christ}, \citenamefont {Hayakawa}, \citenamefont {Izubuchi},
  \citenamefont {Jin}, \citenamefont {Jung},\ and\ \citenamefont
  {Lehner}}]{Blum:2019ugy}%
  \BibitemOpen
  \bibfield  {author} {\bibinfo {author} {\bibfnamefont {Thomas}\ \bibnamefont
  {Blum}}, \bibinfo {author} {\bibfnamefont {Norman}\ \bibnamefont {Christ}},
  \bibinfo {author} {\bibfnamefont {Masashi}\ \bibnamefont {Hayakawa}},
  \bibinfo {author} {\bibfnamefont {Taku}\ \bibnamefont {Izubuchi}}, \bibinfo
  {author} {\bibfnamefont {Luchang}\ \bibnamefont {Jin}}, \bibinfo {author}
  {\bibfnamefont {Chulwoo}\ \bibnamefont {Jung}}, \ and\ \bibinfo {author}
  {\bibfnamefont {Christoph}\ \bibnamefont {Lehner}},\ }\bibfield  {title}
  {\enquote {\bibinfo {title} {{Hadronic Light-by-Light Scattering Contribution
  to the Muon Anomalous Magnetic Moment from Lattice QCD}},}\ }\href {\doibase
  10.1103/PhysRevLett.124.132002} {\bibfield  {journal} {\bibinfo  {journal}
  {Phys. Rev. Lett.}\ }\textbf {\bibinfo {volume} {124}},\ \bibinfo {pages}
  {132002} (\bibinfo {year} {2020})},\ \Eprint
  {http://arxiv.org/abs/1911.08123} {arXiv:1911.08123 [hep-lat]} \BibitemShut
  {NoStop}%
\bibitem [{\citenamefont {Colangelo}\ \emph {et~al.}(2014)\citenamefont
  {Colangelo}, \citenamefont {Hoferichter}, \citenamefont {Nyffeler},
  \citenamefont {Passera},\ and\ \citenamefont {Stoffer}}]{Colangelo:2014qya}%
  \BibitemOpen
  \bibfield  {author} {\bibinfo {author} {\bibfnamefont {Gilberto}\
  \bibnamefont {Colangelo}}, \bibinfo {author} {\bibfnamefont {Martin}\
  \bibnamefont {Hoferichter}}, \bibinfo {author} {\bibfnamefont {Andreas}\
  \bibnamefont {Nyffeler}}, \bibinfo {author} {\bibfnamefont {Massimo}\
  \bibnamefont {Passera}}, \ and\ \bibinfo {author} {\bibfnamefont {Peter}\
  \bibnamefont {Stoffer}},\ }\bibfield  {title} {\enquote {\bibinfo {title}
  {{Remarks on higher-order hadronic corrections to the muon
  g\ensuremath{-}2}},}\ }\href {\doibase 10.1016/j.physletb.2014.06.012}
  {\bibfield  {journal} {\bibinfo  {journal} {Phys. Lett. B}\ }\textbf
  {\bibinfo {volume} {735}},\ \bibinfo {pages} {90--91} (\bibinfo {year}
  {2014})},\ \Eprint {http://arxiv.org/abs/1403.7512} {arXiv:1403.7512
  [hep-ph]} \BibitemShut {NoStop}%
\bibitem [{\citenamefont {Abi}\ \emph {et~al.}(2021)\citenamefont {Abi} \emph
  {et~al.}}]{Abi:2021gix}%
  \BibitemOpen
  \bibfield  {author} {\bibinfo {author} {\bibfnamefont {B.}~\bibnamefont
  {Abi}} \emph {et~al.} (\bibinfo {collaboration} {Muon g-2}),\ }\bibfield
  {title} {\enquote {\bibinfo {title} {{Measurement of the Positive Muon
  Anomalous Magnetic Moment to 0.46 ppm}},}\ }\href {\doibase
  10.1103/PhysRevLett.126.141801} {\bibfield  {journal} {\bibinfo  {journal}
  {Phys. Rev. Lett.}\ }\textbf {\bibinfo {volume} {126}},\ \bibinfo {pages}
  {141801} (\bibinfo {year} {2021})},\ \Eprint
  {http://arxiv.org/abs/2104.03281} {arXiv:2104.03281 [hep-ex]} \BibitemShut
  {NoStop}%
\bibitem [{\citenamefont {Bennett}\ \emph {et~al.}(2006)\citenamefont {Bennett}
  \emph {et~al.}}]{Bennett:2006fi}%
  \BibitemOpen
  \bibfield  {author} {\bibinfo {author} {\bibfnamefont {G.~W.}\ \bibnamefont
  {Bennett}} \emph {et~al.} (\bibinfo {collaboration} {Muon g-2}),\ }\bibfield
  {title} {\enquote {\bibinfo {title} {{Final Report of the Muon E821 Anomalous
  Magnetic Moment Measurement at BNL}},}\ }\href {\doibase
  10.1103/PhysRevD.73.072003} {\bibfield  {journal} {\bibinfo  {journal} {Phys.
  Rev. D}\ }\textbf {\bibinfo {volume} {73}},\ \bibinfo {pages} {072003}
  (\bibinfo {year} {2006})},\ \Eprint {http://arxiv.org/abs/hep-ex/0602035}
  {arXiv:hep-ex/0602035} \BibitemShut {NoStop}%
\bibitem [{\citenamefont {Borsanyi}\ \emph {et~al.}(2021)\citenamefont
  {Borsanyi} \emph {et~al.}}]{Borsanyi:2020mff}%
  \BibitemOpen
  \bibfield  {author} {\bibinfo {author} {\bibfnamefont {Sz.}\ \bibnamefont
  {Borsanyi}} \emph {et~al.},\ }\bibfield  {title} {\enquote {\bibinfo {title}
  {{Leading hadronic contribution to the muon magnetic moment from lattice
  QCD}},}\ }\href {\doibase 10.1038/s41586-021-03418-1} {\bibfield  {journal}
  {\bibinfo  {journal} {Nature}\ }\textbf {\bibinfo {volume} {593}},\ \bibinfo
  {pages} {51--55} (\bibinfo {year} {2021})},\ \Eprint
  {http://arxiv.org/abs/2002.12347} {arXiv:2002.12347 [hep-lat]} \BibitemShut
  {NoStop}%
\bibitem [{\citenamefont {Lehner}\ and\ \citenamefont
  {Meyer}(2020)}]{Lehner:2020crt}%
  \BibitemOpen
  \bibfield  {author} {\bibinfo {author} {\bibfnamefont {Christoph}\
  \bibnamefont {Lehner}}\ and\ \bibinfo {author} {\bibfnamefont {Aaron~S.}\
  \bibnamefont {Meyer}},\ }\bibfield  {title} {\enquote {\bibinfo {title}
  {{Consistency of hadronic vacuum polarization between lattice QCD and the
  R-ratio}},}\ }\href {\doibase 10.1103/PhysRevD.101.074515} {\bibfield
  {journal} {\bibinfo  {journal} {Phys. Rev. D}\ }\textbf {\bibinfo {volume}
  {101}},\ \bibinfo {pages} {074515} (\bibinfo {year} {2020})},\ \Eprint
  {http://arxiv.org/abs/2003.04177} {arXiv:2003.04177 [hep-lat]} \BibitemShut
  {NoStop}%
\bibitem [{\citenamefont {Crivellin}\ \emph {et~al.}(2020)\citenamefont
  {Crivellin}, \citenamefont {Hoferichter}, \citenamefont {Manzari},\ and\
  \citenamefont {Montull}}]{Crivellin:2020zul}%
  \BibitemOpen
  \bibfield  {author} {\bibinfo {author} {\bibfnamefont {Andreas}\ \bibnamefont
  {Crivellin}}, \bibinfo {author} {\bibfnamefont {Martin}\ \bibnamefont
  {Hoferichter}}, \bibinfo {author} {\bibfnamefont {Claudio~Andrea}\
  \bibnamefont {Manzari}}, \ and\ \bibinfo {author} {\bibfnamefont {Marc}\
  \bibnamefont {Montull}},\ }\bibfield  {title} {\enquote {\bibinfo {title}
  {{Hadronic Vacuum Polarization: $(g-2)_\mu$ versus Global Electroweak
  Fits}},}\ }\href {\doibase 10.1103/PhysRevLett.125.091801} {\bibfield
  {journal} {\bibinfo  {journal} {Phys. Rev. Lett.}\ }\textbf {\bibinfo
  {volume} {125}},\ \bibinfo {pages} {091801} (\bibinfo {year} {2020})},\
  \Eprint {http://arxiv.org/abs/2003.04886} {arXiv:2003.04886 [hep-ph]}
  \BibitemShut {NoStop}%
\bibitem [{\citenamefont {Keshavarzi}\ \emph
  {et~al.}(2020{\natexlab{b}})\citenamefont {Keshavarzi}, \citenamefont
  {Marciano}, \citenamefont {Passera},\ and\ \citenamefont
  {Sirlin}}]{Keshavarzi:2020bfy}%
  \BibitemOpen
  \bibfield  {author} {\bibinfo {author} {\bibfnamefont {Alexander}\
  \bibnamefont {Keshavarzi}}, \bibinfo {author} {\bibfnamefont {William~J.}\
  \bibnamefont {Marciano}}, \bibinfo {author} {\bibfnamefont {Massimo}\
  \bibnamefont {Passera}}, \ and\ \bibinfo {author} {\bibfnamefont {Alberto}\
  \bibnamefont {Sirlin}},\ }\bibfield  {title} {\enquote {\bibinfo {title}
  {{Muon $g-2$ and $\Delta \alpha$ connection}},}\ }\href {\doibase
  10.1103/PhysRevD.102.033002} {\bibfield  {journal} {\bibinfo  {journal}
  {Phys. Rev. D}\ }\textbf {\bibinfo {volume} {102}},\ \bibinfo {pages}
  {033002} (\bibinfo {year} {2020}{\natexlab{b}})},\ \Eprint
  {http://arxiv.org/abs/2006.12666} {arXiv:2006.12666 [hep-ph]} \BibitemShut
  {NoStop}%
\bibitem [{\citenamefont {de~Rafael}(2020)}]{deRafael:2020uif}%
  \BibitemOpen
  \bibfield  {author} {\bibinfo {author} {\bibfnamefont {Eduardo}\ \bibnamefont
  {de~Rafael}},\ }\bibfield  {title} {\enquote {\bibinfo {title} {{Constraints
  between $\Delta\alpha_{\rm had}(M_Z^2)$ and $(g_{\mu}-2)_{\rm HVP}$}},}\
  }\href {\doibase 10.1103/PhysRevD.102.056025} {\bibfield  {journal} {\bibinfo
   {journal} {Phys. Rev. D}\ }\textbf {\bibinfo {volume} {102}},\ \bibinfo
  {pages} {056025} (\bibinfo {year} {2020})},\ \Eprint
  {http://arxiv.org/abs/2006.13880} {arXiv:2006.13880 [hep-ph]} \BibitemShut
  {NoStop}%
\bibitem [{\citenamefont {Huang}\ and\ \citenamefont
  {Wagner}(2014)}]{Huang:2014xua}%
  \BibitemOpen
  \bibfield  {author} {\bibinfo {author} {\bibfnamefont {Peisi}\ \bibnamefont
  {Huang}}\ and\ \bibinfo {author} {\bibfnamefont {Carlos E.~M.}\ \bibnamefont
  {Wagner}},\ }\bibfield  {title} {\enquote {\bibinfo {title} {{Blind Spots for
  neutralino Dark Matter in the MSSM with an intermediate $m_A$}},}\ }\href
  {\doibase 10.1103/PhysRevD.90.015018} {\bibfield  {journal} {\bibinfo
  {journal} {Phys. Rev. D}\ }\textbf {\bibinfo {volume} {90}},\ \bibinfo
  {pages} {015018} (\bibinfo {year} {2014})},\ \Eprint
  {http://arxiv.org/abs/1404.0392} {arXiv:1404.0392 [hep-ph]} \BibitemShut
  {NoStop}%
\bibitem [{\citenamefont {Chakraborti}\ \emph {et~al.}(2020)\citenamefont
  {Chakraborti}, \citenamefont {Heinemeyer},\ and\ \citenamefont
  {Saha}}]{Chakraborti:2020vjp}%
  \BibitemOpen
  \bibfield  {author} {\bibinfo {author} {\bibfnamefont {Manimala}\
  \bibnamefont {Chakraborti}}, \bibinfo {author} {\bibfnamefont {Sven}\
  \bibnamefont {Heinemeyer}}, \ and\ \bibinfo {author} {\bibfnamefont {Ipsita}\
  \bibnamefont {Saha}},\ }\bibfield  {title} {\enquote {\bibinfo {title}
  {{Improved $(g-2)_\mu$ Measurements and Supersymmetry}},}\ }\href {\doibase
  10.1140/epjc/s10052-020-08504-8} {\bibfield  {journal} {\bibinfo  {journal}
  {Eur. Phys. J. C}\ }\textbf {\bibinfo {volume} {80}},\ \bibinfo {pages} {984}
  (\bibinfo {year} {2020})},\ \Eprint {http://arxiv.org/abs/2006.15157}
  {arXiv:2006.15157 [hep-ph]} \BibitemShut {NoStop}%
\bibitem [{\citenamefont {Chakraborti}\ \emph {et~al.}(2021)\citenamefont
  {Chakraborti}, \citenamefont {Heinemeyer},\ and\ \citenamefont
  {Saha}}]{Chakraborti:2021kkr}%
  \BibitemOpen
  \bibfield  {author} {\bibinfo {author} {\bibfnamefont {Manimala}\
  \bibnamefont {Chakraborti}}, \bibinfo {author} {\bibfnamefont {Sven}\
  \bibnamefont {Heinemeyer}}, \ and\ \bibinfo {author} {\bibfnamefont {Ipsita}\
  \bibnamefont {Saha}},\ }\bibfield  {title} {\enquote {\bibinfo {title}
  {{Improved ${(g-2)_\mu }$ measurements and wino/higgsino dark matter}},}\
  }\href {\doibase 10.1140/epjc/s10052-021-09814-1} {\bibfield  {journal}
  {\bibinfo  {journal} {Eur. Phys. J. C}\ }\textbf {\bibinfo {volume} {81}},\
  \bibinfo {pages} {1069} (\bibinfo {year} {2021})},\ \Eprint
  {http://arxiv.org/abs/2103.13403} {arXiv:2103.13403 [hep-ph]} \BibitemShut
  {NoStop}%
\bibitem [{\citenamefont {Zhou}\ and\ \citenamefont {Wu}(2003)}]{Zhou:2001ew}%
  \BibitemOpen
  \bibfield  {author} {\bibinfo {author} {\bibfnamefont {Yu-Feng}\ \bibnamefont
  {Zhou}}\ and\ \bibinfo {author} {\bibfnamefont {Yue-Liang}\ \bibnamefont
  {Wu}},\ }\bibfield  {title} {\enquote {\bibinfo {title} {{Lepton flavor
  changing scalar interactions and muon g-2}},}\ }\href {\doibase
  10.1140/epjc/s2003-01137-1} {\bibfield  {journal} {\bibinfo  {journal} {Eur.
  Phys. J. C}\ }\textbf {\bibinfo {volume} {27}},\ \bibinfo {pages} {577--585}
  (\bibinfo {year} {2003})},\ \Eprint {http://arxiv.org/abs/hep-ph/0110302}
  {arXiv:hep-ph/0110302} \BibitemShut {NoStop}%
\bibitem [{\citenamefont {Barger}\ \emph {et~al.}(2011)\citenamefont {Barger},
  \citenamefont {Chiang}, \citenamefont {Keung},\ and\ \citenamefont
  {Marfatia}}]{Barger:2010aj}%
  \BibitemOpen
  \bibfield  {author} {\bibinfo {author} {\bibfnamefont {Vernon}\ \bibnamefont
  {Barger}}, \bibinfo {author} {\bibfnamefont {Cheng-Wei}\ \bibnamefont
  {Chiang}}, \bibinfo {author} {\bibfnamefont {Wai-Yee}\ \bibnamefont {Keung}},
  \ and\ \bibinfo {author} {\bibfnamefont {Danny}\ \bibnamefont {Marfatia}},\
  }\bibfield  {title} {\enquote {\bibinfo {title} {{Proton size anomaly}},}\
  }\href {\doibase 10.1103/PhysRevLett.106.153001} {\bibfield  {journal}
  {\bibinfo  {journal} {Phys. Rev. Lett.}\ }\textbf {\bibinfo {volume} {106}},\
  \bibinfo {pages} {153001} (\bibinfo {year} {2011})},\ \Eprint
  {http://arxiv.org/abs/1011.3519} {arXiv:1011.3519 [hep-ph]} \BibitemShut
  {NoStop}%
\bibitem [{\citenamefont {Chen}\ \emph {et~al.}(2016)\citenamefont {Chen},
  \citenamefont {Davoudiasl}, \citenamefont {Marciano},\ and\ \citenamefont
  {Zhang}}]{Chen:2015vqy}%
  \BibitemOpen
  \bibfield  {author} {\bibinfo {author} {\bibfnamefont {Chien-Yi}\
  \bibnamefont {Chen}}, \bibinfo {author} {\bibfnamefont {Hooman}\ \bibnamefont
  {Davoudiasl}}, \bibinfo {author} {\bibfnamefont {William~J.}\ \bibnamefont
  {Marciano}}, \ and\ \bibinfo {author} {\bibfnamefont {Cen}\ \bibnamefont
  {Zhang}},\ }\bibfield  {title} {\enquote {\bibinfo {title} {{Implications of
  a light \textquotedblleft{}dark Higgs\textquotedblright{} solution to the
  $g_\mu$-2 discrepancy}},}\ }\href {\doibase 10.1103/PhysRevD.93.035006}
  {\bibfield  {journal} {\bibinfo  {journal} {Phys. Rev. D}\ }\textbf {\bibinfo
  {volume} {93}},\ \bibinfo {pages} {035006} (\bibinfo {year} {2016})},\
  \Eprint {http://arxiv.org/abs/1511.04715} {arXiv:1511.04715 [hep-ph]}
  \BibitemShut {NoStop}%
\bibitem [{\citenamefont {Batell}\ \emph {et~al.}(2017)\citenamefont {Batell},
  \citenamefont {Lange}, \citenamefont {McKeen}, \citenamefont {Pospelov},\
  and\ \citenamefont {Ritz}}]{Batell:2016ove}%
  \BibitemOpen
  \bibfield  {author} {\bibinfo {author} {\bibfnamefont {Brian}\ \bibnamefont
  {Batell}}, \bibinfo {author} {\bibfnamefont {Nicholas}\ \bibnamefont
  {Lange}}, \bibinfo {author} {\bibfnamefont {David}\ \bibnamefont {McKeen}},
  \bibinfo {author} {\bibfnamefont {Maxim}\ \bibnamefont {Pospelov}}, \ and\
  \bibinfo {author} {\bibfnamefont {Adam}\ \bibnamefont {Ritz}},\ }\bibfield
  {title} {\enquote {\bibinfo {title} {{Muon anomalous magnetic moment through
  the leptonic Higgs portal}},}\ }\href {\doibase 10.1103/PhysRevD.95.075003}
  {\bibfield  {journal} {\bibinfo  {journal} {Phys. Rev. D}\ }\textbf {\bibinfo
  {volume} {95}},\ \bibinfo {pages} {075003} (\bibinfo {year} {2017})},\
  \Eprint {http://arxiv.org/abs/1606.04943} {arXiv:1606.04943 [hep-ph]}
  \BibitemShut {NoStop}%
\bibitem [{\citenamefont {Davoudiasl}\ and\ \citenamefont
  {Marciano}(2018)}]{Davoudiasl:2018fbb}%
  \BibitemOpen
  \bibfield  {author} {\bibinfo {author} {\bibfnamefont {Hooman}\ \bibnamefont
  {Davoudiasl}}\ and\ \bibinfo {author} {\bibfnamefont {William~J.}\
  \bibnamefont {Marciano}},\ }\bibfield  {title} {\enquote {\bibinfo {title}
  {{Tale of two anomalies}},}\ }\href {\doibase 10.1103/PhysRevD.98.075011}
  {\bibfield  {journal} {\bibinfo  {journal} {Phys. Rev. D}\ }\textbf {\bibinfo
  {volume} {98}},\ \bibinfo {pages} {075011} (\bibinfo {year} {2018})},\
  \Eprint {http://arxiv.org/abs/1806.10252} {arXiv:1806.10252 [hep-ph]}
  \BibitemShut {NoStop}%
\bibitem [{\citenamefont {Liu}\ \emph {et~al.}(2019)\citenamefont {Liu},
  \citenamefont {Wagner},\ and\ \citenamefont {Wang}}]{Liu:2018xkx}%
  \BibitemOpen
  \bibfield  {author} {\bibinfo {author} {\bibfnamefont {Jia}\ \bibnamefont
  {Liu}}, \bibinfo {author} {\bibfnamefont {Carlos E.~M.}\ \bibnamefont
  {Wagner}}, \ and\ \bibinfo {author} {\bibfnamefont {Xiao-Ping}\ \bibnamefont
  {Wang}},\ }\bibfield  {title} {\enquote {\bibinfo {title} {{A light complex
  scalar for the electron and muon anomalous magnetic moments}},}\ }\href
  {\doibase 10.1007/JHEP03(2019)008} {\bibfield  {journal} {\bibinfo  {journal}
  {JHEP}\ }\textbf {\bibinfo {volume} {03}},\ \bibinfo {pages} {008} (\bibinfo
  {year} {2019})},\ \Eprint {http://arxiv.org/abs/1810.11028} {arXiv:1810.11028
  [hep-ph]} \BibitemShut {NoStop}%
\bibitem [{\citenamefont {Ahn}\ \emph {et~al.}(2021)\citenamefont {Ahn} \emph
  {et~al.}}]{Ahn:2020opg}%
  \BibitemOpen
  \bibfield  {author} {\bibinfo {author} {\bibfnamefont {J.~K.}\ \bibnamefont
  {Ahn}} \emph {et~al.} (\bibinfo {collaboration} {KOTO}),\ }\bibfield  {title}
  {\enquote {\bibinfo {title} {{Study of the $K_L \!\to\! \pi^0 \nu
  \overline{\nu}$ Decay at the J-PARC KOTO Experiment}},}\ }\href {\doibase
  10.1103/PhysRevLett.126.121801} {\bibfield  {journal} {\bibinfo  {journal}
  {Phys. Rev. Lett.}\ }\textbf {\bibinfo {volume} {126}},\ \bibinfo {pages}
  {121801} (\bibinfo {year} {2021})},\ \Eprint
  {http://arxiv.org/abs/2012.07571} {arXiv:2012.07571 [hep-ex]} \BibitemShut
  {NoStop}%
\bibitem [{\citenamefont {Liu}\ \emph {et~al.}(2020{\natexlab{a}})\citenamefont
  {Liu}, \citenamefont {McGinnis}, \citenamefont {Wagner},\ and\ \citenamefont
  {Wang}}]{Liu:2020qgx}%
  \BibitemOpen
  \bibfield  {author} {\bibinfo {author} {\bibfnamefont {Jia}\ \bibnamefont
  {Liu}}, \bibinfo {author} {\bibfnamefont {Navin}\ \bibnamefont {McGinnis}},
  \bibinfo {author} {\bibfnamefont {Carlos E.~M.}\ \bibnamefont {Wagner}}, \
  and\ \bibinfo {author} {\bibfnamefont {Xiao-Ping}\ \bibnamefont {Wang}},\
  }\bibfield  {title} {\enquote {\bibinfo {title} {{A light scalar explanation
  of $(g-2)_{\mu}$ and the KOTO anomaly}},}\ }\href {\doibase
  10.1007/JHEP04(2020)197} {\bibfield  {journal} {\bibinfo  {journal} {JHEP}\
  }\textbf {\bibinfo {volume} {04}},\ \bibinfo {pages} {197} (\bibinfo {year}
  {2020}{\natexlab{a}})},\ \Eprint {http://arxiv.org/abs/2001.06522}
  {arXiv:2001.06522 [hep-ph]} \BibitemShut {NoStop}%
\bibitem [{\citenamefont {Aaij}\ \emph
  {et~al.}(2017{\natexlab{b}})\citenamefont {Aaij} \emph
  {et~al.}}]{Aaij:2017vad}%
  \BibitemOpen
  \bibfield  {author} {\bibinfo {author} {\bibfnamefont {Roel}\ \bibnamefont
  {Aaij}} \emph {et~al.} (\bibinfo {collaboration} {LHCb}),\ }\bibfield
  {title} {\enquote {\bibinfo {title} {{Measurement of the $B^0_s\to\mu^+\mu^-$
  branching fraction and effective lifetime and search for $B^0\to\mu^+\mu^-$
  decays}},}\ }\href {\doibase 10.1103/PhysRevLett.118.191801} {\bibfield
  {journal} {\bibinfo  {journal} {Phys. Rev. Lett.}\ }\textbf {\bibinfo
  {volume} {118}},\ \bibinfo {pages} {191801} (\bibinfo {year}
  {2017}{\natexlab{b}})},\ \Eprint {http://arxiv.org/abs/1703.05747}
  {arXiv:1703.05747 [hep-ex]} \BibitemShut {NoStop}%
\bibitem [{\citenamefont {Altmannshofer}\ and\ \citenamefont
  {Stangl}(2021)}]{Altmannshofer:2021qrr}%
  \BibitemOpen
  \bibfield  {author} {\bibinfo {author} {\bibfnamefont {Wolfgang}\
  \bibnamefont {Altmannshofer}}\ and\ \bibinfo {author} {\bibfnamefont {Peter}\
  \bibnamefont {Stangl}},\ }\bibfield  {title} {\enquote {\bibinfo {title}
  {{New physics in rare B decays after Moriond 2021}},}\ }\href {\doibase
  10.1140/epjc/s10052-021-09725-1} {\bibfield  {journal} {\bibinfo  {journal}
  {Eur. Phys. J. C}\ }\textbf {\bibinfo {volume} {81}},\ \bibinfo {pages} {952}
  (\bibinfo {year} {2021})},\ \Eprint {http://arxiv.org/abs/2103.13370}
  {arXiv:2103.13370 [hep-ph]} \BibitemShut {NoStop}%
\bibitem [{\citenamefont {Marciano}\ \emph {et~al.}(2016)\citenamefont
  {Marciano}, \citenamefont {Masiero}, \citenamefont {Paradisi},\ and\
  \citenamefont {Passera}}]{Marciano:2016yhf}%
  \BibitemOpen
  \bibfield  {author} {\bibinfo {author} {\bibfnamefont {W.~J.}\ \bibnamefont
  {Marciano}}, \bibinfo {author} {\bibfnamefont {A.}~\bibnamefont {Masiero}},
  \bibinfo {author} {\bibfnamefont {P.}~\bibnamefont {Paradisi}}, \ and\
  \bibinfo {author} {\bibfnamefont {M.}~\bibnamefont {Passera}},\ }\bibfield
  {title} {\enquote {\bibinfo {title} {{Contributions of axionlike particles to
  lepton dipole moments}},}\ }\href {\doibase 10.1103/PhysRevD.94.115033}
  {\bibfield  {journal} {\bibinfo  {journal} {Phys. Rev.}\ }\textbf {\bibinfo
  {volume} {D94}},\ \bibinfo {pages} {115033} (\bibinfo {year} {2016})},\
  \Eprint {http://arxiv.org/abs/1607.01022} {arXiv:1607.01022 [hep-ph]}
  \BibitemShut {NoStop}%
\bibitem [{\citenamefont {Bauer}\ \emph {et~al.}(2020)\citenamefont {Bauer},
  \citenamefont {Neubert}, \citenamefont {Renner}, \citenamefont {Schnubel},\
  and\ \citenamefont {Thamm}}]{Bauer:2019gfk}%
  \BibitemOpen
  \bibfield  {author} {\bibinfo {author} {\bibfnamefont {Martin}\ \bibnamefont
  {Bauer}}, \bibinfo {author} {\bibfnamefont {Matthias}\ \bibnamefont
  {Neubert}}, \bibinfo {author} {\bibfnamefont {Sophie}\ \bibnamefont
  {Renner}}, \bibinfo {author} {\bibfnamefont {Marvin}\ \bibnamefont
  {Schnubel}}, \ and\ \bibinfo {author} {\bibfnamefont {Andrea}\ \bibnamefont
  {Thamm}},\ }\bibfield  {title} {\enquote {\bibinfo {title} {{Axionlike
  Particles, Lepton-Flavor Violation, and a New Explanation of $a_\mu$ and
  $a_e$}},}\ }\href {\doibase 10.1103/PhysRevLett.124.211803} {\bibfield
  {journal} {\bibinfo  {journal} {Phys. Rev. Lett.}\ }\textbf {\bibinfo
  {volume} {124}},\ \bibinfo {pages} {211803} (\bibinfo {year} {2020})},\
  \Eprint {http://arxiv.org/abs/1908.00008} {arXiv:1908.00008 [hep-ph]}
  \BibitemShut {NoStop}%
\bibitem [{\citenamefont {Liu}\ \emph {et~al.}(2021)\citenamefont {Liu},
  \citenamefont {McGinnis}, \citenamefont {Wagner},\ and\ \citenamefont
  {Wang}}]{Liu:2021wap}%
  \BibitemOpen
  \bibfield  {author} {\bibinfo {author} {\bibfnamefont {Jia}\ \bibnamefont
  {Liu}}, \bibinfo {author} {\bibfnamefont {Navin}\ \bibnamefont {McGinnis}},
  \bibinfo {author} {\bibfnamefont {Carlos E.~M.}\ \bibnamefont {Wagner}}, \
  and\ \bibinfo {author} {\bibfnamefont {Xiao-Ping}\ \bibnamefont {Wang}},\
  }\bibfield  {title} {\enquote {\bibinfo {title} {{Challenges for a QCD Axion
  at the 10 MeV Scale}},}\ }\href {\doibase 10.1007/JHEP05(2021)138} {\bibfield
   {journal} {\bibinfo  {journal} {JHEP}\ }\textbf {\bibinfo {volume} {05}},\
  \bibinfo {pages} {138} (\bibinfo {year} {2021})},\ \Eprint
  {http://arxiv.org/abs/2102.10118} {arXiv:2102.10118 [hep-ph]} \BibitemShut
  {NoStop}%
\bibitem [{\citenamefont {Gunion}(2009)}]{Gunion:2008dg}%
  \BibitemOpen
  \bibfield  {author} {\bibinfo {author} {\bibfnamefont {John~F.}\ \bibnamefont
  {Gunion}},\ }\bibfield  {title} {\enquote {\bibinfo {title} {{A Light CP-odd
  Higgs boson and the muon anomalous magnetic moment}},}\ }\href {\doibase
  10.1088/1126-6708/2009/08/032} {\bibfield  {journal} {\bibinfo  {journal}
  {JHEP}\ }\textbf {\bibinfo {volume} {08}},\ \bibinfo {pages} {032} (\bibinfo
  {year} {2009})},\ \Eprint {http://arxiv.org/abs/0808.2509} {arXiv:0808.2509
  [hep-ph]} \BibitemShut {NoStop}%
\bibitem [{\citenamefont {Wang}\ \emph {et~al.}(2019)\citenamefont {Wang},
  \citenamefont {Yang}, \citenamefont {Zhang},\ and\ \citenamefont
  {Zhang}}]{Wang:2018hnw}%
  \BibitemOpen
  \bibfield  {author} {\bibinfo {author} {\bibfnamefont {Lei}\ \bibnamefont
  {Wang}}, \bibinfo {author} {\bibfnamefont {Jin~Min}\ \bibnamefont {Yang}},
  \bibinfo {author} {\bibfnamefont {Mengchao}\ \bibnamefont {Zhang}}, \ and\
  \bibinfo {author} {\bibfnamefont {Yang}\ \bibnamefont {Zhang}},\ }\bibfield
  {title} {\enquote {\bibinfo {title} {{Revisiting lepton-specific 2HDM in
  light of muon $g-2$ anomaly}},}\ }\href {\doibase
  10.1016/j.physletb.2018.11.045} {\bibfield  {journal} {\bibinfo  {journal}
  {Phys. Lett. B}\ }\textbf {\bibinfo {volume} {788}},\ \bibinfo {pages}
  {519--529} (\bibinfo {year} {2019})},\ \Eprint
  {http://arxiv.org/abs/1809.05857} {arXiv:1809.05857 [hep-ph]} \BibitemShut
  {NoStop}%
\bibitem [{\citenamefont {Kannike}\ \emph {et~al.}(2012)\citenamefont
  {Kannike}, \citenamefont {Raidal}, \citenamefont {Straub},\ and\
  \citenamefont {Strumia}}]{Kannike:2011ng}%
  \BibitemOpen
  \bibfield  {author} {\bibinfo {author} {\bibfnamefont {Kristjan}\
  \bibnamefont {Kannike}}, \bibinfo {author} {\bibfnamefont {Martti}\
  \bibnamefont {Raidal}}, \bibinfo {author} {\bibfnamefont {David~M.}\
  \bibnamefont {Straub}}, \ and\ \bibinfo {author} {\bibfnamefont {Alessandro}\
  \bibnamefont {Strumia}},\ }\bibfield  {title} {\enquote {\bibinfo {title}
  {{Anthropic solution to the magnetic muon anomaly: the charged see-saw}},}\
  }\href {\doibase 10.1007/JHEP02(2012)106} {\bibfield  {journal} {\bibinfo
  {journal} {JHEP}\ }\textbf {\bibinfo {volume} {02}},\ \bibinfo {pages} {106}
  (\bibinfo {year} {2012})},\ \bibinfo {note} {[Erratum: JHEP 10, 136
  (2012)]},\ \Eprint {http://arxiv.org/abs/1111.2551} {arXiv:1111.2551
  [hep-ph]} \BibitemShut {NoStop}%
\bibitem [{\citenamefont {Dermisek}\ and\ \citenamefont
  {Raval}(2013)}]{Dermisek:2013gta}%
  \BibitemOpen
  \bibfield  {author} {\bibinfo {author} {\bibfnamefont {Radovan}\ \bibnamefont
  {Dermisek}}\ and\ \bibinfo {author} {\bibfnamefont {Aditi}\ \bibnamefont
  {Raval}},\ }\bibfield  {title} {\enquote {\bibinfo {title} {{Explanation of
  the Muon g-2 Anomaly with Vectorlike Leptons and its Implications for Higgs
  Decays}},}\ }\href {\doibase 10.1103/PhysRevD.88.013017} {\bibfield
  {journal} {\bibinfo  {journal} {Phys. Rev. D}\ }\textbf {\bibinfo {volume}
  {88}},\ \bibinfo {pages} {013017} (\bibinfo {year} {2013})},\ \Eprint
  {http://arxiv.org/abs/1305.3522} {arXiv:1305.3522 [hep-ph]} \BibitemShut
  {NoStop}%
\bibitem [{\citenamefont {Dermisek}\ \emph {et~al.}(2021)\citenamefont
  {Dermisek}, \citenamefont {Hermanek},\ and\ \citenamefont
  {McGinnis}}]{Dermisek:2021ajd}%
  \BibitemOpen
  \bibfield  {author} {\bibinfo {author} {\bibfnamefont {Radovan}\ \bibnamefont
  {Dermisek}}, \bibinfo {author} {\bibfnamefont {Keith}\ \bibnamefont
  {Hermanek}}, \ and\ \bibinfo {author} {\bibfnamefont {Navin}\ \bibnamefont
  {McGinnis}},\ }\bibfield  {title} {\enquote {\bibinfo {title} {{Muon g-2 in
  two-Higgs-doublet models with vectorlike leptons}},}\ }\href {\doibase
  10.1103/PhysRevD.104.055033} {\bibfield  {journal} {\bibinfo  {journal}
  {Phys. Rev. D}\ }\textbf {\bibinfo {volume} {104}},\ \bibinfo {pages}
  {055033} (\bibinfo {year} {2021})},\ \Eprint
  {http://arxiv.org/abs/2103.05645} {arXiv:2103.05645 [hep-ph]} \BibitemShut
  {NoStop}%
\bibitem [{\citenamefont {Megias}\ \emph {et~al.}(2017)\citenamefont {Megias},
  \citenamefont {Quiros},\ and\ \citenamefont {Salas}}]{Megias:2017dzd}%
  \BibitemOpen
  \bibfield  {author} {\bibinfo {author} {\bibfnamefont {Eugenio}\ \bibnamefont
  {Megias}}, \bibinfo {author} {\bibfnamefont {Mariano}\ \bibnamefont
  {Quiros}}, \ and\ \bibinfo {author} {\bibfnamefont {Lindber}\ \bibnamefont
  {Salas}},\ }\bibfield  {title} {\enquote {\bibinfo {title} {{$g_\mu-2$ from
  Vector-Like Leptons in Warped Space}},}\ }\href {\doibase
  10.1007/JHEP05(2017)016} {\bibfield  {journal} {\bibinfo  {journal} {JHEP}\
  }\textbf {\bibinfo {volume} {05}},\ \bibinfo {pages} {016} (\bibinfo {year}
  {2017})},\ \Eprint {http://arxiv.org/abs/1701.05072} {arXiv:1701.05072
  [hep-ph]} \BibitemShut {NoStop}%
\bibitem [{\citenamefont {Bauer}\ and\ \citenamefont
  {Neubert}(2016)}]{Bauer:2015knc}%
  \BibitemOpen
  \bibfield  {author} {\bibinfo {author} {\bibfnamefont {Martin}\ \bibnamefont
  {Bauer}}\ and\ \bibinfo {author} {\bibfnamefont {Matthias}\ \bibnamefont
  {Neubert}},\ }\bibfield  {title} {\enquote {\bibinfo {title} {{Minimal
  Leptoquark Explanation for the R$_{D^{(*)}}$ , R$_K$ , and $(g-2)_g$
  Anomalies}},}\ }\href {\doibase 10.1103/PhysRevLett.116.141802} {\bibfield
  {journal} {\bibinfo  {journal} {Phys. Rev. Lett.}\ }\textbf {\bibinfo
  {volume} {116}},\ \bibinfo {pages} {141802} (\bibinfo {year} {2016})},\
  \Eprint {http://arxiv.org/abs/1511.01900} {arXiv:1511.01900 [hep-ph]}
  \BibitemShut {NoStop}%
\bibitem [{\citenamefont {Crivellin}\ \emph
  {et~al.}(2021{\natexlab{a}})\citenamefont {Crivellin}, \citenamefont
  {Mueller},\ and\ \citenamefont {Saturnino}}]{Crivellin:2020tsz}%
  \BibitemOpen
  \bibfield  {author} {\bibinfo {author} {\bibfnamefont {Andreas}\ \bibnamefont
  {Crivellin}}, \bibinfo {author} {\bibfnamefont {Dario}\ \bibnamefont
  {Mueller}}, \ and\ \bibinfo {author} {\bibfnamefont {Francesco}\ \bibnamefont
  {Saturnino}},\ }\bibfield  {title} {\enquote {\bibinfo {title} {{Correlating
  h\textrightarrow{}\ensuremath{\mu}+\ensuremath{\mu}- to the Anomalous
  Magnetic Moment of the Muon via Leptoquarks}},}\ }\href {\doibase
  10.1103/PhysRevLett.127.021801} {\bibfield  {journal} {\bibinfo  {journal}
  {Phys. Rev. Lett.}\ }\textbf {\bibinfo {volume} {127}},\ \bibinfo {pages}
  {021801} (\bibinfo {year} {2021}{\natexlab{a}})},\ \Eprint
  {http://arxiv.org/abs/2008.02643} {arXiv:2008.02643 [hep-ph]} \BibitemShut
  {NoStop}%
\bibitem [{\citenamefont {Crivellin}\ \emph
  {et~al.}(2021{\natexlab{b}})\citenamefont {Crivellin}, \citenamefont {Greub},
  \citenamefont {M\"uller},\ and\ \citenamefont
  {Saturnino}}]{Crivellin:2020mjs}%
  \BibitemOpen
  \bibfield  {author} {\bibinfo {author} {\bibfnamefont {Andreas}\ \bibnamefont
  {Crivellin}}, \bibinfo {author} {\bibfnamefont {Christoph}\ \bibnamefont
  {Greub}}, \bibinfo {author} {\bibfnamefont {Dario}\ \bibnamefont {M\"uller}},
  \ and\ \bibinfo {author} {\bibfnamefont {Francesco}\ \bibnamefont
  {Saturnino}},\ }\bibfield  {title} {\enquote {\bibinfo {title} {{Scalar
  Leptoquarks in Leptonic Processes}},}\ }\href {\doibase
  10.1007/JHEP02(2021)182} {\bibfield  {journal} {\bibinfo  {journal} {JHEP}\
  }\textbf {\bibinfo {volume} {02}},\ \bibinfo {pages} {182} (\bibinfo {year}
  {2021}{\natexlab{b}})},\ \Eprint {http://arxiv.org/abs/2010.06593}
  {arXiv:2010.06593 [hep-ph]} \BibitemShut {NoStop}%
\bibitem [{\citenamefont {Hiller}\ \emph {et~al.}(2021)\citenamefont {Hiller},
  \citenamefont {Loose},\ and\ \citenamefont
  {Ni\v{s}and\v{z}i\'c}}]{Hiller:2021pul}%
  \BibitemOpen
  \bibfield  {author} {\bibinfo {author} {\bibfnamefont {Gudrun}\ \bibnamefont
  {Hiller}}, \bibinfo {author} {\bibfnamefont {Dennis}\ \bibnamefont {Loose}},
  \ and\ \bibinfo {author} {\bibfnamefont {Ivan}\ \bibnamefont
  {Ni\v{s}and\v{z}i\'c}},\ }\bibfield  {title} {\enquote {\bibinfo {title}
  {{Flavorful leptoquarks at the LHC and beyond: spin 1}},}\ }\href {\doibase
  10.1007/JHEP06(2021)080} {\bibfield  {journal} {\bibinfo  {journal} {JHEP}\
  }\textbf {\bibinfo {volume} {06}},\ \bibinfo {pages} {080} (\bibinfo {year}
  {2021})},\ \Eprint {http://arxiv.org/abs/2103.12724} {arXiv:2103.12724
  [hep-ph]} \BibitemShut {NoStop}%
\bibitem [{\citenamefont {Altmannshofer}\ \emph {et~al.}(2020)\citenamefont
  {Altmannshofer}, \citenamefont {Dev}, \citenamefont {Soni},\ and\
  \citenamefont {Sui}}]{Altmannshofer:2020axr}%
  \BibitemOpen
  \bibfield  {author} {\bibinfo {author} {\bibfnamefont {Wolfgang}\
  \bibnamefont {Altmannshofer}}, \bibinfo {author} {\bibfnamefont
  {P.~S.~Bhupal}\ \bibnamefont {Dev}}, \bibinfo {author} {\bibfnamefont
  {Amarjit}\ \bibnamefont {Soni}}, \ and\ \bibinfo {author} {\bibfnamefont
  {Yicong}\ \bibnamefont {Sui}},\ }\bibfield  {title} {\enquote {\bibinfo
  {title} {{Addressing R$_{D^{(*)}}$, R$_{K^{(*)}}$, muon $g-2$ and ANITA
  anomalies in a minimal $R$-parity violating supersymmetric framework}},}\
  }\href {\doibase 10.1103/PhysRevD.102.015031} {\bibfield  {journal} {\bibinfo
   {journal} {Phys. Rev. D}\ }\textbf {\bibinfo {volume} {102}},\ \bibinfo
  {pages} {015031} (\bibinfo {year} {2020})},\ \Eprint
  {http://arxiv.org/abs/2002.12910} {arXiv:2002.12910 [hep-ph]} \BibitemShut
  {NoStop}%
\bibitem [{\citenamefont {Heeck}\ and\ \citenamefont
  {Rodejohann}(2011)}]{Heeck:2011wj}%
  \BibitemOpen
  \bibfield  {author} {\bibinfo {author} {\bibfnamefont {Julian}\ \bibnamefont
  {Heeck}}\ and\ \bibinfo {author} {\bibfnamefont {Werner}\ \bibnamefont
  {Rodejohann}},\ }\bibfield  {title} {\enquote {\bibinfo {title} {{Gauged
  $L_\mu - L_\tau$ Symmetry at the Electroweak Scale}},}\ }\href {\doibase
  10.1103/PhysRevD.84.075007} {\bibfield  {journal} {\bibinfo  {journal} {Phys.
  Rev. D}\ }\textbf {\bibinfo {volume} {84}},\ \bibinfo {pages} {075007}
  (\bibinfo {year} {2011})},\ \Eprint {http://arxiv.org/abs/1107.5238}
  {arXiv:1107.5238 [hep-ph]} \BibitemShut {NoStop}%
\bibitem [{\citenamefont {Carena}\ \emph {et~al.}(2019)\citenamefont {Carena},
  \citenamefont {Quir\'os},\ and\ \citenamefont {Zhang}}]{Carena:2018cjh}%
  \BibitemOpen
  \bibfield  {author} {\bibinfo {author} {\bibfnamefont {Marcela}\ \bibnamefont
  {Carena}}, \bibinfo {author} {\bibfnamefont {Mariano}\ \bibnamefont
  {Quir\'os}}, \ and\ \bibinfo {author} {\bibfnamefont {Yue}\ \bibnamefont
  {Zhang}},\ }\bibfield  {title} {\enquote {\bibinfo {title} {{Electroweak
  Baryogenesis from Dark-Sector CP Violation}},}\ }\href {\doibase
  10.1103/PhysRevLett.122.201802} {\bibfield  {journal} {\bibinfo  {journal}
  {Phys. Rev. Lett.}\ }\textbf {\bibinfo {volume} {122}},\ \bibinfo {pages}
  {201802} (\bibinfo {year} {2019})},\ \Eprint
  {http://arxiv.org/abs/1811.09719} {arXiv:1811.09719 [hep-ph]} \BibitemShut
  {NoStop}%
\bibitem [{\citenamefont {Carena}\ \emph {et~al.}(2020)\citenamefont {Carena},
  \citenamefont {Quir\'os},\ and\ \citenamefont {Zhang}}]{Carena:2019xrr}%
  \BibitemOpen
  \bibfield  {author} {\bibinfo {author} {\bibfnamefont {Marcela}\ \bibnamefont
  {Carena}}, \bibinfo {author} {\bibfnamefont {Mariano}\ \bibnamefont
  {Quir\'os}}, \ and\ \bibinfo {author} {\bibfnamefont {Yue}\ \bibnamefont
  {Zhang}},\ }\bibfield  {title} {\enquote {\bibinfo {title} {{Dark CP
  violation and gauged lepton or baryon number for electroweak
  baryogenesis}},}\ }\href {\doibase 10.1103/PhysRevD.101.055014} {\bibfield
  {journal} {\bibinfo  {journal} {Phys. Rev. D}\ }\textbf {\bibinfo {volume}
  {101}},\ \bibinfo {pages} {055014} (\bibinfo {year} {2020})},\ \Eprint
  {http://arxiv.org/abs/1908.04818} {arXiv:1908.04818 [hep-ph]} \BibitemShut
  {NoStop}%
\bibitem [{\citenamefont {Altmannshofer}\ \emph
  {et~al.}(2014{\natexlab{a}})\citenamefont {Altmannshofer}, \citenamefont
  {Gori}, \citenamefont {Pospelov},\ and\ \citenamefont
  {Yavin}}]{Altmannshofer:2014cfa}%
  \BibitemOpen
  \bibfield  {author} {\bibinfo {author} {\bibfnamefont {Wolfgang}\
  \bibnamefont {Altmannshofer}}, \bibinfo {author} {\bibfnamefont {Stefania}\
  \bibnamefont {Gori}}, \bibinfo {author} {\bibfnamefont {Maxim}\ \bibnamefont
  {Pospelov}}, \ and\ \bibinfo {author} {\bibfnamefont {Itay}\ \bibnamefont
  {Yavin}},\ }\bibfield  {title} {\enquote {\bibinfo {title} {{Quark flavor
  transitions in $L_\mu-L_\tau$ models}},}\ }\href {\doibase
  10.1103/PhysRevD.89.095033} {\bibfield  {journal} {\bibinfo  {journal} {Phys.
  Rev. D}\ }\textbf {\bibinfo {volume} {89}},\ \bibinfo {pages} {095033}
  (\bibinfo {year} {2014}{\natexlab{a}})},\ \Eprint
  {http://arxiv.org/abs/1403.1269} {arXiv:1403.1269 [hep-ph]} \BibitemShut
  {NoStop}%
\bibitem [{\citenamefont {Altmannshofer}\ \emph
  {et~al.}(2016{\natexlab{a}})\citenamefont {Altmannshofer}, \citenamefont
  {Carena},\ and\ \citenamefont {Crivellin}}]{Altmannshofer:2016oaq}%
  \BibitemOpen
  \bibfield  {author} {\bibinfo {author} {\bibfnamefont {Wolfgang}\
  \bibnamefont {Altmannshofer}}, \bibinfo {author} {\bibfnamefont {Marcela}\
  \bibnamefont {Carena}}, \ and\ \bibinfo {author} {\bibfnamefont {Andreas}\
  \bibnamefont {Crivellin}},\ }\bibfield  {title} {\enquote {\bibinfo {title}
  {{$L_\mu - L_\tau$ theory of Higgs flavor violation and $(g-2)_\mu$}},}\
  }\href {\doibase 10.1103/PhysRevD.94.095026} {\bibfield  {journal} {\bibinfo
  {journal} {Phys. Rev. D}\ }\textbf {\bibinfo {volume} {94}},\ \bibinfo
  {pages} {095026} (\bibinfo {year} {2016}{\natexlab{a}})},\ \Eprint
  {http://arxiv.org/abs/1604.08221} {arXiv:1604.08221 [hep-ph]} \BibitemShut
  {NoStop}%
\bibitem [{\citenamefont {Altmannshofer}\ \emph
  {et~al.}(2016{\natexlab{b}})\citenamefont {Altmannshofer}, \citenamefont
  {Gori}, \citenamefont {Profumo},\ and\ \citenamefont
  {Queiroz}}]{Altmannshofer:2016jzy}%
  \BibitemOpen
  \bibfield  {author} {\bibinfo {author} {\bibfnamefont {Wolfgang}\
  \bibnamefont {Altmannshofer}}, \bibinfo {author} {\bibfnamefont {Stefania}\
  \bibnamefont {Gori}}, \bibinfo {author} {\bibfnamefont {Stefano}\
  \bibnamefont {Profumo}}, \ and\ \bibinfo {author} {\bibfnamefont
  {Farinaldo~S.}\ \bibnamefont {Queiroz}},\ }\bibfield  {title} {\enquote
  {\bibinfo {title} {{Explaining dark matter and B decay anomalies with an
  $L_\mu - L_\tau$ model}},}\ }\href {\doibase 10.1007/JHEP12(2016)106}
  {\bibfield  {journal} {\bibinfo  {journal} {JHEP}\ }\textbf {\bibinfo
  {volume} {12}},\ \bibinfo {pages} {106} (\bibinfo {year}
  {2016}{\natexlab{b}})},\ \Eprint {http://arxiv.org/abs/1609.04026}
  {arXiv:1609.04026 [hep-ph]} \BibitemShut {NoStop}%
\bibitem [{\citenamefont {Altmannshofer}\ \emph
  {et~al.}(2014{\natexlab{b}})\citenamefont {Altmannshofer}, \citenamefont
  {Gori}, \citenamefont {Pospelov},\ and\ \citenamefont
  {Yavin}}]{Altmannshofer:2014pba}%
  \BibitemOpen
  \bibfield  {author} {\bibinfo {author} {\bibfnamefont {Wolfgang}\
  \bibnamefont {Altmannshofer}}, \bibinfo {author} {\bibfnamefont {Stefania}\
  \bibnamefont {Gori}}, \bibinfo {author} {\bibfnamefont {Maxim}\ \bibnamefont
  {Pospelov}}, \ and\ \bibinfo {author} {\bibfnamefont {Itay}\ \bibnamefont
  {Yavin}},\ }\bibfield  {title} {\enquote {\bibinfo {title} {{Neutrino Trident
  Production: A Powerful Probe of New Physics with Neutrino Beams}},}\ }\href
  {\doibase 10.1103/PhysRevLett.113.091801} {\bibfield  {journal} {\bibinfo
  {journal} {Phys. Rev. Lett.}\ }\textbf {\bibinfo {volume} {113}},\ \bibinfo
  {pages} {091801} (\bibinfo {year} {2014}{\natexlab{b}})},\ \Eprint
  {http://arxiv.org/abs/1406.2332} {arXiv:1406.2332 [hep-ph]} \BibitemShut
  {NoStop}%
\bibitem [{\citenamefont {Geiregat}\ \emph {et~al.}(1990)\citenamefont
  {Geiregat} \emph {et~al.}}]{Geiregat:1990gz}%
  \BibitemOpen
  \bibfield  {author} {\bibinfo {author} {\bibfnamefont {D.}~\bibnamefont
  {Geiregat}} \emph {et~al.} (\bibinfo {collaboration} {CHARM-II}),\ }\bibfield
   {title} {\enquote {\bibinfo {title} {{First observation of neutrino trident
  production}},}\ }\href {\doibase 10.1016/0370-2693(90)90146-W} {\bibfield
  {journal} {\bibinfo  {journal} {Phys. Lett. B}\ }\textbf {\bibinfo {volume}
  {245}},\ \bibinfo {pages} {271--275} (\bibinfo {year} {1990})}\BibitemShut
  {NoStop}%
\bibitem [{\citenamefont {Mishra}\ \emph {et~al.}(1991)\citenamefont {Mishra}
  \emph {et~al.}}]{Mishra:1991bv}%
  \BibitemOpen
  \bibfield  {author} {\bibinfo {author} {\bibfnamefont {S.~R.}\ \bibnamefont
  {Mishra}} \emph {et~al.} (\bibinfo {collaboration} {CCFR}),\ }\bibfield
  {title} {\enquote {\bibinfo {title} {{Neutrino tridents and W Z
  interference}},}\ }\href {\doibase 10.1103/PhysRevLett.66.3117} {\bibfield
  {journal} {\bibinfo  {journal} {Phys. Rev. Lett.}\ }\textbf {\bibinfo
  {volume} {66}},\ \bibinfo {pages} {3117--3120} (\bibinfo {year}
  {1991})}\BibitemShut {NoStop}%
\bibitem [{\citenamefont {Abdullah}\ \emph {et~al.}(2018)\citenamefont
  {Abdullah}, \citenamefont {Dent}, \citenamefont {Dutta}, \citenamefont
  {Kane}, \citenamefont {Liao},\ and\ \citenamefont
  {Strigari}}]{Abdullah:2018ykz}%
  \BibitemOpen
  \bibfield  {author} {\bibinfo {author} {\bibfnamefont {Mohammad}\
  \bibnamefont {Abdullah}}, \bibinfo {author} {\bibfnamefont {James~B.}\
  \bibnamefont {Dent}}, \bibinfo {author} {\bibfnamefont {Bhaskar}\
  \bibnamefont {Dutta}}, \bibinfo {author} {\bibfnamefont {Gordon~L.}\
  \bibnamefont {Kane}}, \bibinfo {author} {\bibfnamefont {Shu}\ \bibnamefont
  {Liao}}, \ and\ \bibinfo {author} {\bibfnamefont {Louis~E.}\ \bibnamefont
  {Strigari}},\ }\bibfield  {title} {\enquote {\bibinfo {title} {{Coherent
  elastic neutrino nucleus scattering as a probe of a Z' through kinetic and
  mass mixing effects}},}\ }\href {\doibase 10.1103/PhysRevD.98.015005}
  {\bibfield  {journal} {\bibinfo  {journal} {Phys. Rev. D}\ }\textbf {\bibinfo
  {volume} {98}},\ \bibinfo {pages} {015005} (\bibinfo {year} {2018})},\
  \Eprint {http://arxiv.org/abs/1803.01224} {arXiv:1803.01224 [hep-ph]}
  \BibitemShut {NoStop}%
\bibitem [{\citenamefont {Amaral}\ \emph {et~al.}(2020)\citenamefont {Amaral},
  \citenamefont {Cerdeno}, \citenamefont {Foldenauer},\ and\ \citenamefont
  {Reid}}]{Amaral:2020tga}%
  \BibitemOpen
  \bibfield  {author} {\bibinfo {author} {\bibfnamefont {Dorian Warren
  Praia~do}\ \bibnamefont {Amaral}}, \bibinfo {author} {\bibfnamefont
  {David~G.}\ \bibnamefont {Cerdeno}}, \bibinfo {author} {\bibfnamefont
  {Patrick}\ \bibnamefont {Foldenauer}}, \ and\ \bibinfo {author}
  {\bibfnamefont {Elliott}\ \bibnamefont {Reid}},\ }\bibfield  {title}
  {\enquote {\bibinfo {title} {{Solar neutrino probes of the muon anomalous
  magnetic moment in the gauged $ \mathrm{U}{(1)}_{L_{\mu }-{L}_{\tau }} $}},}\
  }\href {\doibase 10.1007/JHEP12(2020)155} {\bibfield  {journal} {\bibinfo
  {journal} {JHEP}\ }\textbf {\bibinfo {volume} {12}},\ \bibinfo {pages} {155}
  (\bibinfo {year} {2020})},\ \Eprint {http://arxiv.org/abs/2006.11225}
  {arXiv:2006.11225 [hep-ph]} \BibitemShut {NoStop}%
\bibitem [{\citenamefont {Lees}\ \emph {et~al.}(2016)\citenamefont {Lees} \emph
  {et~al.}}]{TheBABAR:2016rlg}%
  \BibitemOpen
  \bibfield  {author} {\bibinfo {author} {\bibfnamefont {J.~P.}\ \bibnamefont
  {Lees}} \emph {et~al.} (\bibinfo {collaboration} {BaBar}),\ }\bibfield
  {title} {\enquote {\bibinfo {title} {{Search for a muonic dark force at
  BABAR}},}\ }\href {\doibase 10.1103/PhysRevD.94.011102} {\bibfield  {journal}
  {\bibinfo  {journal} {Phys. Rev. D}\ }\textbf {\bibinfo {volume} {94}},\
  \bibinfo {pages} {011102} (\bibinfo {year} {2016})},\ \Eprint
  {http://arxiv.org/abs/1606.03501} {arXiv:1606.03501 [hep-ex]} \BibitemShut
  {NoStop}%
\bibitem [{\citenamefont {Sirunyan}\ \emph {et~al.}(2019)\citenamefont
  {Sirunyan} \emph {et~al.}}]{Sirunyan:2018nnz}%
  \BibitemOpen
  \bibfield  {author} {\bibinfo {author} {\bibfnamefont {Albert~M}\
  \bibnamefont {Sirunyan}} \emph {et~al.} (\bibinfo {collaboration} {CMS}),\
  }\bibfield  {title} {\enquote {\bibinfo {title} {{Search for an
  $L_{\mu}-L_{\tau}$ gauge boson using Z$\to4\mu$ events in proton-proton
  collisions at $\sqrt{s} =$ 13 TeV}},}\ }\href {\doibase
  10.1016/j.physletb.2019.01.072} {\bibfield  {journal} {\bibinfo  {journal}
  {Phys. Lett. B}\ }\textbf {\bibinfo {volume} {792}},\ \bibinfo {pages}
  {345--368} (\bibinfo {year} {2019})},\ \Eprint
  {http://arxiv.org/abs/1808.03684} {arXiv:1808.03684 [hep-ex]} \BibitemShut
  {NoStop}%
\bibitem [{\citenamefont {Escudero}\ \emph {et~al.}(2019)\citenamefont
  {Escudero}, \citenamefont {Hooper}, \citenamefont {Krnjaic},\ and\
  \citenamefont {Pierre}}]{Escudero:2019gzq}%
  \BibitemOpen
  \bibfield  {author} {\bibinfo {author} {\bibfnamefont {Miguel}\ \bibnamefont
  {Escudero}}, \bibinfo {author} {\bibfnamefont {Dan}\ \bibnamefont {Hooper}},
  \bibinfo {author} {\bibfnamefont {Gordan}\ \bibnamefont {Krnjaic}}, \ and\
  \bibinfo {author} {\bibfnamefont {Mathias}\ \bibnamefont {Pierre}},\
  }\bibfield  {title} {\enquote {\bibinfo {title} {{Cosmology with A Very Light
  L$_{\mu}$ \ensuremath{-} L$_{\tau}$ Gauge Boson}},}\ }\href {\doibase
  10.1007/JHEP03(2019)071} {\bibfield  {journal} {\bibinfo  {journal} {JHEP}\
  }\textbf {\bibinfo {volume} {03}},\ \bibinfo {pages} {071} (\bibinfo {year}
  {2019})},\ \Eprint {http://arxiv.org/abs/1901.02010} {arXiv:1901.02010
  [hep-ph]} \BibitemShut {NoStop}%
\bibitem [{\citenamefont {Harnik}\ \emph {et~al.}(2012)\citenamefont {Harnik},
  \citenamefont {Kopp},\ and\ \citenamefont {Machado}}]{Harnik:2012ni}%
  \BibitemOpen
  \bibfield  {author} {\bibinfo {author} {\bibfnamefont {Roni}\ \bibnamefont
  {Harnik}}, \bibinfo {author} {\bibfnamefont {Joachim}\ \bibnamefont {Kopp}},
  \ and\ \bibinfo {author} {\bibfnamefont {Pedro A.~N.}\ \bibnamefont
  {Machado}},\ }\bibfield  {title} {\enquote {\bibinfo {title} {{Exploring nu
  Signals in Dark Matter Detectors}},}\ }\href {\doibase
  10.1088/1475-7516/2012/07/026} {\bibfield  {journal} {\bibinfo  {journal}
  {JCAP}\ }\textbf {\bibinfo {volume} {07}},\ \bibinfo {pages} {026} (\bibinfo
  {year} {2012})},\ \Eprint {http://arxiv.org/abs/1202.6073} {arXiv:1202.6073
  [hep-ph]} \BibitemShut {NoStop}%
\bibitem [{\citenamefont {Bilmis}\ \emph {et~al.}(2015)\citenamefont {Bilmis},
  \citenamefont {Turan}, \citenamefont {Aliev}, \citenamefont {Deniz},
  \citenamefont {Singh},\ and\ \citenamefont {Wong}}]{Bilmis:2015lja}%
  \BibitemOpen
  \bibfield  {author} {\bibinfo {author} {\bibfnamefont {S.}~\bibnamefont
  {Bilmis}}, \bibinfo {author} {\bibfnamefont {I.}~\bibnamefont {Turan}},
  \bibinfo {author} {\bibfnamefont {T.~M.}\ \bibnamefont {Aliev}}, \bibinfo
  {author} {\bibfnamefont {M.}~\bibnamefont {Deniz}}, \bibinfo {author}
  {\bibfnamefont {L.}~\bibnamefont {Singh}}, \ and\ \bibinfo {author}
  {\bibfnamefont {H.~T.}\ \bibnamefont {Wong}},\ }\bibfield  {title} {\enquote
  {\bibinfo {title} {{Constraints on Dark Photon from Neutrino-Electron
  Scattering Experiments}},}\ }\href {\doibase 10.1103/PhysRevD.92.033009}
  {\bibfield  {journal} {\bibinfo  {journal} {Phys. Rev. D}\ }\textbf {\bibinfo
  {volume} {92}},\ \bibinfo {pages} {033009} (\bibinfo {year} {2015})},\
  \Eprint {http://arxiv.org/abs/1502.07763} {arXiv:1502.07763 [hep-ph]}
  \BibitemShut {NoStop}%
\bibitem [{\citenamefont {Yin}\ and\ \citenamefont
  {Yamaguchi}(2020)}]{Yin:2020afe}%
  \BibitemOpen
  \bibfield  {author} {\bibinfo {author} {\bibfnamefont {Wen}\ \bibnamefont
  {Yin}}\ and\ \bibinfo {author} {\bibfnamefont {Masahiro}\ \bibnamefont
  {Yamaguchi}},\ }\bibfield  {title} {\enquote {\bibinfo {title} {{Muon $g-2$
  at multi-TeV muon collider}},}\ }\href@noop {} {\  (\bibinfo {year}
  {2020})},\ \Eprint {http://arxiv.org/abs/2012.03928} {arXiv:2012.03928
  [hep-ph]} \BibitemShut {NoStop}%
\bibitem [{\citenamefont {Capdevilla}\ \emph {et~al.}(2021)\citenamefont
  {Capdevilla}, \citenamefont {Curtin}, \citenamefont {Kahn},\ and\
  \citenamefont {Krnjaic}}]{Capdevilla:2021rwo}%
  \BibitemOpen
  \bibfield  {author} {\bibinfo {author} {\bibfnamefont {Rodolfo}\ \bibnamefont
  {Capdevilla}}, \bibinfo {author} {\bibfnamefont {David}\ \bibnamefont
  {Curtin}}, \bibinfo {author} {\bibfnamefont {Yonatan}\ \bibnamefont {Kahn}},
  \ and\ \bibinfo {author} {\bibfnamefont {Gordan}\ \bibnamefont {Krnjaic}},\
  }\bibfield  {title} {\enquote {\bibinfo {title} {{A No-Lose Theorem for
  Discovering the New Physics of $(g-2)_\mu$ at Muon Colliders}},}\ }\href@noop
  {} {\  (\bibinfo {year} {2021})},\ \Eprint {http://arxiv.org/abs/2101.10334}
  {arXiv:2101.10334 [hep-ph]} \BibitemShut {NoStop}%
\bibitem [{\citenamefont {Nilles}(1984)}]{Nilles:1983ge}%
  \BibitemOpen
  \bibfield  {author} {\bibinfo {author} {\bibfnamefont {Hans~Peter}\
  \bibnamefont {Nilles}},\ }\bibfield  {title} {\enquote {\bibinfo {title}
  {{Supersymmetry, Supergravity and Particle Physics}},}\ }\href {\doibase
  10.1016/0370-1573(84)90008-5} {\bibfield  {journal} {\bibinfo  {journal}
  {Phys. Rept.}\ }\textbf {\bibinfo {volume} {110}},\ \bibinfo {pages} {1--162}
  (\bibinfo {year} {1984})}\BibitemShut {NoStop}%
\bibitem [{\citenamefont {Haber}\ and\ \citenamefont
  {Kane}(1985)}]{Haber:1984rc}%
  \BibitemOpen
  \bibfield  {author} {\bibinfo {author} {\bibfnamefont {Howard~E.}\
  \bibnamefont {Haber}}\ and\ \bibinfo {author} {\bibfnamefont {Gordon~L.}\
  \bibnamefont {Kane}},\ }\bibfield  {title} {\enquote {\bibinfo {title} {{The
  Search for Supersymmetry: Probing Physics Beyond the Standard Model}},}\
  }\href {\doibase 10.1016/0370-1573(85)90051-1} {\bibfield  {journal}
  {\bibinfo  {journal} {Phys. Rept.}\ }\textbf {\bibinfo {volume} {117}},\
  \bibinfo {pages} {75--263} (\bibinfo {year} {1985})}\BibitemShut {NoStop}%
\bibitem [{\citenamefont {Martin}(1998)}]{Martin:1997ns}%
  \BibitemOpen
  \bibfield  {author} {\bibinfo {author} {\bibfnamefont {Stephen~P.}\
  \bibnamefont {Martin}},\ }\bibfield  {title} {\enquote {\bibinfo {title} {{A
  Supersymmetry primer}},}\ }\href {\doibase 10.1142/9789812839657_0001}
  {\bibfield  {journal} {\bibinfo  {journal} {Adv. Ser. Direct. High Energy
  Phys.}\ }\textbf {\bibinfo {volume} {18}},\ \bibinfo {pages} {1--98}
  (\bibinfo {year} {1998})},\ \Eprint {http://arxiv.org/abs/hep-ph/9709356}
  {arXiv:hep-ph/9709356} \BibitemShut {NoStop}%
\bibitem [{\citenamefont {Casas}\ \emph {et~al.}(1995)\citenamefont {Casas},
  \citenamefont {Espinosa}, \citenamefont {Quiros},\ and\ \citenamefont
  {Riotto}}]{Casas:1994us}%
  \BibitemOpen
  \bibfield  {author} {\bibinfo {author} {\bibfnamefont {J.~A.}\ \bibnamefont
  {Casas}}, \bibinfo {author} {\bibfnamefont {J.~R.}\ \bibnamefont {Espinosa}},
  \bibinfo {author} {\bibfnamefont {M.}~\bibnamefont {Quiros}}, \ and\ \bibinfo
  {author} {\bibfnamefont {A.}~\bibnamefont {Riotto}},\ }\bibfield  {title}
  {\enquote {\bibinfo {title} {{The Lightest Higgs boson mass in the minimal
  supersymmetric standard model}},}\ }\href {\doibase
  10.1016/0550-3213(94)00508-C} {\bibfield  {journal} {\bibinfo  {journal}
  {Nucl. Phys. B}\ }\textbf {\bibinfo {volume} {436}},\ \bibinfo {pages}
  {3--29} (\bibinfo {year} {1995})},\ \bibinfo {note} {[Erratum: Nucl.Phys.B
  439, 466--468 (1995)]},\ \Eprint {http://arxiv.org/abs/hep-ph/9407389}
  {arXiv:hep-ph/9407389} \BibitemShut {NoStop}%
\bibitem [{\citenamefont {Carena}\ \emph {et~al.}(1995)\citenamefont {Carena},
  \citenamefont {Espinosa}, \citenamefont {Quiros},\ and\ \citenamefont
  {Wagner}}]{Carena:1995bx}%
  \BibitemOpen
  \bibfield  {author} {\bibinfo {author} {\bibfnamefont {Marcela}\ \bibnamefont
  {Carena}}, \bibinfo {author} {\bibfnamefont {J.~R.}\ \bibnamefont
  {Espinosa}}, \bibinfo {author} {\bibfnamefont {M.}~\bibnamefont {Quiros}}, \
  and\ \bibinfo {author} {\bibfnamefont {C.~E.~M.}\ \bibnamefont {Wagner}},\
  }\bibfield  {title} {\enquote {\bibinfo {title} {{Analytical expressions for
  radiatively corrected Higgs masses and couplings in the MSSM}},}\ }\href
  {\doibase 10.1016/0370-2693(95)00694-G} {\bibfield  {journal} {\bibinfo
  {journal} {Phys. Lett. B}\ }\textbf {\bibinfo {volume} {355}},\ \bibinfo
  {pages} {209--221} (\bibinfo {year} {1995})},\ \Eprint
  {http://arxiv.org/abs/hep-ph/9504316} {arXiv:hep-ph/9504316} \BibitemShut
  {NoStop}%
\bibitem [{\citenamefont {Carena}\ \emph {et~al.}(1996)\citenamefont {Carena},
  \citenamefont {Quiros},\ and\ \citenamefont {Wagner}}]{Carena:1995wu}%
  \BibitemOpen
  \bibfield  {author} {\bibinfo {author} {\bibfnamefont {Marcela}\ \bibnamefont
  {Carena}}, \bibinfo {author} {\bibfnamefont {M.}~\bibnamefont {Quiros}}, \
  and\ \bibinfo {author} {\bibfnamefont {C.~E.~M.}\ \bibnamefont {Wagner}},\
  }\bibfield  {title} {\enquote {\bibinfo {title} {{Effective potential methods
  and the Higgs mass spectrum in the MSSM}},}\ }\href {\doibase
  10.1016/0550-3213(95)00665-6} {\bibfield  {journal} {\bibinfo  {journal}
  {Nucl. Phys. B}\ }\textbf {\bibinfo {volume} {461}},\ \bibinfo {pages}
  {407--436} (\bibinfo {year} {1996})},\ \Eprint
  {http://arxiv.org/abs/hep-ph/9508343} {arXiv:hep-ph/9508343} \BibitemShut
  {NoStop}%
\bibitem [{\citenamefont {Haber}\ \emph {et~al.}(1997)\citenamefont {Haber},
  \citenamefont {Hempfling},\ and\ \citenamefont {Hoang}}]{Haber:1996fp}%
  \BibitemOpen
  \bibfield  {author} {\bibinfo {author} {\bibfnamefont {Howard~E.}\
  \bibnamefont {Haber}}, \bibinfo {author} {\bibfnamefont {Ralf}\ \bibnamefont
  {Hempfling}}, \ and\ \bibinfo {author} {\bibfnamefont {Andre~H.}\
  \bibnamefont {Hoang}},\ }\bibfield  {title} {\enquote {\bibinfo {title}
  {{Approximating the radiatively corrected Higgs mass in the minimal
  supersymmetric model}},}\ }\href {\doibase 10.1007/s002880050498} {\bibfield
  {journal} {\bibinfo  {journal} {Z. Phys. C}\ }\textbf {\bibinfo {volume}
  {75}},\ \bibinfo {pages} {539--554} (\bibinfo {year} {1997})},\ \Eprint
  {http://arxiv.org/abs/hep-ph/9609331} {arXiv:hep-ph/9609331} \BibitemShut
  {NoStop}%
\bibitem [{\citenamefont {Degrassi}\ \emph {et~al.}(2003)\citenamefont
  {Degrassi}, \citenamefont {Heinemeyer}, \citenamefont {Hollik}, \citenamefont
  {Slavich},\ and\ \citenamefont {Weiglein}}]{Degrassi:2002fi}%
  \BibitemOpen
  \bibfield  {author} {\bibinfo {author} {\bibfnamefont {G.}~\bibnamefont
  {Degrassi}}, \bibinfo {author} {\bibfnamefont {S.}~\bibnamefont
  {Heinemeyer}}, \bibinfo {author} {\bibfnamefont {W.}~\bibnamefont {Hollik}},
  \bibinfo {author} {\bibfnamefont {P.}~\bibnamefont {Slavich}}, \ and\
  \bibinfo {author} {\bibfnamefont {G.}~\bibnamefont {Weiglein}},\ }\bibfield
  {title} {\enquote {\bibinfo {title} {{Towards high precision predictions for
  the MSSM Higgs sector}},}\ }\href {\doibase 10.1140/epjc/s2003-01152-2}
  {\bibfield  {journal} {\bibinfo  {journal} {Eur. Phys. J. C}\ }\textbf
  {\bibinfo {volume} {28}},\ \bibinfo {pages} {133--143} (\bibinfo {year}
  {2003})},\ \Eprint {http://arxiv.org/abs/hep-ph/0212020}
  {arXiv:hep-ph/0212020} \BibitemShut {NoStop}%
\bibitem [{\citenamefont {Bagnaschi}\ \emph {et~al.}(2014)\citenamefont
  {Bagnaschi}, \citenamefont {Giudice}, \citenamefont {Slavich},\ and\
  \citenamefont {Strumia}}]{Bagnaschi:2014rsa}%
  \BibitemOpen
  \bibfield  {author} {\bibinfo {author} {\bibfnamefont {Emanuele}\
  \bibnamefont {Bagnaschi}}, \bibinfo {author} {\bibfnamefont {Gian~F.}\
  \bibnamefont {Giudice}}, \bibinfo {author} {\bibfnamefont {Pietro}\
  \bibnamefont {Slavich}}, \ and\ \bibinfo {author} {\bibfnamefont
  {Alessandro}\ \bibnamefont {Strumia}},\ }\bibfield  {title} {\enquote
  {\bibinfo {title} {{Higgs Mass and Unnatural Supersymmetry}},}\ }\href
  {\doibase 10.1007/JHEP09(2014)092} {\bibfield  {journal} {\bibinfo  {journal}
  {JHEP}\ }\textbf {\bibinfo {volume} {09}},\ \bibinfo {pages} {092} (\bibinfo
  {year} {2014})},\ \Eprint {http://arxiv.org/abs/1407.4081} {arXiv:1407.4081
  [hep-ph]} \BibitemShut {NoStop}%
\bibitem [{\citenamefont {Draper}\ \emph {et~al.}(2014)\citenamefont {Draper},
  \citenamefont {Lee},\ and\ \citenamefont {Wagner}}]{Draper:2013oza}%
  \BibitemOpen
  \bibfield  {author} {\bibinfo {author} {\bibfnamefont {Patrick}\ \bibnamefont
  {Draper}}, \bibinfo {author} {\bibfnamefont {Gabriel}\ \bibnamefont {Lee}}, \
  and\ \bibinfo {author} {\bibfnamefont {Carlos E.~M.}\ \bibnamefont
  {Wagner}},\ }\bibfield  {title} {\enquote {\bibinfo {title} {{Precise
  estimates of the Higgs mass in heavy supersymmetry}},}\ }\href {\doibase
  10.1103/PhysRevD.89.055023} {\bibfield  {journal} {\bibinfo  {journal} {Phys.
  Rev.}\ }\textbf {\bibinfo {volume} {D89}},\ \bibinfo {pages} {055023}
  (\bibinfo {year} {2014})},\ \Eprint {http://arxiv.org/abs/1312.5743}
  {arXiv:1312.5743 [hep-ph]} \BibitemShut {NoStop}%
\bibitem [{\citenamefont {Lee}\ and\ \citenamefont
  {Wagner}(2015)}]{Lee:2015uza}%
  \BibitemOpen
  \bibfield  {author} {\bibinfo {author} {\bibfnamefont {Gabriel}\ \bibnamefont
  {Lee}}\ and\ \bibinfo {author} {\bibfnamefont {Carlos E.~M.}\ \bibnamefont
  {Wagner}},\ }\bibfield  {title} {\enquote {\bibinfo {title} {{Higgs bosons in
  heavy supersymmetry with an intermediate m$_A$}},}\ }\href {\doibase
  10.1103/PhysRevD.92.075032} {\bibfield  {journal} {\bibinfo  {journal} {Phys.
  Rev.}\ }\textbf {\bibinfo {volume} {D92}},\ \bibinfo {pages} {075032}
  (\bibinfo {year} {2015})},\ \Eprint {http://arxiv.org/abs/1508.00576}
  {arXiv:1508.00576 [hep-ph]} \BibitemShut {NoStop}%
\bibitem [{\citenamefont {Pardo~Vega}\ and\ \citenamefont
  {Villadoro}(2015)}]{Vega:2015fna}%
  \BibitemOpen
  \bibfield  {author} {\bibinfo {author} {\bibfnamefont {Javier}\ \bibnamefont
  {Pardo~Vega}}\ and\ \bibinfo {author} {\bibfnamefont {Giovanni}\ \bibnamefont
  {Villadoro}},\ }\bibfield  {title} {\enquote {\bibinfo {title} {{SusyHD:
  Higgs mass Determination in Supersymmetry}},}\ }\href {\doibase
  10.1007/JHEP07(2015)159} {\bibfield  {journal} {\bibinfo  {journal} {JHEP}\
  }\textbf {\bibinfo {volume} {07}},\ \bibinfo {pages} {159} (\bibinfo {year}
  {2015})},\ \Eprint {http://arxiv.org/abs/1504.05200} {arXiv:1504.05200
  [hep-ph]} \BibitemShut {NoStop}%
\bibitem [{\citenamefont {Bahl}\ \emph {et~al.}(2018)\citenamefont {Bahl},
  \citenamefont {Heinemeyer}, \citenamefont {Hollik},\ and\ \citenamefont
  {Weiglein}}]{Bahl:2017aev}%
  \BibitemOpen
  \bibfield  {author} {\bibinfo {author} {\bibfnamefont {Henning}\ \bibnamefont
  {Bahl}}, \bibinfo {author} {\bibfnamefont {Sven}\ \bibnamefont {Heinemeyer}},
  \bibinfo {author} {\bibfnamefont {Wolfgang}\ \bibnamefont {Hollik}}, \ and\
  \bibinfo {author} {\bibfnamefont {Georg}\ \bibnamefont {Weiglein}},\
  }\bibfield  {title} {\enquote {\bibinfo {title} {{Reconciling EFT and hybrid
  calculations of the light MSSM Higgs-boson mass}},}\ }\href {\doibase
  10.1140/epjc/s10052-018-5544-3} {\bibfield  {journal} {\bibinfo  {journal}
  {Eur. Phys. J.}\ }\textbf {\bibinfo {volume} {C78}},\ \bibinfo {pages} {57}
  (\bibinfo {year} {2018})},\ \Eprint {http://arxiv.org/abs/1706.00346}
  {arXiv:1706.00346 [hep-ph]} \BibitemShut {NoStop}%
\bibitem [{\citenamefont {Slavich}\ \emph {et~al.}(2021)\citenamefont {Slavich}
  \emph {et~al.}}]{Slavich:2020zjv}%
  \BibitemOpen
  \bibfield  {author} {\bibinfo {author} {\bibfnamefont {P.}~\bibnamefont
  {Slavich}} \emph {et~al.},\ }\bibfield  {title} {\enquote {\bibinfo {title}
  {{Higgs-mass predictions in the MSSM and beyond}},}\ }\href {\doibase
  10.1140/epjc/s10052-021-09198-2} {\bibfield  {journal} {\bibinfo  {journal}
  {Eur. Phys. J. C}\ }\textbf {\bibinfo {volume} {81}},\ \bibinfo {pages} {450}
  (\bibinfo {year} {2021})},\ \Eprint {http://arxiv.org/abs/2012.15629}
  {arXiv:2012.15629 [hep-ph]} \BibitemShut {NoStop}%
\bibitem [{\citenamefont {Okada}\ and\ \citenamefont
  {Tran}(2016)}]{Okada:2016wlm}%
  \BibitemOpen
  \bibfield  {author} {\bibinfo {author} {\bibfnamefont {Nobuchika}\
  \bibnamefont {Okada}}\ and\ \bibinfo {author} {\bibfnamefont {Hieu~Minh}\
  \bibnamefont {Tran}},\ }\bibfield  {title} {\enquote {\bibinfo {title} {{125
  GeV Higgs boson mass and muon $g-2$ in 5D MSSM}},}\ }\href {\doibase
  10.1103/PhysRevD.94.075016} {\bibfield  {journal} {\bibinfo  {journal} {Phys.
  Rev. D}\ }\textbf {\bibinfo {volume} {94}},\ \bibinfo {pages} {075016}
  (\bibinfo {year} {2016})},\ \Eprint {http://arxiv.org/abs/1606.05329}
  {arXiv:1606.05329 [hep-ph]} \BibitemShut {NoStop}%
\bibitem [{\citenamefont {Cox}\ \emph {et~al.}(2018)\citenamefont {Cox},
  \citenamefont {Han},\ and\ \citenamefont {Yanagida}}]{Cox:2018qyi}%
  \BibitemOpen
  \bibfield  {author} {\bibinfo {author} {\bibfnamefont {Peter}\ \bibnamefont
  {Cox}}, \bibinfo {author} {\bibfnamefont {Chengcheng}\ \bibnamefont {Han}}, \
  and\ \bibinfo {author} {\bibfnamefont {Tsutomu~T.}\ \bibnamefont
  {Yanagida}},\ }\bibfield  {title} {\enquote {\bibinfo {title} {{Muon $g-2$
  and dark matter in the minimal supersymmetric standard model}},}\ }\href
  {\doibase 10.1103/PhysRevD.98.055015} {\bibfield  {journal} {\bibinfo
  {journal} {Phys. Rev. D}\ }\textbf {\bibinfo {volume} {98}},\ \bibinfo
  {pages} {055015} (\bibinfo {year} {2018})},\ \Eprint
  {http://arxiv.org/abs/1805.02802} {arXiv:1805.02802 [hep-ph]} \BibitemShut
  {NoStop}%
\bibitem [{\citenamefont {Carena}\ \emph {et~al.}(2018)\citenamefont {Carena},
  \citenamefont {Osborne}, \citenamefont {Shah},\ and\ \citenamefont
  {Wagner}}]{Carena:2018nlf}%
  \BibitemOpen
  \bibfield  {author} {\bibinfo {author} {\bibfnamefont {Marcela}\ \bibnamefont
  {Carena}}, \bibinfo {author} {\bibfnamefont {James}\ \bibnamefont {Osborne}},
  \bibinfo {author} {\bibfnamefont {Nausheen~R.}\ \bibnamefont {Shah}}, \ and\
  \bibinfo {author} {\bibfnamefont {Carlos E.~M.}\ \bibnamefont {Wagner}},\
  }\bibfield  {title} {\enquote {\bibinfo {title} {{Supersymmetry and LHC
  Missing Energy Signals}},}\ }\href {\doibase 10.1103/PhysRevD.98.115010}
  {\bibfield  {journal} {\bibinfo  {journal} {Phys. Rev. D}\ }\textbf {\bibinfo
  {volume} {98}},\ \bibinfo {pages} {115010} (\bibinfo {year} {2018})},\
  \Eprint {http://arxiv.org/abs/1809.11082} {arXiv:1809.11082 [hep-ph]}
  \BibitemShut {NoStop}%
\bibitem [{\citenamefont {Endo}\ and\ \citenamefont
  {Yin}(2019)}]{Endo:2019bcj}%
  \BibitemOpen
  \bibfield  {author} {\bibinfo {author} {\bibfnamefont {Motoi}\ \bibnamefont
  {Endo}}\ and\ \bibinfo {author} {\bibfnamefont {Wen}\ \bibnamefont {Yin}},\
  }\bibfield  {title} {\enquote {\bibinfo {title} {{Explaining electron and
  muon $g-2$ anomaly in SUSY without lepton-flavor mixings}},}\ }\href
  {\doibase 10.1007/JHEP08(2019)122} {\bibfield  {journal} {\bibinfo  {journal}
  {JHEP}\ }\textbf {\bibinfo {volume} {08}},\ \bibinfo {pages} {122} (\bibinfo
  {year} {2019})},\ \Eprint {http://arxiv.org/abs/1906.08768} {arXiv:1906.08768
  [hep-ph]} \BibitemShut {NoStop}%
\bibitem [{\citenamefont {Badziak}\ and\ \citenamefont
  {Sakurai}(2019)}]{Badziak:2019gaf}%
  \BibitemOpen
  \bibfield  {author} {\bibinfo {author} {\bibfnamefont {Marcin}\ \bibnamefont
  {Badziak}}\ and\ \bibinfo {author} {\bibfnamefont {Kazuki}\ \bibnamefont
  {Sakurai}},\ }\bibfield  {title} {\enquote {\bibinfo {title} {{Explanation of
  electron and muon g \ensuremath{-} 2 anomalies in the MSSM}},}\ }\href
  {\doibase 10.1007/JHEP10(2019)024} {\bibfield  {journal} {\bibinfo  {journal}
  {JHEP}\ }\textbf {\bibinfo {volume} {10}},\ \bibinfo {pages} {024} (\bibinfo
  {year} {2019})},\ \Eprint {http://arxiv.org/abs/1908.03607} {arXiv:1908.03607
  [hep-ph]} \BibitemShut {NoStop}%
\bibitem [{\citenamefont {Abdughani}\ \emph {et~al.}(2019)\citenamefont
  {Abdughani}, \citenamefont {Hikasa}, \citenamefont {Wu}, \citenamefont
  {Yang},\ and\ \citenamefont {Zhao}}]{Abdughani:2019wai}%
  \BibitemOpen
  \bibfield  {author} {\bibinfo {author} {\bibfnamefont {Murat}\ \bibnamefont
  {Abdughani}}, \bibinfo {author} {\bibfnamefont {Ken-Ichi}\ \bibnamefont
  {Hikasa}}, \bibinfo {author} {\bibfnamefont {Lei}\ \bibnamefont {Wu}},
  \bibinfo {author} {\bibfnamefont {Jin~Min}\ \bibnamefont {Yang}}, \ and\
  \bibinfo {author} {\bibfnamefont {Jun}\ \bibnamefont {Zhao}},\ }\bibfield
  {title} {\enquote {\bibinfo {title} {{Testing electroweak SUSY for muon $g$
  \ensuremath{-} 2 and dark matter at the LHC and beyond}},}\ }\href {\doibase
  10.1007/JHEP11(2019)095} {\bibfield  {journal} {\bibinfo  {journal} {JHEP}\
  }\textbf {\bibinfo {volume} {11}},\ \bibinfo {pages} {095} (\bibinfo {year}
  {2019})},\ \Eprint {http://arxiv.org/abs/1909.07792} {arXiv:1909.07792
  [hep-ph]} \BibitemShut {NoStop}%
\bibitem [{\citenamefont {Barbieri}\ and\ \citenamefont
  {Maiani}(1982)}]{Barbieri:1982aj}%
  \BibitemOpen
  \bibfield  {author} {\bibinfo {author} {\bibfnamefont {Riccardo}\
  \bibnamefont {Barbieri}}\ and\ \bibinfo {author} {\bibfnamefont
  {L.}~\bibnamefont {Maiani}},\ }\bibfield  {title} {\enquote {\bibinfo {title}
  {{The Muon Anomalous Magnetic Moment in Broken Supersymmetric Theories}},}\
  }\href {\doibase 10.1016/0370-2693(82)90547-0} {\bibfield  {journal}
  {\bibinfo  {journal} {Phys.Lett.}\ }\textbf {\bibinfo {volume} {B117}},\
  \bibinfo {pages} {203} (\bibinfo {year} {1982})}\BibitemShut {NoStop}%
\bibitem [{\citenamefont {Ellis}\ \emph {et~al.}(1982)\citenamefont {Ellis},
  \citenamefont {Hagelin},\ and\ \citenamefont {Nanopoulos}}]{Ellis:1982by}%
  \BibitemOpen
  \bibfield  {author} {\bibinfo {author} {\bibfnamefont {John~R.}\ \bibnamefont
  {Ellis}}, \bibinfo {author} {\bibfnamefont {John~S.}\ \bibnamefont
  {Hagelin}}, \ and\ \bibinfo {author} {\bibfnamefont {Dimitri~V.}\
  \bibnamefont {Nanopoulos}},\ }\bibfield  {title} {\enquote {\bibinfo {title}
  {{Spin 0 Leptons and the Anomalous Magnetic Moment of the Muon}},}\ }\href
  {\doibase 10.1016/0370-2693(82)90343-4} {\bibfield  {journal} {\bibinfo
  {journal} {Phys.Lett.}\ }\textbf {\bibinfo {volume} {B116}},\ \bibinfo
  {pages} {283} (\bibinfo {year} {1982})}\BibitemShut {NoStop}%
\bibitem [{\citenamefont {Kosower}\ \emph {et~al.}(1983)\citenamefont
  {Kosower}, \citenamefont {Krauss},\ and\ \citenamefont
  {Sakai}}]{Kosower:1983yw}%
  \BibitemOpen
  \bibfield  {author} {\bibinfo {author} {\bibfnamefont {David~A.}\
  \bibnamefont {Kosower}}, \bibinfo {author} {\bibfnamefont {Lawrence~M.}\
  \bibnamefont {Krauss}}, \ and\ \bibinfo {author} {\bibfnamefont {Norisuke}\
  \bibnamefont {Sakai}},\ }\bibfield  {title} {\enquote {\bibinfo {title}
  {{Low-Energy Supergravity and the Anomalous Magnetic Moment of the Muon}},}\
  }\href {\doibase 10.1016/0370-2693(83)90152-1} {\bibfield  {journal}
  {\bibinfo  {journal} {Phys.Lett.}\ }\textbf {\bibinfo {volume} {B133}},\
  \bibinfo {pages} {305} (\bibinfo {year} {1983})}\BibitemShut {NoStop}%
\bibitem [{\citenamefont {Moroi}(1996)}]{Moroi:1995yh}%
  \BibitemOpen
  \bibfield  {author} {\bibinfo {author} {\bibfnamefont {Takeo}\ \bibnamefont
  {Moroi}},\ }\bibfield  {title} {\enquote {\bibinfo {title} {{The Muon
  anomalous magnetic dipole moment in the minimal supersymmetric standard
  model}},}\ }\href {\doibase 10.1103/PhysRevD.53.6565} {\bibfield  {journal}
  {\bibinfo  {journal} {Phys. Rev. D}\ }\textbf {\bibinfo {volume} {53}},\
  \bibinfo {pages} {6565--6575} (\bibinfo {year} {1996})},\ \bibinfo {note}
  {[Erratum: Phys.Rev.D 56, 4424 (1997)]},\ \Eprint
  {http://arxiv.org/abs/hep-ph/9512396} {arXiv:hep-ph/9512396} \BibitemShut
  {NoStop}%
\bibitem [{\citenamefont {Carena}\ \emph {et~al.}(1997)\citenamefont {Carena},
  \citenamefont {Giudice},\ and\ \citenamefont {Wagner}}]{Carena:1996qa}%
  \BibitemOpen
  \bibfield  {author} {\bibinfo {author} {\bibfnamefont {Marcela~S.}\
  \bibnamefont {Carena}}, \bibinfo {author} {\bibfnamefont {G.F.}\ \bibnamefont
  {Giudice}}, \ and\ \bibinfo {author} {\bibfnamefont {C.E.M.}\ \bibnamefont
  {Wagner}},\ }\bibfield  {title} {\enquote {\bibinfo {title} {{Constraints on
  supersymmetric models from the muon anomalous magnetic moment}},}\ }\href
  {\doibase 10.1016/S0370-2693(96)01396-2} {\bibfield  {journal} {\bibinfo
  {journal} {Phys.Lett.}\ }\textbf {\bibinfo {volume} {B390}},\ \bibinfo
  {pages} {234--242} (\bibinfo {year} {1997})},\ \Eprint
  {http://arxiv.org/abs/hep-ph/9610233} {arXiv:hep-ph/9610233 [hep-ph]}
  \BibitemShut {NoStop}%
\bibitem [{\citenamefont {Feng}\ and\ \citenamefont
  {Matchev}(2001)}]{Feng:2001tr}%
  \BibitemOpen
  \bibfield  {author} {\bibinfo {author} {\bibfnamefont {Jonathan~L.}\
  \bibnamefont {Feng}}\ and\ \bibinfo {author} {\bibfnamefont {Konstantin~T.}\
  \bibnamefont {Matchev}},\ }\bibfield  {title} {\enquote {\bibinfo {title}
  {{Supersymmetry and the anomalous magnetic moment of the muon}},}\ }\href
  {\doibase 10.1103/PhysRevLett.86.3480} {\bibfield  {journal} {\bibinfo
  {journal} {Phys.Rev.Lett.}\ }\textbf {\bibinfo {volume} {86}},\ \bibinfo
  {pages} {3480--3483} (\bibinfo {year} {2001})},\ \Eprint
  {http://arxiv.org/abs/hep-ph/0102146} {arXiv:hep-ph/0102146 [hep-ph]}
  \BibitemShut {NoStop}%
\bibitem [{\citenamefont {Martin}\ and\ \citenamefont
  {Wells}(2001)}]{Martin:2001st}%
  \BibitemOpen
  \bibfield  {author} {\bibinfo {author} {\bibfnamefont {Stephen~P.}\
  \bibnamefont {Martin}}\ and\ \bibinfo {author} {\bibfnamefont {James~D.}\
  \bibnamefont {Wells}},\ }\bibfield  {title} {\enquote {\bibinfo {title}
  {{Muon Anomalous Magnetic Dipole Moment in Supersymmetric Theories}},}\
  }\href {\doibase 10.1103/PhysRevD.64.035003} {\bibfield  {journal} {\bibinfo
  {journal} {Phys. Rev. D}\ }\textbf {\bibinfo {volume} {64}},\ \bibinfo
  {pages} {035003} (\bibinfo {year} {2001})},\ \Eprint
  {http://arxiv.org/abs/hep-ph/0103067} {arXiv:hep-ph/0103067} \BibitemShut
  {NoStop}%
\bibitem [{\citenamefont {Marchetti}\ \emph {et~al.}(2009)\citenamefont
  {Marchetti}, \citenamefont {Mertens}, \citenamefont {Nierste},\ and\
  \citenamefont {Stockinger}}]{Marchetti:2008hw}%
  \BibitemOpen
  \bibfield  {author} {\bibinfo {author} {\bibfnamefont {Schedar}\ \bibnamefont
  {Marchetti}}, \bibinfo {author} {\bibfnamefont {Susanne}\ \bibnamefont
  {Mertens}}, \bibinfo {author} {\bibfnamefont {Ulrich}\ \bibnamefont
  {Nierste}}, \ and\ \bibinfo {author} {\bibfnamefont {Dominik}\ \bibnamefont
  {Stockinger}},\ }\bibfield  {title} {\enquote {\bibinfo {title}
  {{Tan(beta)-enhanced supersymmetric corrections to the anomalous magnetic
  moment of the muon}},}\ }\href {\doibase 10.1103/PhysRevD.79.013010}
  {\bibfield  {journal} {\bibinfo  {journal} {Phys. Rev. D}\ }\textbf {\bibinfo
  {volume} {79}},\ \bibinfo {pages} {013010} (\bibinfo {year} {2009})},\
  \Eprint {http://arxiv.org/abs/0808.1530} {arXiv:0808.1530 [hep-ph]}
  \BibitemShut {NoStop}%
\bibitem [{\citenamefont {Athron}\ \emph {et~al.}(2016)\citenamefont {Athron},
  \citenamefont {Bach}, \citenamefont {Fargnoli}, \citenamefont {Gnendiger},
  \citenamefont {Greifenhagen}, \citenamefont {Park}, \citenamefont
  {Pa\ss{}ehr}, \citenamefont {St\"ockinger}, \citenamefont
  {St\"ockinger-Kim},\ and\ \citenamefont {Voigt}}]{Athron:2015rva}%
  \BibitemOpen
  \bibfield  {author} {\bibinfo {author} {\bibfnamefont {Peter}\ \bibnamefont
  {Athron}}, \bibinfo {author} {\bibfnamefont {Markus}\ \bibnamefont {Bach}},
  \bibinfo {author} {\bibfnamefont {Helvecio~G.}\ \bibnamefont {Fargnoli}},
  \bibinfo {author} {\bibfnamefont {Christoph}\ \bibnamefont {Gnendiger}},
  \bibinfo {author} {\bibfnamefont {Robert}\ \bibnamefont {Greifenhagen}},
  \bibinfo {author} {\bibfnamefont {Jae-hyeon}\ \bibnamefont {Park}}, \bibinfo
  {author} {\bibfnamefont {Sebastian}\ \bibnamefont {Pa\ss{}ehr}}, \bibinfo
  {author} {\bibfnamefont {Dominik}\ \bibnamefont {St\"ockinger}}, \bibinfo
  {author} {\bibfnamefont {Hyejung}\ \bibnamefont {St\"ockinger-Kim}}, \ and\
  \bibinfo {author} {\bibfnamefont {Alexander}\ \bibnamefont {Voigt}},\
  }\bibfield  {title} {\enquote {\bibinfo {title} {{GM2Calc: Precise MSSM
  prediction for $(g - 2)$ of the muon}},}\ }\href {\doibase
  10.1140/epjc/s10052-015-3870-2} {\bibfield  {journal} {\bibinfo  {journal}
  {Eur. Phys. J. C}\ }\textbf {\bibinfo {volume} {76}},\ \bibinfo {pages} {62}
  (\bibinfo {year} {2016})},\ \Eprint {http://arxiv.org/abs/1510.08071}
  {arXiv:1510.08071 [hep-ph]} \BibitemShut {NoStop}%
\bibitem [{\citenamefont {Carena}\ \emph {et~al.}(2000)\citenamefont {Carena},
  \citenamefont {Garcia}, \citenamefont {Nierste},\ and\ \citenamefont
  {Wagner}}]{Carena:1999py}%
  \BibitemOpen
  \bibfield  {author} {\bibinfo {author} {\bibfnamefont {Marcela}\ \bibnamefont
  {Carena}}, \bibinfo {author} {\bibfnamefont {David}\ \bibnamefont {Garcia}},
  \bibinfo {author} {\bibfnamefont {Ulrich}\ \bibnamefont {Nierste}}, \ and\
  \bibinfo {author} {\bibfnamefont {Carlos E.~M.}\ \bibnamefont {Wagner}},\
  }\bibfield  {title} {\enquote {\bibinfo {title} {{Effective Lagrangian for
  the $\bar{t} b H^{+}$ interaction in the MSSM and charged Higgs
  phenomenology}},}\ }\href {\doibase 10.1016/S0550-3213(00)00146-2} {\bibfield
   {journal} {\bibinfo  {journal} {Nucl. Phys. B}\ }\textbf {\bibinfo {volume}
  {577}},\ \bibinfo {pages} {88--120} (\bibinfo {year} {2000})},\ \Eprint
  {http://arxiv.org/abs/hep-ph/9912516} {arXiv:hep-ph/9912516} \BibitemShut
  {NoStop}%
\bibitem [{\citenamefont {Drees}\ and\ \citenamefont
  {Ghaffari}(2021)}]{Drees:2021pbh}%
  \BibitemOpen
  \bibfield  {author} {\bibinfo {author} {\bibfnamefont {Manuel}\ \bibnamefont
  {Drees}}\ and\ \bibinfo {author} {\bibfnamefont {Ghazaal}\ \bibnamefont
  {Ghaffari}},\ }\bibfield  {title} {\enquote {\bibinfo {title} {{Impact of the
  bounds on the direct search for neutralino dark matter on naturalness}},}\
  }\href {\doibase 10.1103/PhysRevD.104.075031} {\bibfield  {journal} {\bibinfo
   {journal} {Phys. Rev. D}\ }\textbf {\bibinfo {volume} {104}},\ \bibinfo
  {pages} {075031} (\bibinfo {year} {2021})},\ \Eprint
  {http://arxiv.org/abs/2103.15617} {arXiv:2103.15617 [hep-ph]} \BibitemShut
  {NoStop}%
\bibitem [{\citenamefont {Han}\ \emph {et~al.}(2019)\citenamefont {Han},
  \citenamefont {Liu}, \citenamefont {Mukhopadhyay},\ and\ \citenamefont
  {Wang}}]{Han:2018gej}%
  \BibitemOpen
  \bibfield  {author} {\bibinfo {author} {\bibfnamefont {Tao}\ \bibnamefont
  {Han}}, \bibinfo {author} {\bibfnamefont {Hongkai}\ \bibnamefont {Liu}},
  \bibinfo {author} {\bibfnamefont {Satyanarayan}\ \bibnamefont
  {Mukhopadhyay}}, \ and\ \bibinfo {author} {\bibfnamefont {Xing}\ \bibnamefont
  {Wang}},\ }\bibfield  {title} {\enquote {\bibinfo {title} {{Dark Matter Blind
  Spots at One-Loop}},}\ }\href {\doibase 10.1007/JHEP03(2019)080} {\bibfield
  {journal} {\bibinfo  {journal} {JHEP}\ }\textbf {\bibinfo {volume} {03}},\
  \bibinfo {pages} {080} (\bibinfo {year} {2019})},\ \Eprint
  {http://arxiv.org/abs/1810.04679} {arXiv:1810.04679 [hep-ph]} \BibitemShut
  {NoStop}%
\bibitem [{\citenamefont {Aaboud}\ \emph
  {et~al.}(2018{\natexlab{a}})\citenamefont {Aaboud} \emph
  {et~al.}}]{Aaboud:2017sjh}%
  \BibitemOpen
  \bibfield  {author} {\bibinfo {author} {\bibfnamefont {Morad}\ \bibnamefont
  {Aaboud}} \emph {et~al.} (\bibinfo {collaboration} {ATLAS}),\ }\bibfield
  {title} {\enquote {\bibinfo {title} {{Search for additional heavy neutral
  Higgs and gauge bosons in the ditau final state produced in 36 fb$^{-1}$ of
  pp collisions at $ \sqrt{s}=13 $ TeV with the ATLAS detector}},}\ }\href
  {\doibase 10.1007/JHEP01(2018)055} {\bibfield  {journal} {\bibinfo  {journal}
  {JHEP}\ }\textbf {\bibinfo {volume} {01}},\ \bibinfo {pages} {055} (\bibinfo
  {year} {2018}{\natexlab{a}})},\ \Eprint {http://arxiv.org/abs/1709.07242}
  {arXiv:1709.07242 [hep-ex]} \BibitemShut {NoStop}%
\bibitem [{\citenamefont {Sirunyan}\ \emph
  {et~al.}(2018{\natexlab{a}})\citenamefont {Sirunyan} \emph
  {et~al.}}]{Sirunyan:2018zut}%
  \BibitemOpen
  \bibfield  {author} {\bibinfo {author} {\bibfnamefont {Albert~M}\
  \bibnamefont {Sirunyan}} \emph {et~al.} (\bibinfo {collaboration} {CMS}),\
  }\bibfield  {title} {\enquote {\bibinfo {title} {{Search for additional
  neutral MSSM Higgs bosons in the $\tau\tau$ final state in proton-proton
  collisions at $\sqrt{s}=$ 13 TeV}},}\ }\href {\doibase
  10.1007/JHEP09(2018)007} {\bibfield  {journal} {\bibinfo  {journal} {JHEP}\
  }\textbf {\bibinfo {volume} {09}},\ \bibinfo {pages} {007} (\bibinfo {year}
  {2018}{\natexlab{a}})},\ \Eprint {http://arxiv.org/abs/1803.06553}
  {arXiv:1803.06553 [hep-ex]} \BibitemShut {NoStop}%
\bibitem [{\citenamefont {Sirunyan}\ \emph
  {et~al.}(2020{\natexlab{a}})\citenamefont {Sirunyan} \emph
  {et~al.}}]{Sirunyan:2019xjg}%
  \BibitemOpen
  \bibfield  {author} {\bibinfo {author} {\bibfnamefont {Albert~M}\
  \bibnamefont {Sirunyan}} \emph {et~al.} (\bibinfo {collaboration} {CMS}),\
  }\bibfield  {title} {\enquote {\bibinfo {title} {{Search for a heavy
  pseudoscalar Higgs boson decaying into a 125 GeV Higgs boson and a Z boson in
  final states with two tau and two light leptons at $\sqrt{s}=$ 13 TeV}},}\
  }\href {\doibase 10.1007/JHEP03(2020)065} {\bibfield  {journal} {\bibinfo
  {journal} {JHEP}\ }\textbf {\bibinfo {volume} {03}},\ \bibinfo {pages} {065}
  (\bibinfo {year} {2020}{\natexlab{a}})},\ \Eprint
  {http://arxiv.org/abs/1910.11634} {arXiv:1910.11634 [hep-ex]} \BibitemShut
  {NoStop}%
\bibitem [{\citenamefont {Aad}\ \emph {et~al.}(2020{\natexlab{a}})\citenamefont
  {Aad} \emph {et~al.}}]{Aad:2020zxo}%
  \BibitemOpen
  \bibfield  {author} {\bibinfo {author} {\bibfnamefont {Georges}\ \bibnamefont
  {Aad}} \emph {et~al.} (\bibinfo {collaboration} {ATLAS}),\ }\bibfield
  {title} {\enquote {\bibinfo {title} {{Search for heavy Higgs bosons decaying
  into two tau leptons with the ATLAS detector using $pp$ collisions at
  $\sqrt{s}=13$ TeV}},}\ }\href {\doibase 10.1103/PhysRevLett.125.051801}
  {\bibfield  {journal} {\bibinfo  {journal} {Phys. Rev. Lett.}\ }\textbf
  {\bibinfo {volume} {125}},\ \bibinfo {pages} {051801} (\bibinfo {year}
  {2020}{\natexlab{a}})},\ \Eprint {http://arxiv.org/abs/2002.12223}
  {arXiv:2002.12223 [hep-ex]} \BibitemShut {NoStop}%
\bibitem [{\citenamefont {Behnke}\ \emph {et~al.}(2017)\citenamefont {Behnke}
  \emph {et~al.}}]{Behnke:2016lsk}%
  \BibitemOpen
  \bibfield  {author} {\bibinfo {author} {\bibfnamefont {E.}~\bibnamefont
  {Behnke}} \emph {et~al.},\ }\bibfield  {title} {\enquote {\bibinfo {title}
  {{Final Results of the PICASSO Dark Matter Search Experiment}},}\ }\href
  {\doibase 10.1016/j.astropartphys.2017.02.005} {\bibfield  {journal}
  {\bibinfo  {journal} {Astropart. Phys.}\ }\textbf {\bibinfo {volume} {90}},\
  \bibinfo {pages} {85--92} (\bibinfo {year} {2017})},\ \Eprint
  {http://arxiv.org/abs/1611.01499} {arXiv:1611.01499 [hep-ex]} \BibitemShut
  {NoStop}%
\bibitem [{\citenamefont {Fu}\ \emph {et~al.}(2017)\citenamefont {Fu} \emph
  {et~al.}}]{Fu:2016ega}%
  \BibitemOpen
  \bibfield  {author} {\bibinfo {author} {\bibfnamefont {Changbo}\ \bibnamefont
  {Fu}} \emph {et~al.} (\bibinfo {collaboration} {PandaX-II}),\ }\bibfield
  {title} {\enquote {\bibinfo {title} {{Spin-Dependent
  Weakly-Interacting-Massive-Particle\textendash{}Nucleon Cross Section Limits
  from First Data of PandaX-II Experiment}},}\ }\href {\doibase
  10.1103/PhysRevLett.118.071301} {\bibfield  {journal} {\bibinfo  {journal}
  {Phys. Rev. Lett.}\ }\textbf {\bibinfo {volume} {118}},\ \bibinfo {pages}
  {071301} (\bibinfo {year} {2017})},\ \bibinfo {note} {[Erratum:
  Phys.Rev.Lett. 120, 049902 (2018)]},\ \Eprint
  {http://arxiv.org/abs/1611.06553} {arXiv:1611.06553 [hep-ex]} \BibitemShut
  {NoStop}%
\bibitem [{\citenamefont {Aprile}\ \emph
  {et~al.}(2019{\natexlab{a}})\citenamefont {Aprile} \emph
  {et~al.}}]{Aprile:2019dbj}%
  \BibitemOpen
  \bibfield  {author} {\bibinfo {author} {\bibfnamefont {E.}~\bibnamefont
  {Aprile}} \emph {et~al.} (\bibinfo {collaboration} {XENON}),\ }\bibfield
  {title} {\enquote {\bibinfo {title} {{Constraining the spin-dependent
  WIMP-nucleon cross sections with XENON1T}},}\ }\href {\doibase
  10.1103/PhysRevLett.122.141301} {\bibfield  {journal} {\bibinfo  {journal}
  {Phys. Rev. Lett.}\ }\textbf {\bibinfo {volume} {122}},\ \bibinfo {pages}
  {141301} (\bibinfo {year} {2019}{\natexlab{a}})},\ \Eprint
  {http://arxiv.org/abs/1902.03234} {arXiv:1902.03234 [astro-ph.CO]}
  \BibitemShut {NoStop}%
\bibitem [{\citenamefont {Amole}\ \emph {et~al.}(2019)\citenamefont {Amole}
  \emph {et~al.}}]{Amole:2019fdf}%
  \BibitemOpen
  \bibfield  {author} {\bibinfo {author} {\bibfnamefont {C.}~\bibnamefont
  {Amole}} \emph {et~al.} (\bibinfo {collaboration} {PICO}),\ }\bibfield
  {title} {\enquote {\bibinfo {title} {{Dark Matter Search Results from the
  Complete Exposure of the PICO-60 C$_3$F$_8$ Bubble Chamber}},}\ }\href
  {\doibase 10.1103/PhysRevD.100.022001} {\bibfield  {journal} {\bibinfo
  {journal} {Phys. Rev. D}\ }\textbf {\bibinfo {volume} {100}},\ \bibinfo
  {pages} {022001} (\bibinfo {year} {2019})},\ \Eprint
  {http://arxiv.org/abs/1902.04031} {arXiv:1902.04031 [astro-ph.CO]}
  \BibitemShut {NoStop}%
\bibitem [{\citenamefont {Sirunyan}\ \emph
  {et~al.}(2020{\natexlab{b}})\citenamefont {Sirunyan} \emph
  {et~al.}}]{CMS:2019zmn}%
  \BibitemOpen
  \bibfield  {author} {\bibinfo {author} {\bibfnamefont {Albert~M}\
  \bibnamefont {Sirunyan}} \emph {et~al.} (\bibinfo {collaboration} {CMS}),\
  }\bibfield  {title} {\enquote {\bibinfo {title} {{Search for Supersymmetry
  with a Compressed Mass Spectrum in Events with a Soft $\tau$ Lepton, a Highly
  Energetic Jet, and Large Missing Transverse Momentum in Proton-Proton
  Collisions at $\sqrt{s}=$ TeV}},}\ }\href {\doibase
  10.1103/PhysRevLett.124.041803} {\bibfield  {journal} {\bibinfo  {journal}
  {Phys. Rev. Lett.}\ }\textbf {\bibinfo {volume} {124}},\ \bibinfo {pages}
  {041803} (\bibinfo {year} {2020}{\natexlab{b}})},\ \Eprint
  {http://arxiv.org/abs/1910.01185} {arXiv:1910.01185 [hep-ex]} \BibitemShut
  {NoStop}%
\bibitem [{\citenamefont {Aad}\ \emph {et~al.}(2014{\natexlab{a}})\citenamefont
  {Aad} \emph {et~al.}}]{Aad:2014vma}%
  \BibitemOpen
  \bibfield  {author} {\bibinfo {author} {\bibfnamefont {Georges}\ \bibnamefont
  {Aad}} \emph {et~al.} (\bibinfo {collaboration} {ATLAS}),\ }\bibfield
  {title} {\enquote {\bibinfo {title} {{Search for direct production of
  charginos, neutralinos and sleptons in final states with two leptons and
  missing transverse momentum in $pp$ collisions at $\sqrt{s} =$ 8 TeV with the
  ATLAS detector}},}\ }\href {\doibase 10.1007/JHEP05(2014)071} {\bibfield
  {journal} {\bibinfo  {journal} {JHEP}\ }\textbf {\bibinfo {volume} {05}},\
  \bibinfo {pages} {071} (\bibinfo {year} {2014}{\natexlab{a}})},\ \Eprint
  {http://arxiv.org/abs/1403.5294} {arXiv:1403.5294 [hep-ex]} \BibitemShut
  {NoStop}%
\bibitem [{\citenamefont {Aad}\ \emph {et~al.}(2020{\natexlab{b}})\citenamefont
  {Aad} \emph {et~al.}}]{Aad:2019vnb}%
  \BibitemOpen
  \bibfield  {author} {\bibinfo {author} {\bibfnamefont {Georges}\ \bibnamefont
  {Aad}} \emph {et~al.} (\bibinfo {collaboration} {ATLAS}),\ }\bibfield
  {title} {\enquote {\bibinfo {title} {{Search for electroweak production of
  charginos and sleptons decaying into final states with two leptons and
  missing transverse momentum in $\sqrt{s}=13$ TeV $pp$ collisions using the
  ATLAS detector}},}\ }\href {\doibase 10.1140/epjc/s10052-019-7594-6}
  {\bibfield  {journal} {\bibinfo  {journal} {Eur. Phys. J. C}\ }\textbf
  {\bibinfo {volume} {80}},\ \bibinfo {pages} {123} (\bibinfo {year}
  {2020}{\natexlab{b}})},\ \Eprint {http://arxiv.org/abs/1908.08215}
  {arXiv:1908.08215 [hep-ex]} \BibitemShut {NoStop}%
\bibitem [{\citenamefont {Sirunyan}\ \emph
  {et~al.}(2020{\natexlab{c}})\citenamefont {Sirunyan} \emph
  {et~al.}}]{CMS:2019eln}%
  \BibitemOpen
  \bibfield  {author} {\bibinfo {author} {\bibfnamefont {Albert~M}\
  \bibnamefont {Sirunyan}} \emph {et~al.} (\bibinfo {collaboration} {CMS}),\
  }\bibfield  {title} {\enquote {\bibinfo {title} {{Search for direct pair
  production of supersymmetric partners to the $\tau$ lepton in proton-proton
  collisions at $\sqrt{s}=$ 13 TeV}},}\ }\href {\doibase
  10.1140/epjc/s10052-020-7739-7} {\bibfield  {journal} {\bibinfo  {journal}
  {Eur. Phys. J. C}\ }\textbf {\bibinfo {volume} {80}},\ \bibinfo {pages} {189}
  (\bibinfo {year} {2020}{\natexlab{c}})},\ \Eprint
  {http://arxiv.org/abs/1907.13179} {arXiv:1907.13179 [hep-ex]} \BibitemShut
  {NoStop}%
\bibitem [{\citenamefont {Aad}\ \emph {et~al.}(2020{\natexlab{c}})\citenamefont
  {Aad} \emph {et~al.}}]{Aad:2019byo}%
  \BibitemOpen
  \bibfield  {author} {\bibinfo {author} {\bibfnamefont {Georges}\ \bibnamefont
  {Aad}} \emph {et~al.} (\bibinfo {collaboration} {ATLAS}),\ }\bibfield
  {title} {\enquote {\bibinfo {title} {{Search for direct stau production in
  events with two hadronic $\tau$-leptons in $\sqrt{s} = 13$ TeV $pp$
  collisions with the ATLAS detector}},}\ }\href {\doibase
  10.1103/PhysRevD.101.032009} {\bibfield  {journal} {\bibinfo  {journal}
  {Phys. Rev. D}\ }\textbf {\bibinfo {volume} {101}},\ \bibinfo {pages}
  {032009} (\bibinfo {year} {2020}{\natexlab{c}})},\ \Eprint
  {http://arxiv.org/abs/1911.06660} {arXiv:1911.06660 [hep-ex]} \BibitemShut
  {NoStop}%
\bibitem [{\citenamefont {Aad}\ \emph {et~al.}(2020{\natexlab{d}})\citenamefont
  {Aad} \emph {et~al.}}]{Aad:2019qnd}%
  \BibitemOpen
  \bibfield  {author} {\bibinfo {author} {\bibfnamefont {Georges}\ \bibnamefont
  {Aad}} \emph {et~al.} (\bibinfo {collaboration} {ATLAS}),\ }\bibfield
  {title} {\enquote {\bibinfo {title} {{Searches for electroweak production of
  supersymmetric particles with compressed mass spectra in $\sqrt{s}=$ 13 TeV
  $pp$ collisions with the ATLAS detector}},}\ }\href {\doibase
  10.1103/PhysRevD.101.052005} {\bibfield  {journal} {\bibinfo  {journal}
  {Phys. Rev. D}\ }\textbf {\bibinfo {volume} {101}},\ \bibinfo {pages}
  {052005} (\bibinfo {year} {2020}{\natexlab{d}})},\ \Eprint
  {http://arxiv.org/abs/1911.12606} {arXiv:1911.12606 [hep-ex]} \BibitemShut
  {NoStop}%
\bibitem [{\citenamefont {Aaboud}\ \emph
  {et~al.}(2018{\natexlab{b}})\citenamefont {Aaboud} \emph
  {et~al.}}]{Aaboud:2017nhr}%
  \BibitemOpen
  \bibfield  {author} {\bibinfo {author} {\bibfnamefont {Morad}\ \bibnamefont
  {Aaboud}} \emph {et~al.} (\bibinfo {collaboration} {ATLAS}),\ }\bibfield
  {title} {\enquote {\bibinfo {title} {{Search for the direct production of
  charginos and neutralinos in final states with tau leptons in $\sqrt{s} = $
  13 TeV $pp$ collisions with the ATLAS detector}},}\ }\href {\doibase
  10.1140/epjc/s10052-018-5583-9} {\bibfield  {journal} {\bibinfo  {journal}
  {Eur. Phys. J. C}\ }\textbf {\bibinfo {volume} {78}},\ \bibinfo {pages} {154}
  (\bibinfo {year} {2018}{\natexlab{b}})},\ \Eprint
  {http://arxiv.org/abs/1708.07875} {arXiv:1708.07875 [hep-ex]} \BibitemShut
  {NoStop}%
\bibitem [{\citenamefont {Sirunyan}\ \emph
  {et~al.}(2018{\natexlab{b}})\citenamefont {Sirunyan} \emph
  {et~al.}}]{Sirunyan:2018ubx}%
  \BibitemOpen
  \bibfield  {author} {\bibinfo {author} {\bibfnamefont {A.~M.}\ \bibnamefont
  {Sirunyan}} \emph {et~al.} (\bibinfo {collaboration} {CMS}),\ }\bibfield
  {title} {\enquote {\bibinfo {title} {{Combined search for electroweak
  production of charginos and neutralinos in proton-proton collisions at
  $\sqrt{s} =$ 13 TeV}},}\ }\href {\doibase 10.1007/JHEP03(2018)160} {\bibfield
   {journal} {\bibinfo  {journal} {JHEP}\ }\textbf {\bibinfo {volume} {03}},\
  \bibinfo {pages} {160} (\bibinfo {year} {2018}{\natexlab{b}})},\ \Eprint
  {http://arxiv.org/abs/1801.03957} {arXiv:1801.03957 [hep-ex]} \BibitemShut
  {NoStop}%
\bibitem [{\citenamefont {Aaboud}\ \emph
  {et~al.}(2018{\natexlab{c}})\citenamefont {Aaboud} \emph
  {et~al.}}]{Aaboud:2018jiw}%
  \BibitemOpen
  \bibfield  {author} {\bibinfo {author} {\bibfnamefont {M.}~\bibnamefont
  {Aaboud}} \emph {et~al.} (\bibinfo {collaboration} {ATLAS}),\ }\bibfield
  {title} {\enquote {\bibinfo {title} {{Search for electroweak production of
  supersymmetric particles in final states with two or three leptons at
  $\sqrt{s}=13\,$TeV with the ATLAS detector}},}\ }\href {\doibase
  10.1140/epjc/s10052-018-6423-7} {\bibfield  {journal} {\bibinfo  {journal}
  {Eur. Phys. J. C}\ }\textbf {\bibinfo {volume} {78}},\ \bibinfo {pages} {995}
  (\bibinfo {year} {2018}{\natexlab{c}})},\ \Eprint
  {http://arxiv.org/abs/1803.02762} {arXiv:1803.02762 [hep-ex]} \BibitemShut
  {NoStop}%
\bibitem [{\citenamefont {Aaboud}\ \emph
  {et~al.}(2018{\natexlab{d}})\citenamefont {Aaboud} \emph
  {et~al.}}]{Aaboud:2018sua}%
  \BibitemOpen
  \bibfield  {author} {\bibinfo {author} {\bibfnamefont {Morad}\ \bibnamefont
  {Aaboud}} \emph {et~al.} (\bibinfo {collaboration} {ATLAS}),\ }\bibfield
  {title} {\enquote {\bibinfo {title} {{Search for chargino-neutralino
  production using recursive jigsaw reconstruction in final states with two or
  three charged leptons in proton-proton collisions at $\sqrt{s}=13$ TeV with
  the ATLAS detector}},}\ }\href {\doibase 10.1103/PhysRevD.98.092012}
  {\bibfield  {journal} {\bibinfo  {journal} {Phys. Rev. D}\ }\textbf {\bibinfo
  {volume} {98}},\ \bibinfo {pages} {092012} (\bibinfo {year}
  {2018}{\natexlab{d}})},\ \Eprint {http://arxiv.org/abs/1806.02293}
  {arXiv:1806.02293 [hep-ex]} \BibitemShut {NoStop}%
\bibitem [{\citenamefont {Aaboud}\ \emph
  {et~al.}(2019{\natexlab{a}})\citenamefont {Aaboud} \emph
  {et~al.}}]{Aaboud:2018ngk}%
  \BibitemOpen
  \bibfield  {author} {\bibinfo {author} {\bibfnamefont {Morad}\ \bibnamefont
  {Aaboud}} \emph {et~al.} (\bibinfo {collaboration} {ATLAS}),\ }\bibfield
  {title} {\enquote {\bibinfo {title} {{Search for chargino and neutralino
  production in final states with a Higgs boson and missing transverse momentum
  at $\sqrt{s} = 13$ TeV with the ATLAS detector}},}\ }\href {\doibase
  10.1103/PhysRevD.100.012006} {\bibfield  {journal} {\bibinfo  {journal}
  {Phys. Rev. D}\ }\textbf {\bibinfo {volume} {100}},\ \bibinfo {pages}
  {012006} (\bibinfo {year} {2019}{\natexlab{a}})},\ \Eprint
  {http://arxiv.org/abs/1812.09432} {arXiv:1812.09432 [hep-ex]} \BibitemShut
  {NoStop}%
\bibitem [{\citenamefont
  {Collaboration}(2019{\natexlab{a}})}]{ATLAS-CONF-2019-008}%
  \BibitemOpen
  \bibfield  {author} {\bibinfo {author} {\bibfnamefont {ATLAS}\ \bibnamefont
  {Collaboration}},\ }\href {https://cds.cern.ch/record/2668387} {\emph
  {\bibinfo {title} {{Search for electroweak production of charginos and
  sleptons decaying in final states with two leptons and missing transverse
  momentum in $\sqrt{s}=13$ TeV $pp$ collisions using the ATLAS detector}}}},\
  \bibinfo {type} {Tech. Rep.}\ \bibinfo {number} {ATLAS-CONF-2019-008}\
  (\bibinfo  {institution} {CERN},\ \bibinfo {address} {Geneva},\ \bibinfo
  {year} {2019})\BibitemShut {NoStop}%
\bibitem [{\citenamefont {Aad}\ \emph {et~al.}(2021{\natexlab{a}})\citenamefont
  {Aad} \emph {et~al.}}]{ATLAS:2021moa}%
  \BibitemOpen
  \bibfield  {author} {\bibinfo {author} {\bibfnamefont {Georges}\ \bibnamefont
  {Aad}} \emph {et~al.} (\bibinfo {collaboration} {ATLAS}),\ }\bibfield
  {title} {\enquote {\bibinfo {title} {{Search for chargino--neutralino pair
  production in final states with three leptons and missing transverse momentum
  in $\sqrt{s} = 13$ TeV $pp$ collisions with the ATLAS detector}},}\ }\href
  {\doibase 10.1140/epjc/s10052-021-09749-7} {\bibfield  {journal} {\bibinfo
  {journal} {Eur. Phys. J. C}\ }\textbf {\bibinfo {volume} {81}},\ \bibinfo
  {pages} {1118} (\bibinfo {year} {2021}{\natexlab{a}})},\ \Eprint
  {http://arxiv.org/abs/2106.01676} {arXiv:2106.01676 [hep-ex]} \BibitemShut
  {NoStop}%
\bibitem [{\citenamefont {Angloher}\ \emph {et~al.}(2016)\citenamefont
  {Angloher} \emph {et~al.}}]{Angloher:2015ewa}%
  \BibitemOpen
  \bibfield  {author} {\bibinfo {author} {\bibfnamefont {G.}~\bibnamefont
  {Angloher}} \emph {et~al.} (\bibinfo {collaboration} {CRESST}),\ }\bibfield
  {title} {\enquote {\bibinfo {title} {{Results on light dark matter particles
  with a low-threshold CRESST-II detector}},}\ }\href {\doibase
  10.1140/epjc/s10052-016-3877-3} {\bibfield  {journal} {\bibinfo  {journal}
  {Eur. Phys. J. C}\ }\textbf {\bibinfo {volume} {76}},\ \bibinfo {pages} {25}
  (\bibinfo {year} {2016})},\ \Eprint {http://arxiv.org/abs/1509.01515}
  {arXiv:1509.01515 [astro-ph.CO]} \BibitemShut {NoStop}%
\bibitem [{\citenamefont {Agnes}\ \emph {et~al.}(2018)\citenamefont {Agnes}
  \emph {et~al.}}]{Agnes:2018ves}%
  \BibitemOpen
  \bibfield  {author} {\bibinfo {author} {\bibfnamefont {P.}~\bibnamefont
  {Agnes}} \emph {et~al.} (\bibinfo {collaboration} {DarkSide}),\ }\bibfield
  {title} {\enquote {\bibinfo {title} {{Low-Mass Dark Matter Search with the
  DarkSide-50 Experiment}},}\ }\href {\doibase 10.1103/PhysRevLett.121.081307}
  {\bibfield  {journal} {\bibinfo  {journal} {Phys. Rev. Lett.}\ }\textbf
  {\bibinfo {volume} {121}},\ \bibinfo {pages} {081307} (\bibinfo {year}
  {2018})},\ \Eprint {http://arxiv.org/abs/1802.06994} {arXiv:1802.06994
  [astro-ph.HE]} \BibitemShut {NoStop}%
\bibitem [{\citenamefont {Aprile}\ \emph {et~al.}(2018)\citenamefont {Aprile}
  \emph {et~al.}}]{Aprile:2018dbl}%
  \BibitemOpen
  \bibfield  {author} {\bibinfo {author} {\bibfnamefont {E.}~\bibnamefont
  {Aprile}} \emph {et~al.} (\bibinfo {collaboration} {XENON}),\ }\bibfield
  {title} {\enquote {\bibinfo {title} {{Dark Matter Search Results from a One
  Ton-Year Exposure of XENON1T}},}\ }\href {\doibase
  10.1103/PhysRevLett.121.111302} {\bibfield  {journal} {\bibinfo  {journal}
  {Phys. Rev. Lett.}\ }\textbf {\bibinfo {volume} {121}},\ \bibinfo {pages}
  {111302} (\bibinfo {year} {2018})},\ \Eprint
  {http://arxiv.org/abs/1805.12562} {arXiv:1805.12562 [astro-ph.CO]}
  \BibitemShut {NoStop}%
\bibitem [{\citenamefont {Aprile}\ \emph
  {et~al.}(2019{\natexlab{b}})\citenamefont {Aprile} \emph
  {et~al.}}]{Aprile:2019xxb}%
  \BibitemOpen
  \bibfield  {author} {\bibinfo {author} {\bibfnamefont {E.}~\bibnamefont
  {Aprile}} \emph {et~al.} (\bibinfo {collaboration} {XENON}),\ }\bibfield
  {title} {\enquote {\bibinfo {title} {{Light Dark Matter Search with
  Ionization Signals in XENON1T}},}\ }\href {\doibase
  10.1103/PhysRevLett.123.251801} {\bibfield  {journal} {\bibinfo  {journal}
  {Phys. Rev. Lett.}\ }\textbf {\bibinfo {volume} {123}},\ \bibinfo {pages}
  {251801} (\bibinfo {year} {2019}{\natexlab{b}})},\ \Eprint
  {http://arxiv.org/abs/1907.11485} {arXiv:1907.11485 [hep-ex]} \BibitemShut
  {NoStop}%
\bibitem [{\citenamefont {Ellis}\ \emph {et~al.}(1998)\citenamefont {Ellis},
  \citenamefont {Falk},\ and\ \citenamefont {Olive}}]{Ellis:1998kh}%
  \BibitemOpen
  \bibfield  {author} {\bibinfo {author} {\bibfnamefont {John~R.}\ \bibnamefont
  {Ellis}}, \bibinfo {author} {\bibfnamefont {Toby}\ \bibnamefont {Falk}}, \
  and\ \bibinfo {author} {\bibfnamefont {Keith~A.}\ \bibnamefont {Olive}},\
  }\bibfield  {title} {\enquote {\bibinfo {title} {{Neutralino - Stau
  coannihilation and the cosmological upper limit on the mass of the lightest
  supersymmetric particle}},}\ }\href {\doibase 10.1016/S0370-2693(98)01392-6}
  {\bibfield  {journal} {\bibinfo  {journal} {Phys. Lett. B}\ }\textbf
  {\bibinfo {volume} {444}},\ \bibinfo {pages} {367--372} (\bibinfo {year}
  {1998})},\ \Eprint {http://arxiv.org/abs/hep-ph/9810360}
  {arXiv:hep-ph/9810360} \BibitemShut {NoStop}%
\bibitem [{\citenamefont {Ellis}\ \emph {et~al.}(2000)\citenamefont {Ellis},
  \citenamefont {Falk}, \citenamefont {Olive},\ and\ \citenamefont
  {Srednicki}}]{Ellis:1999mm}%
  \BibitemOpen
  \bibfield  {author} {\bibinfo {author} {\bibfnamefont {John~R.}\ \bibnamefont
  {Ellis}}, \bibinfo {author} {\bibfnamefont {Toby}\ \bibnamefont {Falk}},
  \bibinfo {author} {\bibfnamefont {Keith~A.}\ \bibnamefont {Olive}}, \ and\
  \bibinfo {author} {\bibfnamefont {Mark}\ \bibnamefont {Srednicki}},\
  }\bibfield  {title} {\enquote {\bibinfo {title} {{Calculations of
  neutralino-stau coannihilation channels and the cosmologically relevant
  region of MSSM parameter space}},}\ }\href {\doibase
  10.1016/S0927-6505(99)00104-8} {\bibfield  {journal} {\bibinfo  {journal}
  {Astropart. Phys.}\ }\textbf {\bibinfo {volume} {13}},\ \bibinfo {pages}
  {181--213} (\bibinfo {year} {2000})},\ \bibinfo {note} {[Erratum:
  Astropart.Phys. 15, 413--414 (2001)]},\ \Eprint
  {http://arxiv.org/abs/hep-ph/9905481} {arXiv:hep-ph/9905481} \BibitemShut
  {NoStop}%
\bibitem [{\citenamefont {Buckley}\ \emph {et~al.}(2013)\citenamefont
  {Buckley}, \citenamefont {Hooper},\ and\ \citenamefont
  {Kumar}}]{Buckley:2013sca}%
  \BibitemOpen
  \bibfield  {author} {\bibinfo {author} {\bibfnamefont {Matthew~R.}\
  \bibnamefont {Buckley}}, \bibinfo {author} {\bibfnamefont {Dan}\ \bibnamefont
  {Hooper}}, \ and\ \bibinfo {author} {\bibfnamefont {Jason}\ \bibnamefont
  {Kumar}},\ }\bibfield  {title} {\enquote {\bibinfo {title} {{Phenomenology of
  Dirac Neutralino Dark Matter}},}\ }\href {\doibase
  10.1103/PhysRevD.88.063532} {\bibfield  {journal} {\bibinfo  {journal} {Phys.
  Rev. D}\ }\textbf {\bibinfo {volume} {88}},\ \bibinfo {pages} {063532}
  (\bibinfo {year} {2013})},\ \Eprint {http://arxiv.org/abs/1307.3561}
  {arXiv:1307.3561 [hep-ph]} \BibitemShut {NoStop}%
\bibitem [{\citenamefont {Han}\ \emph {et~al.}(2013)\citenamefont {Han},
  \citenamefont {Liu},\ and\ \citenamefont {Natarajan}}]{Han:2013gba}%
  \BibitemOpen
  \bibfield  {author} {\bibinfo {author} {\bibfnamefont {Tao}\ \bibnamefont
  {Han}}, \bibinfo {author} {\bibfnamefont {Zhen}\ \bibnamefont {Liu}}, \ and\
  \bibinfo {author} {\bibfnamefont {Aravind}\ \bibnamefont {Natarajan}},\
  }\bibfield  {title} {\enquote {\bibinfo {title} {{Dark matter and Higgs
  bosons in the MSSM}},}\ }\href {\doibase 10.1007/JHEP11(2013)008} {\bibfield
  {journal} {\bibinfo  {journal} {JHEP}\ }\textbf {\bibinfo {volume} {11}},\
  \bibinfo {pages} {008} (\bibinfo {year} {2013})},\ \Eprint
  {http://arxiv.org/abs/1303.3040} {arXiv:1303.3040 [hep-ph]} \BibitemShut
  {NoStop}%
\bibitem [{\citenamefont {Cabrera}\ \emph {et~al.}(2016)\citenamefont
  {Cabrera}, \citenamefont {Casas}, \citenamefont {Delgado}, \citenamefont
  {Robles},\ and\ \citenamefont {Ruiz~de Austri}}]{Cabrera:2016wwr}%
  \BibitemOpen
  \bibfield  {author} {\bibinfo {author} {\bibfnamefont {Mar\'\i{}a~Eugenia}\
  \bibnamefont {Cabrera}}, \bibinfo {author} {\bibfnamefont {J.~Alberto}\
  \bibnamefont {Casas}}, \bibinfo {author} {\bibfnamefont {Antonio}\
  \bibnamefont {Delgado}}, \bibinfo {author} {\bibfnamefont {Sandra}\
  \bibnamefont {Robles}}, \ and\ \bibinfo {author} {\bibfnamefont {Roberto}\
  \bibnamefont {Ruiz~de Austri}},\ }\bibfield  {title} {\enquote {\bibinfo
  {title} {{Naturalness of MSSM dark matter}},}\ }\href {\doibase
  10.1007/JHEP08(2016)058} {\bibfield  {journal} {\bibinfo  {journal} {JHEP}\
  }\textbf {\bibinfo {volume} {08}},\ \bibinfo {pages} {058} (\bibinfo {year}
  {2016})},\ \Eprint {http://arxiv.org/abs/1604.02102} {arXiv:1604.02102
  [hep-ph]} \BibitemShut {NoStop}%
\bibitem [{\citenamefont {Baker}\ and\ \citenamefont
  {Thamm}(2018)}]{Baker:2018uox}%
  \BibitemOpen
  \bibfield  {author} {\bibinfo {author} {\bibfnamefont {Michael~J.}\
  \bibnamefont {Baker}}\ and\ \bibinfo {author} {\bibfnamefont {Andrea}\
  \bibnamefont {Thamm}},\ }\bibfield  {title} {\enquote {\bibinfo {title}
  {{Leptonic WIMP Coannihilation and the Current Dark Matter Search
  Strategy}},}\ }\href {\doibase 10.1007/JHEP10(2018)187} {\bibfield  {journal}
  {\bibinfo  {journal} {JHEP}\ }\textbf {\bibinfo {volume} {10}},\ \bibinfo
  {pages} {187} (\bibinfo {year} {2018})},\ \Eprint
  {http://arxiv.org/abs/1806.07896} {arXiv:1806.07896 [hep-ph]} \BibitemShut
  {NoStop}%
\bibitem [{\citenamefont {Pierce}\ \emph {et~al.}(2013)\citenamefont {Pierce},
  \citenamefont {Shah},\ and\ \citenamefont {Freese}}]{Pierce:2013rda}%
  \BibitemOpen
  \bibfield  {author} {\bibinfo {author} {\bibfnamefont {Aaron}\ \bibnamefont
  {Pierce}}, \bibinfo {author} {\bibfnamefont {Nausheen~R.}\ \bibnamefont
  {Shah}}, \ and\ \bibinfo {author} {\bibfnamefont {Katherine}\ \bibnamefont
  {Freese}},\ }\bibfield  {title} {\enquote {\bibinfo {title} {{Neutralino Dark
  Matter with Light Staus}},}\ }\href@noop {} {\  (\bibinfo {year} {2013})},\
  \Eprint {http://arxiv.org/abs/1309.7351} {arXiv:1309.7351 [hep-ph]}
  \BibitemShut {NoStop}%
\bibitem [{\citenamefont {Fukushima}\ \emph {et~al.}(2014)\citenamefont
  {Fukushima}, \citenamefont {Kelso}, \citenamefont {Kumar}, \citenamefont
  {Sandick},\ and\ \citenamefont {Yamamoto}}]{Fukushima:2014yia}%
  \BibitemOpen
  \bibfield  {author} {\bibinfo {author} {\bibfnamefont {Keita}\ \bibnamefont
  {Fukushima}}, \bibinfo {author} {\bibfnamefont {Chris}\ \bibnamefont
  {Kelso}}, \bibinfo {author} {\bibfnamefont {Jason}\ \bibnamefont {Kumar}},
  \bibinfo {author} {\bibfnamefont {Pearl}\ \bibnamefont {Sandick}}, \ and\
  \bibinfo {author} {\bibfnamefont {Takahiro}\ \bibnamefont {Yamamoto}},\
  }\bibfield  {title} {\enquote {\bibinfo {title} {{MSSM dark matter and a
  light slepton sector: The incredible bulk}},}\ }\href {\doibase
  10.1103/PhysRevD.90.095007} {\bibfield  {journal} {\bibinfo  {journal} {Phys.
  Rev. D}\ }\textbf {\bibinfo {volume} {90}},\ \bibinfo {pages} {095007}
  (\bibinfo {year} {2014})},\ \Eprint {http://arxiv.org/abs/1406.4903}
  {arXiv:1406.4903 [hep-ph]} \BibitemShut {NoStop}%
\bibitem [{\citenamefont {Vami}(2019)}]{Vami:2019slp}%
  \BibitemOpen
  \bibfield  {author} {\bibinfo {author} {\bibfnamefont {Tamas~Almos}\
  \bibnamefont {Vami}} (\bibinfo {collaboration} {ATLAS, CMS}),\ }\bibfield
  {title} {\enquote {\bibinfo {title} {{Searches for gluinos and squarks}},}\
  }\href {\doibase 10.22323/1.350.0168} {\bibfield  {journal} {\bibinfo
  {journal} {PoS}\ }\textbf {\bibinfo {volume} {LHCP2019}},\ \bibinfo {pages}
  {168} (\bibinfo {year} {2019})},\ \Eprint {http://arxiv.org/abs/1909.11753}
  {arXiv:1909.11753 [hep-ex]} \BibitemShut {NoStop}%
\bibitem [{\citenamefont {Aad}\ \emph {et~al.}(2021{\natexlab{b}})\citenamefont
  {Aad} \emph {et~al.}}]{Aad:2020aze}%
  \BibitemOpen
  \bibfield  {author} {\bibinfo {author} {\bibfnamefont {Georges}\ \bibnamefont
  {Aad}} \emph {et~al.} (\bibinfo {collaboration} {ATLAS}),\ }\bibfield
  {title} {\enquote {\bibinfo {title} {{Search for squarks and gluinos in final
  states with jets and missing transverse momentum using 139 fb$^{-1}$ of
  $\sqrt{s}$ =13 TeV $pp$ collision data with the ATLAS detector}},}\ }\href
  {\doibase 10.1007/JHEP02(2021)143} {\bibfield  {journal} {\bibinfo  {journal}
  {JHEP}\ }\textbf {\bibinfo {volume} {02}},\ \bibinfo {pages} {143} (\bibinfo
  {year} {2021}{\natexlab{b}})},\ \Eprint {http://arxiv.org/abs/2010.14293}
  {arXiv:2010.14293 [hep-ex]} \BibitemShut {NoStop}%
\bibitem [{\citenamefont {Aad}\ \emph {et~al.}(2021{\natexlab{c}})\citenamefont
  {Aad} \emph {et~al.}}]{Aad:2020aob}%
  \BibitemOpen
  \bibfield  {author} {\bibinfo {author} {\bibfnamefont {Georges}\ \bibnamefont
  {Aad}} \emph {et~al.} (\bibinfo {collaboration} {ATLAS}),\ }\bibfield
  {title} {\enquote {\bibinfo {title} {{Search for new phenomena with top quark
  pairs in final states with one lepton, jets, and missing transverse momentum
  in $pp$ collisions at $ \sqrt{s} $ = 13 TeV with the ATLAS detector}},}\
  }\href {\doibase 10.1007/JHEP04(2021)174} {\bibfield  {journal} {\bibinfo
  {journal} {JHEP}\ }\textbf {\bibinfo {volume} {04}},\ \bibinfo {pages} {174}
  (\bibinfo {year} {2021}{\natexlab{c}})},\ \Eprint
  {http://arxiv.org/abs/2012.03799} {arXiv:2012.03799 [hep-ex]} \BibitemShut
  {NoStop}%
\bibitem [{\citenamefont {Aad}\ \emph {et~al.}(2021{\natexlab{d}})\citenamefont
  {Aad} \emph {et~al.}}]{Aad:2021zyy}%
  \BibitemOpen
  \bibfield  {author} {\bibinfo {author} {\bibfnamefont {Georges}\ \bibnamefont
  {Aad}} \emph {et~al.} (\bibinfo {collaboration} {ATLAS}),\ }\bibfield
  {title} {\enquote {\bibinfo {title} {{Search for squarks and gluinos in final
  states with one isolated lepton, jets, and missing transverse momentum at
  $\sqrt{s}=13$ TeV with the ATLAS detector}},}\ }\href@noop {} {\  (\bibinfo
  {year} {2021}{\natexlab{d}})},\ \Eprint {http://arxiv.org/abs/2101.01629}
  {arXiv:2101.01629 [hep-ex]} \BibitemShut {NoStop}%
\bibitem [{\citenamefont {Martin}(2009)}]{Martin:2009ad}%
  \BibitemOpen
  \bibfield  {author} {\bibinfo {author} {\bibfnamefont {Stephen~P.}\
  \bibnamefont {Martin}},\ }\bibfield  {title} {\enquote {\bibinfo {title}
  {{Non-universal gaugino masses from non-singlet F-terms in non-minimal
  unified models}},}\ }\href {\doibase 10.1103/PhysRevD.79.095019} {\bibfield
  {journal} {\bibinfo  {journal} {Phys. Rev. D}\ }\textbf {\bibinfo {volume}
  {79}},\ \bibinfo {pages} {095019} (\bibinfo {year} {2009})},\ \Eprint
  {http://arxiv.org/abs/0903.3568} {arXiv:0903.3568 [hep-ph]} \BibitemShut
  {NoStop}%
\bibitem [{\citenamefont {Liu}\ \emph {et~al.}(2020{\natexlab{b}})\citenamefont
  {Liu}, \citenamefont {McGinnis}, \citenamefont {Wagner},\ and\ \citenamefont
  {Wang}}]{Liu:2020muv}%
  \BibitemOpen
  \bibfield  {author} {\bibinfo {author} {\bibfnamefont {Jia}\ \bibnamefont
  {Liu}}, \bibinfo {author} {\bibfnamefont {Navin}\ \bibnamefont {McGinnis}},
  \bibinfo {author} {\bibfnamefont {Carlos E.~M.}\ \bibnamefont {Wagner}}, \
  and\ \bibinfo {author} {\bibfnamefont {Xiao-Ping}\ \bibnamefont {Wang}},\
  }\bibfield  {title} {\enquote {\bibinfo {title} {{Searching for the
  Higgsino-Bino Sector at the LHC}},}\ }\href {\doibase
  10.1007/JHEP09(2020)073} {\bibfield  {journal} {\bibinfo  {journal} {JHEP}\
  }\textbf {\bibinfo {volume} {09}},\ \bibinfo {pages} {073} (\bibinfo {year}
  {2020}{\natexlab{b}})},\ \Eprint {http://arxiv.org/abs/2006.07389}
  {arXiv:2006.07389 [hep-ph]} \BibitemShut {NoStop}%
\bibitem [{\citenamefont {Liu}\ \emph {et~al.}(2020{\natexlab{c}})\citenamefont
  {Liu}, \citenamefont {McGinnis}, \citenamefont {Wagner},\ and\ \citenamefont
  {Wang}}]{Liu:2020ctf}%
  \BibitemOpen
  \bibfield  {author} {\bibinfo {author} {\bibfnamefont {Jia}\ \bibnamefont
  {Liu}}, \bibinfo {author} {\bibfnamefont {Navin}\ \bibnamefont {McGinnis}},
  \bibinfo {author} {\bibfnamefont {Carlos E.~M.}\ \bibnamefont {Wagner}}, \
  and\ \bibinfo {author} {\bibfnamefont {Xiao-Ping}\ \bibnamefont {Wang}},\
  }\bibfield  {title} {\enquote {\bibinfo {title} {{The scale of superpartner
  masses and electroweakino searches at the high-luminosity LHC}},}\ }\href
  {\doibase 10.1007/JHEP12(2020)087} {\bibfield  {journal} {\bibinfo  {journal}
  {JHEP}\ }\textbf {\bibinfo {volume} {12}},\ \bibinfo {pages} {087} (\bibinfo
  {year} {2020}{\natexlab{c}})},\ \Eprint {http://arxiv.org/abs/2008.11847}
  {arXiv:2008.11847 [hep-ph]} \BibitemShut {NoStop}%
\bibitem [{\citenamefont {Belanger}\ \emph {et~al.}(2009)\citenamefont
  {Belanger}, \citenamefont {Boudjema}, \citenamefont {Pukhov},\ and\
  \citenamefont {Semenov}}]{Belanger:2008sj}%
  \BibitemOpen
  \bibfield  {author} {\bibinfo {author} {\bibfnamefont {G.}~\bibnamefont
  {Belanger}}, \bibinfo {author} {\bibfnamefont {F.}~\bibnamefont {Boudjema}},
  \bibinfo {author} {\bibfnamefont {A.}~\bibnamefont {Pukhov}}, \ and\ \bibinfo
  {author} {\bibfnamefont {A.}~\bibnamefont {Semenov}},\ }\bibfield  {title}
  {\enquote {\bibinfo {title} {{Dark matter direct detection rate in a generic
  model with micrOMEGAs 2.2}},}\ }\href {\doibase 10.1016/j.cpc.2008.11.019}
  {\bibfield  {journal} {\bibinfo  {journal} {Comput. Phys. Commun.}\ }\textbf
  {\bibinfo {volume} {180}},\ \bibinfo {pages} {747--767} (\bibinfo {year}
  {2009})},\ \Eprint {http://arxiv.org/abs/0803.2360} {arXiv:0803.2360
  [hep-ph]} \BibitemShut {NoStop}%
\bibitem [{\citenamefont {B\'elanger}\ \emph {et~al.}(2018)\citenamefont
  {B\'elanger}, \citenamefont {Boudjema}, \citenamefont {Goudelis},
  \citenamefont {Pukhov},\ and\ \citenamefont {Zaldivar}}]{Belanger:2018ccd}%
  \BibitemOpen
  \bibfield  {author} {\bibinfo {author} {\bibfnamefont {Genevi\`eve}\
  \bibnamefont {B\'elanger}}, \bibinfo {author} {\bibfnamefont {Fawzi}\
  \bibnamefont {Boudjema}}, \bibinfo {author} {\bibfnamefont {Andreas}\
  \bibnamefont {Goudelis}}, \bibinfo {author} {\bibfnamefont {Alexander}\
  \bibnamefont {Pukhov}}, \ and\ \bibinfo {author} {\bibfnamefont {Bryan}\
  \bibnamefont {Zaldivar}},\ }\bibfield  {title} {\enquote {\bibinfo {title}
  {{micrOMEGAs5.0 : Freeze-in}},}\ }\href {\doibase 10.1016/j.cpc.2018.04.027}
  {\bibfield  {journal} {\bibinfo  {journal} {Comput. Phys. Commun.}\ }\textbf
  {\bibinfo {volume} {231}},\ \bibinfo {pages} {173--186} (\bibinfo {year}
  {2018})},\ \Eprint {http://arxiv.org/abs/1801.03509} {arXiv:1801.03509
  [hep-ph]} \BibitemShut {NoStop}%
\bibitem [{\citenamefont {B\'elanger}\ \emph {et~al.}(2021)\citenamefont
  {B\'elanger}, \citenamefont {Mjallal},\ and\ \citenamefont
  {Pukhov}}]{Belanger:2020gnr}%
  \BibitemOpen
  \bibfield  {author} {\bibinfo {author} {\bibfnamefont {G.}~\bibnamefont
  {B\'elanger}}, \bibinfo {author} {\bibfnamefont {A.}~\bibnamefont {Mjallal}},
  \ and\ \bibinfo {author} {\bibfnamefont {A.}~\bibnamefont {Pukhov}},\
  }\bibfield  {title} {\enquote {\bibinfo {title} {{Recasting direct detection
  limits within micrOMEGAs and implication for non-standard Dark Matter
  scenarios}},}\ }\href {\doibase 10.1140/epjc/s10052-021-09012-z} {\bibfield
  {journal} {\bibinfo  {journal} {Eur. Phys. J. C}\ }\textbf {\bibinfo {volume}
  {81}},\ \bibinfo {pages} {239} (\bibinfo {year} {2021})},\ \Eprint
  {http://arxiv.org/abs/2003.08621} {arXiv:2003.08621 [hep-ph]} \BibitemShut
  {NoStop}%
\bibitem [{\citenamefont {Djouadi}\ \emph {et~al.}(2007)\citenamefont
  {Djouadi}, \citenamefont {Muhlleitner},\ and\ \citenamefont
  {Spira}}]{Djouadi:2006bz}%
  \BibitemOpen
  \bibfield  {author} {\bibinfo {author} {\bibfnamefont {A.}~\bibnamefont
  {Djouadi}}, \bibinfo {author} {\bibfnamefont {M.~M.}\ \bibnamefont
  {Muhlleitner}}, \ and\ \bibinfo {author} {\bibfnamefont {M.}~\bibnamefont
  {Spira}},\ }\bibfield  {title} {\enquote {\bibinfo {title} {{Decays of
  supersymmetric particles: The Program SUSY-HIT
  (SUspect-SdecaY-Hdecay-InTerface)}},}\ }\href@noop {} {\bibfield  {journal}
  {\bibinfo  {journal} {Acta Phys. Polon. B}\ }\textbf {\bibinfo {volume}
  {38}},\ \bibinfo {pages} {635--644} (\bibinfo {year} {2007})},\ \Eprint
  {http://arxiv.org/abs/hep-ph/0609292} {arXiv:hep-ph/0609292} \BibitemShut
  {NoStop}%
\bibitem [{\citenamefont {Dercks}\ \emph {et~al.}(2017)\citenamefont {Dercks},
  \citenamefont {Desai}, \citenamefont {Kim}, \citenamefont {Rolbiecki},
  \citenamefont {Tattersall},\ and\ \citenamefont {Weber}}]{Dercks:2016npn}%
  \BibitemOpen
  \bibfield  {author} {\bibinfo {author} {\bibfnamefont {Daniel}\ \bibnamefont
  {Dercks}}, \bibinfo {author} {\bibfnamefont {Nishita}\ \bibnamefont {Desai}},
  \bibinfo {author} {\bibfnamefont {Jong~Soo}\ \bibnamefont {Kim}}, \bibinfo
  {author} {\bibfnamefont {Krzysztof}\ \bibnamefont {Rolbiecki}}, \bibinfo
  {author} {\bibfnamefont {Jamie}\ \bibnamefont {Tattersall}}, \ and\ \bibinfo
  {author} {\bibfnamefont {Torsten}\ \bibnamefont {Weber}},\ }\bibfield
  {title} {\enquote {\bibinfo {title} {{CheckMATE 2: From the model to the
  limit}},}\ }\href {\doibase 10.1016/j.cpc.2017.08.021} {\bibfield  {journal}
  {\bibinfo  {journal} {Comput. Phys. Commun.}\ }\textbf {\bibinfo {volume}
  {221}},\ \bibinfo {pages} {383--418} (\bibinfo {year} {2017})},\ \Eprint
  {http://arxiv.org/abs/1611.09856} {arXiv:1611.09856 [hep-ph]} \BibitemShut
  {NoStop}%
\bibitem [{\citenamefont {Alwall}\ \emph {et~al.}(2014)\citenamefont {Alwall},
  \citenamefont {Frederix}, \citenamefont {Frixione}, \citenamefont {Hirschi},
  \citenamefont {Maltoni}, \citenamefont {Mattelaer}, \citenamefont {Shao},
  \citenamefont {Stelzer}, \citenamefont {Torrielli},\ and\ \citenamefont
  {Zaro}}]{Alwall:2014hca}%
  \BibitemOpen
  \bibfield  {author} {\bibinfo {author} {\bibfnamefont {J.}~\bibnamefont
  {Alwall}}, \bibinfo {author} {\bibfnamefont {R.}~\bibnamefont {Frederix}},
  \bibinfo {author} {\bibfnamefont {S.}~\bibnamefont {Frixione}}, \bibinfo
  {author} {\bibfnamefont {V.}~\bibnamefont {Hirschi}}, \bibinfo {author}
  {\bibfnamefont {F.}~\bibnamefont {Maltoni}}, \bibinfo {author} {\bibfnamefont
  {O.}~\bibnamefont {Mattelaer}}, \bibinfo {author} {\bibfnamefont {H.~S.}\
  \bibnamefont {Shao}}, \bibinfo {author} {\bibfnamefont {T.}~\bibnamefont
  {Stelzer}}, \bibinfo {author} {\bibfnamefont {P.}~\bibnamefont {Torrielli}},
  \ and\ \bibinfo {author} {\bibfnamefont {M.}~\bibnamefont {Zaro}},\
  }\bibfield  {title} {\enquote {\bibinfo {title} {{The automated computation
  of tree-level and next-to-leading order differential cross sections, and
  their matching to parton shower simulations}},}\ }\href {\doibase
  10.1007/JHEP07(2014)079} {\bibfield  {journal} {\bibinfo  {journal} {JHEP}\
  }\textbf {\bibinfo {volume} {07}},\ \bibinfo {pages} {079} (\bibinfo {year}
  {2014})},\ \Eprint {http://arxiv.org/abs/1405.0301} {arXiv:1405.0301
  [hep-ph]} \BibitemShut {NoStop}%
\bibitem [{\citenamefont {Sj\"ostrand}\ \emph {et~al.}(2015)\citenamefont
  {Sj\"ostrand}, \citenamefont {Ask}, \citenamefont {Christiansen},
  \citenamefont {Corke}, \citenamefont {Desai}, \citenamefont {Ilten},
  \citenamefont {Mrenna}, \citenamefont {Prestel}, \citenamefont {Rasmussen},\
  and\ \citenamefont {Skands}}]{Sjostrand:2014zea}%
  \BibitemOpen
  \bibfield  {author} {\bibinfo {author} {\bibfnamefont {Torbj\"orn}\
  \bibnamefont {Sj\"ostrand}}, \bibinfo {author} {\bibfnamefont {Stefan}\
  \bibnamefont {Ask}}, \bibinfo {author} {\bibfnamefont {Jesper~R.}\
  \bibnamefont {Christiansen}}, \bibinfo {author} {\bibfnamefont {Richard}\
  \bibnamefont {Corke}}, \bibinfo {author} {\bibfnamefont {Nishita}\
  \bibnamefont {Desai}}, \bibinfo {author} {\bibfnamefont {Philip}\
  \bibnamefont {Ilten}}, \bibinfo {author} {\bibfnamefont {Stephen}\
  \bibnamefont {Mrenna}}, \bibinfo {author} {\bibfnamefont {Stefan}\
  \bibnamefont {Prestel}}, \bibinfo {author} {\bibfnamefont {Christine~O.}\
  \bibnamefont {Rasmussen}}, \ and\ \bibinfo {author} {\bibfnamefont
  {Peter~Z.}\ \bibnamefont {Skands}},\ }\bibfield  {title} {\enquote {\bibinfo
  {title} {{An introduction to PYTHIA 8.2}},}\ }\href {\doibase
  10.1016/j.cpc.2015.01.024} {\bibfield  {journal} {\bibinfo  {journal}
  {Comput. Phys. Commun.}\ }\textbf {\bibinfo {volume} {191}},\ \bibinfo
  {pages} {159--177} (\bibinfo {year} {2015})},\ \Eprint
  {http://arxiv.org/abs/1410.3012} {arXiv:1410.3012 [hep-ph]} \BibitemShut
  {NoStop}%
\bibitem [{\citenamefont {de~Favereau}\ \emph {et~al.}(2014)\citenamefont
  {de~Favereau}, \citenamefont {Delaere}, \citenamefont {Demin}, \citenamefont
  {Giammanco}, \citenamefont {Lema\^\i{}tre}, \citenamefont {Mertens},\ and\
  \citenamefont {Selvaggi}}]{deFavereau:2013fsa}%
  \BibitemOpen
  \bibfield  {author} {\bibinfo {author} {\bibfnamefont {J.}~\bibnamefont
  {de~Favereau}}, \bibinfo {author} {\bibfnamefont {C.}~\bibnamefont
  {Delaere}}, \bibinfo {author} {\bibfnamefont {P.}~\bibnamefont {Demin}},
  \bibinfo {author} {\bibfnamefont {A.}~\bibnamefont {Giammanco}}, \bibinfo
  {author} {\bibfnamefont {V.}~\bibnamefont {Lema\^\i{}tre}}, \bibinfo {author}
  {\bibfnamefont {A.}~\bibnamefont {Mertens}}, \ and\ \bibinfo {author}
  {\bibfnamefont {M.}~\bibnamefont {Selvaggi}} (\bibinfo {collaboration}
  {DELPHES 3}),\ }\bibfield  {title} {\enquote {\bibinfo {title} {{DELPHES 3, A
  modular framework for fast simulation of a generic collider experiment}},}\
  }\href {\doibase 10.1007/JHEP02(2014)057} {\bibfield  {journal} {\bibinfo
  {journal} {JHEP}\ }\textbf {\bibinfo {volume} {02}},\ \bibinfo {pages} {057}
  (\bibinfo {year} {2014})},\ \Eprint {http://arxiv.org/abs/1307.6346}
  {arXiv:1307.6346 [hep-ex]} \BibitemShut {NoStop}%
\bibitem [{\citenamefont {Aad}\ \emph {et~al.}(2013{\natexlab{a}})\citenamefont
  {Aad} \emph {et~al.}}]{ATLAS:2013qzt}%
  \BibitemOpen
  \bibfield  {author} {\bibinfo {author} {\bibfnamefont {Georges}\ \bibnamefont
  {Aad}} \emph {et~al.} (\bibinfo {collaboration} {ATLAS}),\ }\bibfield
  {title} {\enquote {\bibinfo {title} {{Search for new phenomena in final
  states with large jet multiplicities and missing transverse momentum at
  $\sqrt{s}$=8 TeV proton-proton collisions using the ATLAS experiment}},}\
  }\href {\doibase 10.1007/JHEP10(2013)130} {\bibfield  {journal} {\bibinfo
  {journal} {JHEP}\ }\textbf {\bibinfo {volume} {10}},\ \bibinfo {pages} {130}
  (\bibinfo {year} {2013}{\natexlab{a}})},\ \bibinfo {note} {[Erratum: JHEP 01,
  109 (2014)]},\ \Eprint {http://arxiv.org/abs/1308.1841} {arXiv:1308.1841
  [hep-ex]} \BibitemShut {NoStop}%
\bibitem [{\citenamefont {Aad}\ \emph {et~al.}(2013{\natexlab{b}})\citenamefont
  {Aad} \emph {et~al.}}]{ATLAS:2013lcn}%
  \BibitemOpen
  \bibfield  {author} {\bibinfo {author} {\bibfnamefont {Georges}\ \bibnamefont
  {Aad}} \emph {et~al.} (\bibinfo {collaboration} {ATLAS}),\ }\bibfield
  {title} {\enquote {\bibinfo {title} {{Search for direct third-generation
  squark pair production in final states with missing transverse momentum and
  two $b$-jets in $\sqrt{s} =$ 8 TeV $pp$ collisions with the ATLAS
  detector}},}\ }\href {\doibase 10.1007/JHEP10(2013)189} {\bibfield  {journal}
  {\bibinfo  {journal} {JHEP}\ }\textbf {\bibinfo {volume} {10}},\ \bibinfo
  {pages} {189} (\bibinfo {year} {2013}{\natexlab{b}})},\ \Eprint
  {http://arxiv.org/abs/1308.2631} {arXiv:1308.2631 [hep-ex]} \BibitemShut
  {NoStop}%
\bibitem [{\citenamefont {Aad}\ \emph {et~al.}(2014{\natexlab{b}})\citenamefont
  {Aad} \emph {et~al.}}]{ATLAS:2014hzd}%
  \BibitemOpen
  \bibfield  {author} {\bibinfo {author} {\bibfnamefont {Georges}\ \bibnamefont
  {Aad}} \emph {et~al.} (\bibinfo {collaboration} {ATLAS}),\ }\bibfield
  {title} {\enquote {\bibinfo {title} {{Search for Invisible Decays of a Higgs
  Boson Produced in Association with a Z Boson in ATLAS}},}\ }\href {\doibase
  10.1103/PhysRevLett.112.201802} {\bibfield  {journal} {\bibinfo  {journal}
  {Phys. Rev. Lett.}\ }\textbf {\bibinfo {volume} {112}},\ \bibinfo {pages}
  {201802} (\bibinfo {year} {2014}{\natexlab{b}})},\ \Eprint
  {http://arxiv.org/abs/1402.3244} {arXiv:1402.3244 [hep-ex]} \BibitemShut
  {NoStop}%
\bibitem [{\citenamefont {Aad}\ \emph {et~al.}(2014{\natexlab{c}})\citenamefont
  {Aad} \emph {et~al.}}]{ATLAS:2014ikz}%
  \BibitemOpen
  \bibfield  {author} {\bibinfo {author} {\bibfnamefont {Georges}\ \bibnamefont
  {Aad}} \emph {et~al.} (\bibinfo {collaboration} {ATLAS}),\ }\bibfield
  {title} {\enquote {\bibinfo {title} {{Search for direct production of
  charginos and neutralinos in events with three leptons and missing transverse
  momentum in $\sqrt{s} =$ 8TeV $pp$ collisions with the ATLAS detector}},}\
  }\href {\doibase 10.1007/JHEP04(2014)169} {\bibfield  {journal} {\bibinfo
  {journal} {JHEP}\ }\textbf {\bibinfo {volume} {04}},\ \bibinfo {pages} {169}
  (\bibinfo {year} {2014}{\natexlab{c}})},\ \Eprint
  {http://arxiv.org/abs/1402.7029} {arXiv:1402.7029 [hep-ex]} \BibitemShut
  {NoStop}%
\bibitem [{\citenamefont {Aad}\ \emph {et~al.}(2014{\natexlab{d}})\citenamefont
  {Aad} \emph {et~al.}}]{ATLAS:2014gmw}%
  \BibitemOpen
  \bibfield  {author} {\bibinfo {author} {\bibfnamefont {Georges}\ \bibnamefont
  {Aad}} \emph {et~al.} (\bibinfo {collaboration} {ATLAS}),\ }\bibfield
  {title} {\enquote {\bibinfo {title} {{Search for direct top-squark pair
  production in final states with two leptons in pp collisions at $\sqrt{s} =$
  8TeV with the ATLAS detector}},}\ }\href {\doibase 10.1007/JHEP06(2014)124}
  {\bibfield  {journal} {\bibinfo  {journal} {JHEP}\ }\textbf {\bibinfo
  {volume} {06}},\ \bibinfo {pages} {124} (\bibinfo {year}
  {2014}{\natexlab{d}})},\ \Eprint {http://arxiv.org/abs/1403.4853}
  {arXiv:1403.4853 [hep-ex]} \BibitemShut {NoStop}%
\bibitem [{\citenamefont {Aad}\ \emph {et~al.}(2014{\natexlab{e}})\citenamefont
  {Aad} \emph {et~al.}}]{ATLAS:2014bgg}%
  \BibitemOpen
  \bibfield  {author} {\bibinfo {author} {\bibfnamefont {Georges}\ \bibnamefont
  {Aad}} \emph {et~al.} (\bibinfo {collaboration} {ATLAS}),\ }\bibfield
  {title} {\enquote {\bibinfo {title} {{Search for direct top squark pair
  production in events with a Z boson, b-jets and missing transverse momentum
  in sqrt(s)=8 TeV pp collisions with the ATLAS detector}},}\ }\href {\doibase
  10.1140/epjc/s10052-014-2883-6} {\bibfield  {journal} {\bibinfo  {journal}
  {Eur. Phys. J. C}\ }\textbf {\bibinfo {volume} {74}},\ \bibinfo {pages}
  {2883} (\bibinfo {year} {2014}{\natexlab{e}})},\ \Eprint
  {http://arxiv.org/abs/1403.5222} {arXiv:1403.5222 [hep-ex]} \BibitemShut
  {NoStop}%
\bibitem [{\citenamefont {Aad}\ \emph {et~al.}(2014{\natexlab{f}})\citenamefont
  {Aad} \emph {et~al.}}]{ATLAS:2014zve}%
  \BibitemOpen
  \bibfield  {author} {\bibinfo {author} {\bibfnamefont {Georges}\ \bibnamefont
  {Aad}} \emph {et~al.} (\bibinfo {collaboration} {ATLAS}),\ }\bibfield
  {title} {\enquote {\bibinfo {title} {{Search for direct production of
  charginos, neutralinos and sleptons in final states with two leptons and
  missing transverse momentum in $pp$ collisions at $\sqrt{s} =$ 8 TeV with the
  ATLAS detector}},}\ }\href {\doibase 10.1007/JHEP05(2014)071} {\bibfield
  {journal} {\bibinfo  {journal} {JHEP}\ }\textbf {\bibinfo {volume} {05}},\
  \bibinfo {pages} {071} (\bibinfo {year} {2014}{\natexlab{f}})},\ \Eprint
  {http://arxiv.org/abs/1403.5294} {arXiv:1403.5294 [hep-ex]} \BibitemShut
  {NoStop}%
\bibitem [{\citenamefont {Aad}\ \emph {et~al.}(2014{\natexlab{g}})\citenamefont
  {Aad} \emph {et~al.}}]{ATLAS:2014kpx}%
  \BibitemOpen
  \bibfield  {author} {\bibinfo {author} {\bibfnamefont {Georges}\ \bibnamefont
  {Aad}} \emph {et~al.} (\bibinfo {collaboration} {ATLAS}),\ }\bibfield
  {title} {\enquote {\bibinfo {title} {{Search for supersymmetry at
  $\sqrt{s}$=8 TeV in final states with jets and two same-sign leptons or three
  leptons with the ATLAS detector}},}\ }\href {\doibase
  10.1007/JHEP06(2014)035} {\bibfield  {journal} {\bibinfo  {journal} {JHEP}\
  }\textbf {\bibinfo {volume} {06}},\ \bibinfo {pages} {035} (\bibinfo {year}
  {2014}{\natexlab{g}})},\ \Eprint {http://arxiv.org/abs/1404.2500}
  {arXiv:1404.2500 [hep-ex]} \BibitemShut {NoStop}%
\bibitem [{\citenamefont {Aad}\ \emph {et~al.}(2014{\natexlab{h}})\citenamefont
  {Aad} \emph {et~al.}}]{ATLAS:2014jxt}%
  \BibitemOpen
  \bibfield  {author} {\bibinfo {author} {\bibfnamefont {Georges}\ \bibnamefont
  {Aad}} \emph {et~al.} (\bibinfo {collaboration} {ATLAS}),\ }\bibfield
  {title} {\enquote {\bibinfo {title} {{Search for squarks and gluinos with the
  ATLAS detector in final states with jets and missing transverse momentum
  using $\sqrt{s}=8$ TeV proton--proton collision data}},}\ }\href {\doibase
  10.1007/JHEP09(2014)176} {\bibfield  {journal} {\bibinfo  {journal} {JHEP}\
  }\textbf {\bibinfo {volume} {09}},\ \bibinfo {pages} {176} (\bibinfo {year}
  {2014}{\natexlab{h}})},\ \Eprint {http://arxiv.org/abs/1405.7875}
  {arXiv:1405.7875 [hep-ex]} \BibitemShut {NoStop}%
\bibitem [{\citenamefont {Aad}\ \emph {et~al.}(2014{\natexlab{i}})\citenamefont
  {Aad} \emph {et~al.}}]{ATLAS:2014mmo}%
  \BibitemOpen
  \bibfield  {author} {\bibinfo {author} {\bibfnamefont {Georges}\ \bibnamefont
  {Aad}} \emph {et~al.} (\bibinfo {collaboration} {ATLAS}),\ }\bibfield
  {title} {\enquote {\bibinfo {title} {{Search for top squark pair production
  in final states with one isolated lepton, jets, and missing transverse
  momentum in $\sqrt s =$8 TeV $pp$ collisions with the ATLAS detector}},}\
  }\href {\doibase 10.1007/JHEP11(2014)118} {\bibfield  {journal} {\bibinfo
  {journal} {JHEP}\ }\textbf {\bibinfo {volume} {11}},\ \bibinfo {pages} {118}
  (\bibinfo {year} {2014}{\natexlab{i}})},\ \Eprint
  {http://arxiv.org/abs/1407.0583} {arXiv:1407.0583 [hep-ex]} \BibitemShut
  {NoStop}%
\bibitem [{\citenamefont {Aad}\ \emph {et~al.}(2014{\natexlab{j}})\citenamefont
  {Aad} \emph {et~al.}}]{ATLAS:2014hqe}%
  \BibitemOpen
  \bibfield  {author} {\bibinfo {author} {\bibfnamefont {Georges}\ \bibnamefont
  {Aad}} \emph {et~al.} (\bibinfo {collaboration} {ATLAS}),\ }\bibfield
  {title} {\enquote {\bibinfo {title} {{Search for pair-produced
  third-generation squarks decaying via charm quarks or in compressed
  supersymmetric scenarios in $pp$ collisions at $\sqrt{s}=8~$TeV with the
  ATLAS detector}},}\ }\href {\doibase 10.1103/PhysRevD.90.052008} {\bibfield
  {journal} {\bibinfo  {journal} {Phys. Rev. D}\ }\textbf {\bibinfo {volume}
  {90}},\ \bibinfo {pages} {052008} (\bibinfo {year} {2014}{\natexlab{j}})},\
  \Eprint {http://arxiv.org/abs/1407.0608} {arXiv:1407.0608 [hep-ex]}
  \BibitemShut {NoStop}%
\bibitem [{\citenamefont {Aad}\ \emph {et~al.}(2015{\natexlab{a}})\citenamefont
  {Aad} \emph {et~al.}}]{ATLAS:2014kci}%
  \BibitemOpen
  \bibfield  {author} {\bibinfo {author} {\bibfnamefont {Georges}\ \bibnamefont
  {Aad}} \emph {et~al.} (\bibinfo {collaboration} {ATLAS}),\ }\bibfield
  {title} {\enquote {\bibinfo {title} {{Search for new phenomena in events with
  a photon and missing transverse momentum in $pp$ collisions at $\sqrt{s}=8$
  TeV with the ATLAS detector}},}\ }\href {\doibase 10.1103/PhysRevD.91.012008}
  {\bibfield  {journal} {\bibinfo  {journal} {Phys. Rev. D}\ }\textbf {\bibinfo
  {volume} {91}},\ \bibinfo {pages} {012008} (\bibinfo {year}
  {2015}{\natexlab{a}})},\ \bibinfo {note} {[Erratum: Phys.Rev.D 92, 059903
  (2015)]},\ \Eprint {http://arxiv.org/abs/1411.1559} {arXiv:1411.1559
  [hep-ex]} \BibitemShut {NoStop}%
\bibitem [{\citenamefont {Aad}\ \emph {et~al.}(2015{\natexlab{b}})\citenamefont
  {Aad} \emph {et~al.}}]{ATLAS:2015tfi}%
  \BibitemOpen
  \bibfield  {author} {\bibinfo {author} {\bibfnamefont {Georges}\ \bibnamefont
  {Aad}} \emph {et~al.} (\bibinfo {collaboration} {ATLAS}),\ }\bibfield
  {title} {\enquote {\bibinfo {title} {{Search for direct pair production of a
  chargino and a neutralino decaying to the 125 GeV Higgs boson in $\sqrt{s} =
  8$ TeV ${pp}$ collisions with the ATLAS detector}},}\ }\href {\doibase
  10.1140/epjc/s10052-015-3408-7} {\bibfield  {journal} {\bibinfo  {journal}
  {Eur. Phys. J. C}\ }\textbf {\bibinfo {volume} {75}},\ \bibinfo {pages} {208}
  (\bibinfo {year} {2015}{\natexlab{b}})},\ \Eprint
  {http://arxiv.org/abs/1501.07110} {arXiv:1501.07110 [hep-ex]} \BibitemShut
  {NoStop}%
\bibitem [{\citenamefont {Aad}\ \emph {et~al.}(2015{\natexlab{c}})\citenamefont
  {Aad} \emph {et~al.}}]{ATLAS:2015qlt}%
  \BibitemOpen
  \bibfield  {author} {\bibinfo {author} {\bibfnamefont {Georges}\ \bibnamefont
  {Aad}} \emph {et~al.} (\bibinfo {collaboration} {ATLAS}),\ }\bibfield
  {title} {\enquote {\bibinfo {title} {{Search for new phenomena in final
  states with an energetic jet and large missing transverse momentum in pp
  collisions at $\sqrt{s}=$8 TeV with the ATLAS detector}},}\ }\href {\doibase
  10.1140/epjc/s10052-015-3517-3} {\bibfield  {journal} {\bibinfo  {journal}
  {Eur. Phys. J. C}\ }\textbf {\bibinfo {volume} {75}},\ \bibinfo {pages} {299}
  (\bibinfo {year} {2015}{\natexlab{c}})},\ \bibinfo {note} {[Erratum:
  Eur.Phys.J.C 75, 408 (2015)]},\ \Eprint {http://arxiv.org/abs/1502.01518}
  {arXiv:1502.01518 [hep-ex]} \BibitemShut {NoStop}%
\bibitem [{\citenamefont {Aad}\ \emph {et~al.}(2015{\natexlab{d}})\citenamefont
  {Aad} \emph {et~al.}}]{ATLAS:2015xmt}%
  \BibitemOpen
  \bibfield  {author} {\bibinfo {author} {\bibfnamefont {Georges}\ \bibnamefont
  {Aad}} \emph {et~al.} (\bibinfo {collaboration} {ATLAS}),\ }\bibfield
  {title} {\enquote {\bibinfo {title} {{Search for massive supersymmetric
  particles decaying to many jets using the ATLAS detector in $pp$ collisions
  at $\sqrt{s} = 8$ TeV}},}\ }\href {\doibase 10.1103/PhysRevD.91.112016}
  {\bibfield  {journal} {\bibinfo  {journal} {Phys. Rev. D}\ }\textbf {\bibinfo
  {volume} {91}},\ \bibinfo {pages} {112016} (\bibinfo {year}
  {2015}{\natexlab{d}})},\ \bibinfo {note} {[Erratum: Phys.Rev.D 93, 039901
  (2016)]},\ \Eprint {http://arxiv.org/abs/1502.05686} {arXiv:1502.05686
  [hep-ex]} \BibitemShut {NoStop}%
\bibitem [{\citenamefont {Aad}\ \emph {et~al.}(2015{\natexlab{e}})\citenamefont
  {Aad} \emph {et~al.}}]{ATLAS:2015vch}%
  \BibitemOpen
  \bibfield  {author} {\bibinfo {author} {\bibfnamefont {Georges}\ \bibnamefont
  {Aad}} \emph {et~al.} (\bibinfo {collaboration} {ATLAS}),\ }\bibfield
  {title} {\enquote {\bibinfo {title} {{Search for supersymmetry in events
  containing a same-flavour opposite-sign dilepton pair, jets, and large
  missing transverse momentum in $\sqrt{s}=8$ TeV pp collisions with the ATLAS
  detector}},}\ }\href {\doibase 10.1140/epjc/s10052-015-3518-2} {\bibfield
  {journal} {\bibinfo  {journal} {Eur. Phys. J. C}\ }\textbf {\bibinfo {volume}
  {75}},\ \bibinfo {pages} {318} (\bibinfo {year} {2015}{\natexlab{e}})},\
  \bibinfo {note} {[Erratum: Eur.Phys.J.C 75, 463 (2015)]},\ \Eprint
  {http://arxiv.org/abs/1503.03290} {arXiv:1503.03290 [hep-ex]} \BibitemShut
  {NoStop}%
\bibitem [{\citenamefont {Aad}\ \emph {et~al.}(2015{\natexlab{f}})\citenamefont
  {Aad} \emph {et~al.}}]{ATLAS:2015zwy}%
  \BibitemOpen
  \bibfield  {author} {\bibinfo {author} {\bibfnamefont {Georges}\ \bibnamefont
  {Aad}} \emph {et~al.} (\bibinfo {collaboration} {ATLAS}),\ }\bibfield
  {title} {\enquote {\bibinfo {title} {{ATLAS Run 1 searches for direct pair
  production of third-generation squarks at the Large Hadron Collider}},}\
  }\href {\doibase 10.1140/epjc/s10052-015-3726-9} {\bibfield  {journal}
  {\bibinfo  {journal} {Eur. Phys. J. C}\ }\textbf {\bibinfo {volume} {75}},\
  \bibinfo {pages} {510} (\bibinfo {year} {2015}{\natexlab{f}})},\ \bibinfo
  {note} {[Erratum: Eur.Phys.J.C 76, 153 (2016)]},\ \Eprint
  {http://arxiv.org/abs/1506.08616} {arXiv:1506.08616 [hep-ex]} \BibitemShut
  {NoStop}%
\bibitem [{\citenamefont {Aad}\ \emph {et~al.}(2015{\natexlab{g}})\citenamefont
  {Aad} \emph {et~al.}}]{ATLAS:2015dcr}%
  \BibitemOpen
  \bibfield  {author} {\bibinfo {author} {\bibfnamefont {Georges}\ \bibnamefont
  {Aad}} \emph {et~al.} (\bibinfo {collaboration} {ATLAS}),\ }\bibfield
  {title} {\enquote {\bibinfo {title} {{Search for photonic signatures of
  gauge-mediated supersymmetry in 8 TeV pp collisions with the ATLAS
  detector}},}\ }\href {\doibase 10.1103/PhysRevD.92.072001} {\bibfield
  {journal} {\bibinfo  {journal} {Phys. Rev. D}\ }\textbf {\bibinfo {volume}
  {92}},\ \bibinfo {pages} {072001} (\bibinfo {year} {2015}{\natexlab{g}})},\
  \Eprint {http://arxiv.org/abs/1507.05493} {arXiv:1507.05493 [hep-ex]}
  \BibitemShut {NoStop}%
\bibitem [{\citenamefont {Collaboration}(2012)}]{ATLAS-CONF-2012-104}%
  \BibitemOpen
  \bibfield  {author} {\bibinfo {author} {\bibfnamefont {ATLAS}\ \bibnamefont
  {Collaboration}},\ }\href {https://cds.cern.ch/record/1472673} {\emph
  {\bibinfo {title} {{Search for supersymmetry at $\sqrt{s} = 8$ TeV in final
  states with jets, missing transverse momentum and one isolated lepton}}}},\
  \bibinfo {type} {Tech. Rep.}\ \bibinfo {number} {ATLAS-CONF-2012-104}\
  (\bibinfo  {institution} {CERN},\ \bibinfo {address} {Geneva},\ \bibinfo
  {year} {2012})\BibitemShut {NoStop}%
\bibitem [{\citenamefont
  {Collaboration}(2013{\natexlab{a}})}]{ATLAS-CONF-2013-024}%
  \BibitemOpen
  \bibfield  {author} {\bibinfo {author} {\bibfnamefont {ATLAS}\ \bibnamefont
  {Collaboration}},\ }\href {https://cds.cern.ch/record/1525880} {\emph
  {\bibinfo {title} {{Search for direct production of the top squark in the
  all-hadronic ttbar + etmiss final state in 21 fb-1 of p-pcollisions at
  sqrt(s)=8 TeV with the ATLAS detector}}}},\ \bibinfo {type} {Tech. Rep.}\
  \bibinfo {number} {ATLAS-CONF-2013-024}\ (\bibinfo  {institution} {CERN},\
  \bibinfo {address} {Geneva},\ \bibinfo {year} {2013})\BibitemShut {NoStop}%
\bibitem [{\citenamefont
  {Collaboration}(2013{\natexlab{b}})}]{ATLAS-CONF-2013-049}%
  \BibitemOpen
  \bibfield  {author} {\bibinfo {author} {\bibfnamefont {ATLAS}\ \bibnamefont
  {Collaboration}},\ }\href {https://cds.cern.ch/record/1547565} {\emph
  {\bibinfo {title} {{Search for direct-slepton and direct-chargino production
  in final states with two opposite-sign leptons, missing transverse momentum
  and no jets in 20/fb of pp collisions at sqrt(s) = 8 TeV with the ATLAS
  detector}}}},\ \bibinfo {type} {Tech. Rep.}\ \bibinfo {number}
  {ATLAS-CONF-2013-049}\ (\bibinfo  {institution} {CERN},\ \bibinfo {address}
  {Geneva},\ \bibinfo {year} {2013})\BibitemShut {NoStop}%
\bibitem [{\citenamefont
  {Collaboration}(2013{\natexlab{c}})}]{ATLAS-CONF-2013-061}%
  \BibitemOpen
  \bibfield  {author} {\bibinfo {author} {\bibfnamefont {ATLAS}\ \bibnamefont
  {Collaboration}},\ }\href {https://cds.cern.ch/record/1557778} {\emph
  {\bibinfo {title} {{Search for strong production of supersymmetric particles
  in final states with missing transverse momentum and at least three b-jets
  using 20.1 fb-1 of pp collisions at sqrt(s) = 8 TeV with the ATLAS
  Detector.}}}},\ \bibinfo {type} {Tech. Rep.}\ \bibinfo {number}
  {ATLAS-CONF-2013-061}\ (\bibinfo  {institution} {CERN},\ \bibinfo {address}
  {Geneva},\ \bibinfo {year} {2013})\BibitemShut {NoStop}%
\bibitem [{\citenamefont
  {Collaboration}(2013{\natexlab{d}})}]{ATLAS-CONF-2013-089}%
  \BibitemOpen
  \bibfield  {author} {\bibinfo {author} {\bibfnamefont {ATLAS}\ \bibnamefont
  {Collaboration}},\ }\href {https://cds.cern.ch/record/1595272} {\emph
  {\bibinfo {title} {{Search for strongly produced supersymmetric particles in
  decays with two leptons at $\sqrt{s}$ = 8 TeV}}}},\ \bibinfo {type} {Tech.
  Rep.}\ \bibinfo {number} {ATLAS-CONF-2013-089}\ (\bibinfo  {institution}
  {CERN},\ \bibinfo {address} {Geneva},\ \bibinfo {year} {2013})\BibitemShut
  {NoStop}%
\bibitem [{\citenamefont
  {Collaboration}(2015{\natexlab{a}})}]{ATLAS-CONF-2015-004}%
  \BibitemOpen
  \bibfield  {author} {\bibinfo {author} {\bibfnamefont {ATLAS}\ \bibnamefont
  {Collaboration}},\ }\href {https://cds.cern.ch/record/2002121} {\emph
  {\bibinfo {title} {{Search for an Invisibly Decaying Higgs Boson Produced via
  Vector Boson Fusion in $pp$ Collisions at $\sqrt{s}=8$ TeV using the ATLAS
  Detector at the LHC}}}},\ \bibinfo {type} {Tech. Rep.}\ \bibinfo {number}
  {ATLAS-CONF-2015-004}\ (\bibinfo  {institution} {CERN},\ \bibinfo {address}
  {Geneva},\ \bibinfo {year} {2015})\BibitemShut {NoStop}%
\bibitem [{\citenamefont {Chatrchyan}\ \emph {et~al.}(2013)\citenamefont
  {Chatrchyan} \emph {et~al.}}]{CMS:2013cgf}%
  \BibitemOpen
  \bibfield  {author} {\bibinfo {author} {\bibfnamefont {Serguei}\ \bibnamefont
  {Chatrchyan}} \emph {et~al.} (\bibinfo {collaboration} {CMS}),\ }\bibfield
  {title} {\enquote {\bibinfo {title} {{Search for Supersymmetry in Hadronic
  Final States with Missing Transverse Energy Using the Variables $\alpha_T$
  and b-Quark Multiplicity in $pp$ collisions at $\sqrt{s}= 8$ TeV}},}\ }\href
  {\doibase 10.1140/epjc/s10052-013-2568-6} {\bibfield  {journal} {\bibinfo
  {journal} {Eur. Phys. J. C}\ }\textbf {\bibinfo {volume} {73}},\ \bibinfo
  {pages} {2568} (\bibinfo {year} {2013})},\ \Eprint
  {http://arxiv.org/abs/1303.2985} {arXiv:1303.2985 [hep-ex]} \BibitemShut
  {NoStop}%
\bibitem [{\citenamefont {Khachatryan}\ \emph
  {et~al.}(2015{\natexlab{a}})\citenamefont {Khachatryan} \emph
  {et~al.}}]{CMS:2014jvv}%
  \BibitemOpen
  \bibfield  {author} {\bibinfo {author} {\bibfnamefont {Vardan}\ \bibnamefont
  {Khachatryan}} \emph {et~al.} (\bibinfo {collaboration} {CMS}),\ }\bibfield
  {title} {\enquote {\bibinfo {title} {{Search for dark matter, extra
  dimensions, and unparticles in monojet events in proton\textendash{}proton
  collisions at $\sqrt{s} = 8$ TeV}},}\ }\href {\doibase
  10.1140/epjc/s10052-015-3451-4} {\bibfield  {journal} {\bibinfo  {journal}
  {Eur. Phys. J. C}\ }\textbf {\bibinfo {volume} {75}},\ \bibinfo {pages} {235}
  (\bibinfo {year} {2015}{\natexlab{a}})},\ \Eprint
  {http://arxiv.org/abs/1408.3583} {arXiv:1408.3583 [hep-ex]} \BibitemShut
  {NoStop}%
\bibitem [{\citenamefont {Khachatryan}\ \emph
  {et~al.}(2015{\natexlab{b}})\citenamefont {Khachatryan} \emph
  {et~al.}}]{CMS:2015kjt}%
  \BibitemOpen
  \bibfield  {author} {\bibinfo {author} {\bibfnamefont {Vardan}\ \bibnamefont
  {Khachatryan}} \emph {et~al.} (\bibinfo {collaboration} {CMS}),\ }\bibfield
  {title} {\enquote {\bibinfo {title} {{Search for Physics Beyond the Standard
  Model in Events with Two Leptons, Jets, and Missing Transverse Momentum in pp
  Collisions at sqrt(s) = 8 TeV}},}\ }\href {\doibase 10.1007/JHEP04(2015)124}
  {\bibfield  {journal} {\bibinfo  {journal} {JHEP}\ }\textbf {\bibinfo
  {volume} {04}},\ \bibinfo {pages} {124} (\bibinfo {year}
  {2015}{\natexlab{b}})},\ \Eprint {http://arxiv.org/abs/1502.06031}
  {arXiv:1502.06031 [hep-ex]} \BibitemShut {NoStop}%
\bibitem [{\citenamefont {Khachatryan}\ \emph
  {et~al.}(2015{\natexlab{c}})\citenamefont {Khachatryan} \emph
  {et~al.}}]{CMS:2015zwg}%
  \BibitemOpen
  \bibfield  {author} {\bibinfo {author} {\bibfnamefont {Vardan}\ \bibnamefont
  {Khachatryan}} \emph {et~al.} (\bibinfo {collaboration} {CMS}),\ }\bibfield
  {title} {\enquote {\bibinfo {title} {{Search for the production of dark
  matter in association with top-quark pairs in the single-lepton final state
  in proton-proton collisions at sqrt(s) = 8 TeV}},}\ }\href {\doibase
  10.1007/JHEP06(2015)121} {\bibfield  {journal} {\bibinfo  {journal} {JHEP}\
  }\textbf {\bibinfo {volume} {06}},\ \bibinfo {pages} {121} (\bibinfo {year}
  {2015}{\natexlab{c}})},\ \Eprint {http://arxiv.org/abs/1504.03198}
  {arXiv:1504.03198 [hep-ex]} \BibitemShut {NoStop}%
\bibitem [{\citenamefont {Khachatryan}\ \emph {et~al.}(2016)\citenamefont
  {Khachatryan} \emph {et~al.}}]{CMS:2016aro}%
  \BibitemOpen
  \bibfield  {author} {\bibinfo {author} {\bibfnamefont {Vardan}\ \bibnamefont
  {Khachatryan}} \emph {et~al.} (\bibinfo {collaboration} {CMS}),\ }\bibfield
  {title} {\enquote {\bibinfo {title} {{Search for heavy Majorana neutrinos in
  e$^{±}$e$^{±}$+ jets and e$^{±}$ $\mu^{±}$+ jets events in proton-proton
  collisions at $ \sqrt{s}=8 $ TeV}},}\ }\href {\doibase
  10.1007/JHEP04(2016)169} {\bibfield  {journal} {\bibinfo  {journal} {JHEP}\
  }\textbf {\bibinfo {volume} {04}},\ \bibinfo {pages} {169} (\bibinfo {year}
  {2016})},\ \Eprint {http://arxiv.org/abs/1603.02248} {arXiv:1603.02248
  [hep-ex]} \BibitemShut {NoStop}%
\bibitem [{\citenamefont {Aaboud}\ \emph {et~al.}(2016)\citenamefont {Aaboud}
  \emph {et~al.}}]{ATLAS:2016zxj}%
  \BibitemOpen
  \bibfield  {author} {\bibinfo {author} {\bibfnamefont {Morad}\ \bibnamefont
  {Aaboud}} \emph {et~al.} (\bibinfo {collaboration} {ATLAS}),\ }\bibfield
  {title} {\enquote {\bibinfo {title} {{Search for new phenomena in events with
  a photon and missing transverse momentum in $pp$ collisions at $\sqrt{s}=13$
  TeV with the ATLAS detector}},}\ }\href {\doibase 10.1007/JHEP06(2016)059}
  {\bibfield  {journal} {\bibinfo  {journal} {JHEP}\ }\textbf {\bibinfo
  {volume} {06}},\ \bibinfo {pages} {059} (\bibinfo {year} {2016})},\ \Eprint
  {http://arxiv.org/abs/1604.01306} {arXiv:1604.01306 [hep-ex]} \BibitemShut
  {NoStop}%
\bibitem [{\citenamefont {Aad}\ \emph {et~al.}(2016)\citenamefont {Aad} \emph
  {et~al.}}]{ATLAS:2016gty}%
  \BibitemOpen
  \bibfield  {author} {\bibinfo {author} {\bibfnamefont {Georges}\ \bibnamefont
  {Aad}} \emph {et~al.} (\bibinfo {collaboration} {ATLAS}),\ }\bibfield
  {title} {\enquote {\bibinfo {title} {{Search for pair production of gluinos
  decaying via stop and sbottom in events with $b$-jets and large missing
  transverse momentum in $pp$ collisions at $\sqrt{s} = 13$ TeV with the ATLAS
  detector}},}\ }\href {\doibase 10.1103/PhysRevD.94.032003} {\bibfield
  {journal} {\bibinfo  {journal} {Phys. Rev. D}\ }\textbf {\bibinfo {volume}
  {94}},\ \bibinfo {pages} {032003} (\bibinfo {year} {2016})},\ \Eprint
  {http://arxiv.org/abs/1605.09318} {arXiv:1605.09318 [hep-ex]} \BibitemShut
  {NoStop}%
\bibitem [{\citenamefont {Aaboud}\ \emph
  {et~al.}(2017{\natexlab{a}})\citenamefont {Aaboud} \emph
  {et~al.}}]{ATLAS:2016wgc}%
  \BibitemOpen
  \bibfield  {author} {\bibinfo {author} {\bibfnamefont {Morad}\ \bibnamefont
  {Aaboud}} \emph {et~al.} (\bibinfo {collaboration} {ATLAS}),\ }\bibfield
  {title} {\enquote {\bibinfo {title} {{Measurement of the $t\bar{t}Z$ and
  $t\bar{t}W$ production cross sections in multilepton final states using 3.2
  fb$^{-1}$ of $pp$ collisions at $\sqrt{s}$ = 13 TeV with the ATLAS
  detector}},}\ }\href {\doibase 10.1140/epjc/s10052-016-4574-y} {\bibfield
  {journal} {\bibinfo  {journal} {Eur. Phys. J. C}\ }\textbf {\bibinfo {volume}
  {77}},\ \bibinfo {pages} {40} (\bibinfo {year} {2017}{\natexlab{a}})},\
  \Eprint {http://arxiv.org/abs/1609.01599} {arXiv:1609.01599 [hep-ex]}
  \BibitemShut {NoStop}%
\bibitem [{\citenamefont {Aaboud}\ \emph
  {et~al.}(2017{\natexlab{b}})\citenamefont {Aaboud} \emph
  {et~al.}}]{ATLAS:2017nga}%
  \BibitemOpen
  \bibfield  {author} {\bibinfo {author} {\bibfnamefont {Morad}\ \bibnamefont
  {Aaboud}} \emph {et~al.} (\bibinfo {collaboration} {ATLAS}),\ }\bibfield
  {title} {\enquote {\bibinfo {title} {{Search for dark matter at $\sqrt{s}=13$
  TeV in final states containing an energetic photon and large missing
  transverse momentum with the ATLAS detector}},}\ }\href {\doibase
  10.1140/epjc/s10052-017-4965-8} {\bibfield  {journal} {\bibinfo  {journal}
  {Eur. Phys. J. C}\ }\textbf {\bibinfo {volume} {77}},\ \bibinfo {pages} {393}
  (\bibinfo {year} {2017}{\natexlab{b}})},\ \Eprint
  {http://arxiv.org/abs/1704.03848} {arXiv:1704.03848 [hep-ex]} \BibitemShut
  {NoStop}%
\bibitem [{\citenamefont {Aaboud}\ \emph
  {et~al.}(2017{\natexlab{c}})\citenamefont {Aaboud} \emph
  {et~al.}}]{ATLAS:2017tmw}%
  \BibitemOpen
  \bibfield  {author} {\bibinfo {author} {\bibfnamefont {Morad}\ \bibnamefont
  {Aaboud}} \emph {et~al.} (\bibinfo {collaboration} {ATLAS}),\ }\bibfield
  {title} {\enquote {\bibinfo {title} {{Search for supersymmetry in final
  states with two same-sign or three leptons and jets using 36 fb$^{-1}$ of
  $\sqrt{s}=13$ TeV $pp$ collision data with the ATLAS detector}},}\ }\href
  {\doibase 10.1007/JHEP09(2017)084} {\bibfield  {journal} {\bibinfo  {journal}
  {JHEP}\ }\textbf {\bibinfo {volume} {09}},\ \bibinfo {pages} {084} (\bibinfo
  {year} {2017}{\natexlab{c}})},\ \bibinfo {note} {[Erratum: JHEP 08, 121
  (2019)]},\ \Eprint {http://arxiv.org/abs/1706.03731} {arXiv:1706.03731
  [hep-ex]} \BibitemShut {NoStop}%
\bibitem [{\citenamefont {Aaboud}\ \emph
  {et~al.}(2018{\natexlab{e}})\citenamefont {Aaboud} \emph
  {et~al.}}]{ATLAS:2017qwn}%
  \BibitemOpen
  \bibfield  {author} {\bibinfo {author} {\bibfnamefont {Morad}\ \bibnamefont
  {Aaboud}} \emph {et~al.} (\bibinfo {collaboration} {ATLAS}),\ }\bibfield
  {title} {\enquote {\bibinfo {title} {{Search for the direct production of
  charginos and neutralinos in final states with tau leptons in $\sqrt{s} = $
  13 TeV $pp$ collisions with the ATLAS detector}},}\ }\href {\doibase
  10.1140/epjc/s10052-018-5583-9} {\bibfield  {journal} {\bibinfo  {journal}
  {Eur. Phys. J. C}\ }\textbf {\bibinfo {volume} {78}},\ \bibinfo {pages} {154}
  (\bibinfo {year} {2018}{\natexlab{e}})},\ \Eprint
  {http://arxiv.org/abs/1708.07875} {arXiv:1708.07875 [hep-ex]} \BibitemShut
  {NoStop}%
\bibitem [{\citenamefont {Aaboud}\ \emph
  {et~al.}(2017{\natexlab{d}})\citenamefont {Aaboud} \emph
  {et~al.}}]{ATLAS:2017drc}%
  \BibitemOpen
  \bibfield  {author} {\bibinfo {author} {\bibfnamefont {Morad}\ \bibnamefont
  {Aaboud}} \emph {et~al.} (\bibinfo {collaboration} {ATLAS}),\ }\bibfield
  {title} {\enquote {\bibinfo {title} {{Search for a scalar partner of the top
  quark in the jets plus missing transverse momentum final state at
  $\sqrt{s}$=13 TeV with the ATLAS detector}},}\ }\href {\doibase
  10.1007/JHEP12(2017)085} {\bibfield  {journal} {\bibinfo  {journal} {JHEP}\
  }\textbf {\bibinfo {volume} {12}},\ \bibinfo {pages} {085} (\bibinfo {year}
  {2017}{\natexlab{d}})},\ \Eprint {http://arxiv.org/abs/1709.04183}
  {arXiv:1709.04183 [hep-ex]} \BibitemShut {NoStop}%
\bibitem [{\citenamefont {Aaboud}\ \emph
  {et~al.}(2018{\natexlab{f}})\citenamefont {Aaboud} \emph
  {et~al.}}]{ATLAS:2017vat}%
  \BibitemOpen
  \bibfield  {author} {\bibinfo {author} {\bibfnamefont {Morad}\ \bibnamefont
  {Aaboud}} \emph {et~al.} (\bibinfo {collaboration} {ATLAS}),\ }\bibfield
  {title} {\enquote {\bibinfo {title} {{Search for electroweak production of
  supersymmetric states in scenarios with compressed mass spectra at
  $\sqrt{s}=13$ TeV with the ATLAS detector}},}\ }\href {\doibase
  10.1103/PhysRevD.97.052010} {\bibfield  {journal} {\bibinfo  {journal} {Phys.
  Rev. D}\ }\textbf {\bibinfo {volume} {97}},\ \bibinfo {pages} {052010}
  (\bibinfo {year} {2018}{\natexlab{f}})},\ \Eprint
  {http://arxiv.org/abs/1712.08119} {arXiv:1712.08119 [hep-ex]} \BibitemShut
  {NoStop}%
\bibitem [{\citenamefont {Aaboud}\ \emph
  {et~al.}(2018{\natexlab{g}})\citenamefont {Aaboud} \emph
  {et~al.}}]{ATLAS:2017mjy}%
  \BibitemOpen
  \bibfield  {author} {\bibinfo {author} {\bibfnamefont {Morad}\ \bibnamefont
  {Aaboud}} \emph {et~al.} (\bibinfo {collaboration} {ATLAS}),\ }\bibfield
  {title} {\enquote {\bibinfo {title} {{Search for squarks and gluinos in final
  states with jets and missing transverse momentum using 36 fb$^{-1}$ of
  $\sqrt{s}=13$ TeV pp collision data with the ATLAS detector}},}\ }\href
  {\doibase 10.1103/PhysRevD.97.112001} {\bibfield  {journal} {\bibinfo
  {journal} {Phys. Rev. D}\ }\textbf {\bibinfo {volume} {97}},\ \bibinfo
  {pages} {112001} (\bibinfo {year} {2018}{\natexlab{g}})},\ \Eprint
  {http://arxiv.org/abs/1712.02332} {arXiv:1712.02332 [hep-ex]} \BibitemShut
  {NoStop}%
\bibitem [{\citenamefont {Aaboud}\ \emph
  {et~al.}(2018{\natexlab{h}})\citenamefont {Aaboud} \emph
  {et~al.}}]{ATLAS:2018nud}%
  \BibitemOpen
  \bibfield  {author} {\bibinfo {author} {\bibfnamefont {Morad}\ \bibnamefont
  {Aaboud}} \emph {et~al.} (\bibinfo {collaboration} {ATLAS}),\ }\bibfield
  {title} {\enquote {\bibinfo {title} {{Search for photonic signatures of
  gauge-mediated supersymmetry in 13 TeV $pp$ collisions with the ATLAS
  detector}},}\ }\href {\doibase 10.1103/PhysRevD.97.092006} {\bibfield
  {journal} {\bibinfo  {journal} {Phys. Rev. D}\ }\textbf {\bibinfo {volume}
  {97}},\ \bibinfo {pages} {092006} (\bibinfo {year} {2018}{\natexlab{h}})},\
  \Eprint {http://arxiv.org/abs/1802.03158} {arXiv:1802.03158 [hep-ex]}
  \BibitemShut {NoStop}%
\bibitem [{\citenamefont {Aaboud}\ \emph
  {et~al.}(2018{\natexlab{i}})\citenamefont {Aaboud} \emph
  {et~al.}}]{ATLAS:2018ojr}%
  \BibitemOpen
  \bibfield  {author} {\bibinfo {author} {\bibfnamefont {M.}~\bibnamefont
  {Aaboud}} \emph {et~al.} (\bibinfo {collaboration} {ATLAS}),\ }\bibfield
  {title} {\enquote {\bibinfo {title} {{Search for electroweak production of
  supersymmetric particles in final states with two or three leptons at
  $\sqrt{s}=13\,$TeV with the ATLAS detector}},}\ }\href {\doibase
  10.1140/epjc/s10052-018-6423-7} {\bibfield  {journal} {\bibinfo  {journal}
  {Eur. Phys. J. C}\ }\textbf {\bibinfo {volume} {78}},\ \bibinfo {pages} {995}
  (\bibinfo {year} {2018}{\natexlab{i}})},\ \Eprint
  {http://arxiv.org/abs/1803.02762} {arXiv:1803.02762 [hep-ex]} \BibitemShut
  {NoStop}%
\bibitem [{\citenamefont {Aaboud}\ \emph
  {et~al.}(2019{\natexlab{b}})\citenamefont {Aaboud} \emph
  {et~al.}}]{ATLAS:2018zdn}%
  \BibitemOpen
  \bibfield  {author} {\bibinfo {author} {\bibfnamefont {Morad}\ \bibnamefont
  {Aaboud}} \emph {et~al.} (\bibinfo {collaboration} {ATLAS}),\ }\bibfield
  {title} {\enquote {\bibinfo {title} {{A strategy for a general search for new
  phenomena using data-derived signal regions and its application within the
  ATLAS experiment}},}\ }\href {\doibase 10.1140/epjc/s10052-019-6540-y}
  {\bibfield  {journal} {\bibinfo  {journal} {Eur. Phys. J. C}\ }\textbf
  {\bibinfo {volume} {79}},\ \bibinfo {pages} {120} (\bibinfo {year}
  {2019}{\natexlab{b}})},\ \Eprint {http://arxiv.org/abs/1807.07447}
  {arXiv:1807.07447 [hep-ex]} \BibitemShut {NoStop}%
\bibitem [{\citenamefont {Aad}\ \emph {et~al.}(2019)\citenamefont {Aad} \emph
  {et~al.}}]{ATLAS:2019gdh}%
  \BibitemOpen
  \bibfield  {author} {\bibinfo {author} {\bibfnamefont {Georges}\ \bibnamefont
  {Aad}} \emph {et~al.} (\bibinfo {collaboration} {ATLAS}),\ }\bibfield
  {title} {\enquote {\bibinfo {title} {{Search for bottom-squark pair
  production with the ATLAS detector in final states containing Higgs bosons,
  $b$-jets and missing transverse momentum}},}\ }\href {\doibase
  10.1007/JHEP12(2019)060} {\bibfield  {journal} {\bibinfo  {journal} {JHEP}\
  }\textbf {\bibinfo {volume} {12}},\ \bibinfo {pages} {060} (\bibinfo {year}
  {2019})},\ \Eprint {http://arxiv.org/abs/1908.03122} {arXiv:1908.03122
  [hep-ex]} \BibitemShut {NoStop}%
\bibitem [{\citenamefont {Aad}\ \emph {et~al.}(2020{\natexlab{e}})\citenamefont
  {Aad} \emph {et~al.}}]{ATLAS:2019lff}%
  \BibitemOpen
  \bibfield  {author} {\bibinfo {author} {\bibfnamefont {Georges}\ \bibnamefont
  {Aad}} \emph {et~al.} (\bibinfo {collaboration} {ATLAS}),\ }\bibfield
  {title} {\enquote {\bibinfo {title} {{Search for electroweak production of
  charginos and sleptons decaying into final states with two leptons and
  missing transverse momentum in $\sqrt{s}=13$ TeV $pp$ collisions using the
  ATLAS detector}},}\ }\href {\doibase 10.1140/epjc/s10052-019-7594-6}
  {\bibfield  {journal} {\bibinfo  {journal} {Eur. Phys. J. C}\ }\textbf
  {\bibinfo {volume} {80}},\ \bibinfo {pages} {123} (\bibinfo {year}
  {2020}{\natexlab{e}})},\ \Eprint {http://arxiv.org/abs/1908.08215}
  {arXiv:1908.08215 [hep-ex]} \BibitemShut {NoStop}%
\bibitem [{\citenamefont {Aad}\ \emph {et~al.}(2020{\natexlab{f}})\citenamefont
  {Aad} \emph {et~al.}}]{ATLAS:2019fag}%
  \BibitemOpen
  \bibfield  {author} {\bibinfo {author} {\bibfnamefont {Georges}\ \bibnamefont
  {Aad}} \emph {et~al.} (\bibinfo {collaboration} {ATLAS}),\ }\bibfield
  {title} {\enquote {\bibinfo {title} {{Search for squarks and gluinos in final
  states with same-sign leptons and jets using 139 fb$^{-1}$ of data collected
  with the ATLAS detector}},}\ }\href {\doibase 10.1007/JHEP06(2020)046}
  {\bibfield  {journal} {\bibinfo  {journal} {JHEP}\ }\textbf {\bibinfo
  {volume} {06}},\ \bibinfo {pages} {046} (\bibinfo {year}
  {2020}{\natexlab{f}})},\ \Eprint {http://arxiv.org/abs/1909.08457}
  {arXiv:1909.08457 [hep-ex]} \BibitemShut {NoStop}%
\bibitem [{\citenamefont {Aad}\ \emph {et~al.}(2020{\natexlab{g}})\citenamefont
  {Aad} \emph {et~al.}}]{ATLAS:2019gti}%
  \BibitemOpen
  \bibfield  {author} {\bibinfo {author} {\bibfnamefont {Georges}\ \bibnamefont
  {Aad}} \emph {et~al.} (\bibinfo {collaboration} {ATLAS}),\ }\bibfield
  {title} {\enquote {\bibinfo {title} {{Search for direct stau production in
  events with two hadronic $\tau$-leptons in $\sqrt{s} = 13$ TeV $pp$
  collisions with the ATLAS detector}},}\ }\href {\doibase
  10.1103/PhysRevD.101.032009} {\bibfield  {journal} {\bibinfo  {journal}
  {Phys. Rev. D}\ }\textbf {\bibinfo {volume} {101}},\ \bibinfo {pages}
  {032009} (\bibinfo {year} {2020}{\natexlab{g}})},\ \Eprint
  {http://arxiv.org/abs/1911.06660} {arXiv:1911.06660 [hep-ex]} \BibitemShut
  {NoStop}%
\bibitem [{\citenamefont {Aad}\ \emph {et~al.}(2020{\natexlab{h}})\citenamefont
  {Aad} \emph {et~al.}}]{ATLAS:2019lng}%
  \BibitemOpen
  \bibfield  {author} {\bibinfo {author} {\bibfnamefont {Georges}\ \bibnamefont
  {Aad}} \emph {et~al.} (\bibinfo {collaboration} {ATLAS}),\ }\bibfield
  {title} {\enquote {\bibinfo {title} {{Searches for electroweak production of
  supersymmetric particles with compressed mass spectra in $\sqrt{s}=$ 13 TeV
  $pp$ collisions with the ATLAS detector}},}\ }\href {\doibase
  10.1103/PhysRevD.101.052005} {\bibfield  {journal} {\bibinfo  {journal}
  {Phys. Rev. D}\ }\textbf {\bibinfo {volume} {101}},\ \bibinfo {pages}
  {052005} (\bibinfo {year} {2020}{\natexlab{h}})},\ \Eprint
  {http://arxiv.org/abs/1911.12606} {arXiv:1911.12606 [hep-ex]} \BibitemShut
  {NoStop}%
\bibitem [{\citenamefont {Aad}\ \emph {et~al.}(2020{\natexlab{i}})\citenamefont
  {Aad} \emph {et~al.}}]{ATLAS:2020qlk}%
  \BibitemOpen
  \bibfield  {author} {\bibinfo {author} {\bibfnamefont {Georges}\ \bibnamefont
  {Aad}} \emph {et~al.} (\bibinfo {collaboration} {ATLAS}),\ }\bibfield
  {title} {\enquote {\bibinfo {title} {{Search for direct production of
  electroweakinos in final states with missing transverse momentum and a Higgs
  boson decaying into photons in pp collisions at $ \sqrt{s} $ = 13 TeV with
  the ATLAS detector}},}\ }\href {\doibase 10.1007/JHEP10(2020)005} {\bibfield
  {journal} {\bibinfo  {journal} {JHEP}\ }\textbf {\bibinfo {volume} {10}},\
  \bibinfo {pages} {005} (\bibinfo {year} {2020}{\natexlab{i}})},\ \Eprint
  {http://arxiv.org/abs/2004.10894} {arXiv:2004.10894 [hep-ex]} \BibitemShut
  {NoStop}%
\bibitem [{\citenamefont {Aad}\ \emph {et~al.}(2020{\natexlab{j}})\citenamefont
  {Aad} \emph {et~al.}}]{ATLAS:2020dsf}%
  \BibitemOpen
  \bibfield  {author} {\bibinfo {author} {\bibfnamefont {Georges}\ \bibnamefont
  {Aad}} \emph {et~al.} (\bibinfo {collaboration} {ATLAS}),\ }\bibfield
  {title} {\enquote {\bibinfo {title} {{Search for a scalar partner of the top
  quark in the all-hadronic $t{\bar{t}}$ plus missing transverse momentum final
  state at $\sqrt{s}=13$ TeV with the ATLAS detector}},}\ }\href {\doibase
  10.1140/epjc/s10052-020-8102-8} {\bibfield  {journal} {\bibinfo  {journal}
  {Eur. Phys. J. C}\ }\textbf {\bibinfo {volume} {80}},\ \bibinfo {pages} {737}
  (\bibinfo {year} {2020}{\natexlab{j}})},\ \Eprint
  {http://arxiv.org/abs/2004.14060} {arXiv:2004.14060 [hep-ex]} \BibitemShut
  {NoStop}%
\bibitem [{\citenamefont {Aad}\ \emph {et~al.}(2021{\natexlab{e}})\citenamefont
  {Aad} \emph {et~al.}}]{ATLAS:2021twp}%
  \BibitemOpen
  \bibfield  {author} {\bibinfo {author} {\bibfnamefont {Georges}\ \bibnamefont
  {Aad}} \emph {et~al.} (\bibinfo {collaboration} {ATLAS}),\ }\bibfield
  {title} {\enquote {\bibinfo {title} {{Search for squarks and gluinos in final
  states with one isolated lepton, jets, and missing transverse momentum at
  $\sqrt{s}=13$~ with the ATLAS detector}},}\ }\href {\doibase
  10.1140/epjc/s10052-021-09344-w} {\bibfield  {journal} {\bibinfo  {journal}
  {Eur. Phys. J. C}\ }\textbf {\bibinfo {volume} {81}},\ \bibinfo {pages} {600}
  (\bibinfo {year} {2021}{\natexlab{e}})},\ \Eprint
  {http://arxiv.org/abs/2101.01629} {arXiv:2101.01629 [hep-ex]} \BibitemShut
  {NoStop}%
\bibitem [{\citenamefont {Aad}\ \emph {et~al.}(2021{\natexlab{f}})\citenamefont
  {Aad} \emph {et~al.}}]{ATLAS:2021yyr}%
  \BibitemOpen
  \bibfield  {author} {\bibinfo {author} {\bibfnamefont {Georges}\ \bibnamefont
  {Aad}} \emph {et~al.} (\bibinfo {collaboration} {ATLAS}),\ }\bibfield
  {title} {\enquote {\bibinfo {title} {{Search for supersymmetry in events with
  four or more charged leptons in $139\,\mbox{fb\(^{-1}\)}$ of $\sqrt{s}=13$
  TeV $pp$ collisions with the ATLAS detector}},}\ }\href {\doibase
  10.1007/JHEP07(2021)167} {\  (\bibinfo {year} {2021}{\natexlab{f}}),\
  10.1007/JHEP07(2021)167},\ \Eprint {http://arxiv.org/abs/2103.11684}
  {arXiv:2103.11684 [hep-ex]} \BibitemShut {NoStop}%
\bibitem [{\citenamefont {Aad}\ \emph {et~al.}(2021{\natexlab{g}})\citenamefont
  {Aad} \emph {et~al.}}]{ATLAS:2021fbt}%
  \BibitemOpen
  \bibfield  {author} {\bibinfo {author} {\bibfnamefont {Georges}\ \bibnamefont
  {Aad}} \emph {et~al.} (\bibinfo {collaboration} {ATLAS}),\ }\bibfield
  {title} {\enquote {\bibinfo {title} {{Search for R-parity violating
  supersymmetry in a final state containing leptons and many jets with the
  ATLAS experiment using $\sqrt{s} = 13$ TeV proton-proton collision data}},}\
  }\href@noop {} {\  (\bibinfo {year} {2021}{\natexlab{g}})},\ \Eprint
  {http://arxiv.org/abs/2106.09609} {arXiv:2106.09609 [hep-ex]} \BibitemShut
  {NoStop}%
\bibitem [{\citenamefont
  {Collaboration}(2015{\natexlab{b}})}]{ATLAS-CONF-2015-082}%
  \BibitemOpen
  \bibfield  {author} {\bibinfo {author} {\bibfnamefont {ATLAS}\ \bibnamefont
  {Collaboration}},\ }\href {https://cds.cern.ch/record/2114854} {\emph
  {\bibinfo {title} {{A search for Supersymmetry in events containing a
  leptonically decaying $Z$ boson, jets and missing transverse momentum in
  $\sqrt{s}=13~$TeV $pp$ collisions with the ATLAS detector}}}},\ \bibinfo
  {type} {Tech. Rep.}\ \bibinfo {number} {ATLAS-CONF-2015-082}\ (\bibinfo
  {institution} {CERN},\ \bibinfo {address} {Geneva},\ \bibinfo {year}
  {2015})\BibitemShut {NoStop}%
\bibitem [{\citenamefont
  {Collaboration}(2016{\natexlab{a}})}]{ATLAS-CONF-2016-013}%
  \BibitemOpen
  \bibfield  {author} {\bibinfo {author} {\bibfnamefont {ATLAS}\ \bibnamefont
  {Collaboration}},\ }\href {https://cds.cern.ch/record/2140998} {\emph
  {\bibinfo {title} {{Search for production of vector-like top quark pairs and
  of four top quarks in the lepton-plus-jets final state in $pp$ collisions at
  $\sqrt{s}=13$ TeV with the ATLAS detector}}}},\ \bibinfo {type} {Tech. Rep.}\
  \bibinfo {number} {ATLAS-CONF-2016-013}\ (\bibinfo  {institution} {CERN},\
  \bibinfo {address} {Geneva},\ \bibinfo {year} {2016})\BibitemShut {NoStop}%
\bibitem [{\citenamefont
  {Collaboration}(2016{\natexlab{b}})}]{ATLAS-CONF-2016-050}%
  \BibitemOpen
  \bibfield  {author} {\bibinfo {author} {\bibfnamefont {ATLAS}\ \bibnamefont
  {Collaboration}},\ }\href {https://cds.cern.ch/record/2206132} {\emph
  {\bibinfo {title} {{Search for top squarks in final states with one isolated
  lepton, jets, and missing transverse momentum in $\sqrt{s}$ = 13 TeV pp
  collisions with the ATLAS detector}}}},\ \bibinfo {type} {Tech. Rep.}\
  \bibinfo {number} {ATLAS-CONF-2016-050}\ (\bibinfo  {institution} {CERN},\
  \bibinfo {address} {Geneva},\ \bibinfo {year} {2016})\BibitemShut {NoStop}%
\bibitem [{\citenamefont
  {Collaboration}(2016{\natexlab{c}})}]{ATLAS-CONF-2016-054}%
  \BibitemOpen
  \bibfield  {author} {\bibinfo {author} {\bibfnamefont {ATLAS}\ \bibnamefont
  {Collaboration}},\ }\href {https://cds.cern.ch/record/2206136} {\emph
  {\bibinfo {title} {{Search for squarks and gluinos in events with an isolated
  lepton, jets and missing transverse momentum at $\sqrt{s}$ = 13 TeV with the
  ATLAS detector}}}},\ \bibinfo {type} {Tech. Rep.}\ \bibinfo {number}
  {ATLAS-CONF-2016-054}\ (\bibinfo  {institution} {CERN},\ \bibinfo {address}
  {Geneva},\ \bibinfo {year} {2016})\BibitemShut {NoStop}%
\bibitem [{\citenamefont
  {Collaboration}(2016{\natexlab{d}})}]{ATLAS-CONF-2016-066}%
  \BibitemOpen
  \bibfield  {author} {\bibinfo {author} {\bibfnamefont {ATLAS}\ \bibnamefont
  {Collaboration}},\ }\href {https://cds.cern.ch/record/2206209} {\emph
  {\bibinfo {title} {{Search for Supersymmetry in events with photons, jets and
  missing transverse energy with the ATLAS detector in 13 TeV pp
  collisions}}}},\ \bibinfo {type} {Tech. Rep.}\ \bibinfo {number}
  {ATLAS-CONF-2016-066}\ (\bibinfo  {institution} {CERN},\ \bibinfo {address}
  {Geneva},\ \bibinfo {year} {2016})\BibitemShut {NoStop}%
\bibitem [{\citenamefont
  {Collaboration}(2016{\natexlab{e}})}]{ATLAS-CONF-2016-076}%
  \BibitemOpen
  \bibfield  {author} {\bibinfo {author} {\bibfnamefont {ATLAS}\ \bibnamefont
  {Collaboration}},\ }\href {https://cds.cern.ch/record/2206249} {\emph
  {\bibinfo {title} {{Search for direct top squark pair production and dark
  matter production in final states with two leptons in $\sqrt{s} = 13$ TeV
  $pp$ collisions using 13.3 fb$^{-1}$ of ATLAS data}}}},\ \bibinfo {type}
  {Tech. Rep.}\ \bibinfo {number} {ATLAS-CONF-2016-076}\ (\bibinfo
  {institution} {CERN},\ \bibinfo {address} {Geneva},\ \bibinfo {year}
  {2016})\BibitemShut {NoStop}%
\bibitem [{\citenamefont
  {Collaboration}(2016{\natexlab{f}})}]{ATLAS-CONF-2016-096}%
  \BibitemOpen
  \bibfield  {author} {\bibinfo {author} {\bibfnamefont {ATLAS}\ \bibnamefont
  {Collaboration}},\ }\href {https://cds.cern.ch/record/2212162} {\emph
  {\bibinfo {title} {{Search for supersymmetry with two and three leptons and
  missing transverse momentum in the final state at $\sqrt{s}=13$ TeV with the
  ATLAS detector}}}},\ \bibinfo {type} {Tech. Rep.}\ \bibinfo {number}
  {ATLAS-CONF-2016-096}\ (\bibinfo  {institution} {CERN},\ \bibinfo {address}
  {Geneva},\ \bibinfo {year} {2016})\BibitemShut {NoStop}%
\bibitem [{\citenamefont {Collaboration}(2017)}]{ATLAS-CONF-2017-060}%
  \BibitemOpen
  \bibfield  {author} {\bibinfo {author} {\bibfnamefont {ATLAS}\ \bibnamefont
  {Collaboration}},\ }\href {https://cds.cern.ch/record/2273876} {\emph
  {\bibinfo {title} {{Search for dark matter and other new phenomena in events
  with an energetic jet and large missing transverse momentum using the ATLAS
  detector}}}},\ \bibinfo {type} {Tech. Rep.}\ \bibinfo {number}
  {ATLAS-CONF-2017-060}\ (\bibinfo  {institution} {CERN},\ \bibinfo {address}
  {Geneva},\ \bibinfo {year} {2017})\BibitemShut {NoStop}%
\bibitem [{\citenamefont
  {Collaboration}(2018{\natexlab{a}})}]{ATLAS-CONF-2018-041}%
  \BibitemOpen
  \bibfield  {author} {\bibinfo {author} {\bibfnamefont {ATLAS}\ \bibnamefont
  {Collaboration}},\ }\href {https://cds.cern.ch/record/2632347} {\emph
  {\bibinfo {title} {{Search for supersymmetry in final states with missing
  transverse momentum and multiple $b$-jets in proton-proton collisions at
  $\sqrt{s} = 13$ TeV with the ATLAS detector}}}},\ \bibinfo {type} {Tech.
  Rep.}\ \bibinfo {number} {ATLAS-CONF-2018-041}\ (\bibinfo  {institution}
  {CERN},\ \bibinfo {address} {Geneva},\ \bibinfo {year} {2018})\BibitemShut
  {NoStop}%
\bibitem [{\citenamefont
  {Collaboration}(2019{\natexlab{b}})}]{ATLAS-CONF-2019-040}%
  \BibitemOpen
  \bibfield  {author} {\bibinfo {author} {\bibfnamefont {ATLAS}\ \bibnamefont
  {Collaboration}},\ }\href {https://cds.cern.ch/record/2686254} {\emph
  {\bibinfo {title} {{Search for squarks and gluinos in final states with jets
  and missing transverse momentum using 139 fb$^{-1}$ of $\sqrt{s}$ =13 TeV
  $pp$ collision data with the ATLAS detector}}}},\ \bibinfo {type} {Tech.
  Rep.}\ \bibinfo {number} {ATLAS-CONF-2019-040}\ (\bibinfo  {institution}
  {CERN},\ \bibinfo {address} {Geneva},\ \bibinfo {year} {2019})\BibitemShut
  {NoStop}%
\bibitem [{\citenamefont
  {Collaboration}(2019{\natexlab{c}})}]{ATLAS-CONF-2019-020}%
  \BibitemOpen
  \bibfield  {author} {\bibinfo {author} {\bibfnamefont {ATLAS}\ \bibnamefont
  {Collaboration}},\ }\href {https://cds.cern.ch/record/2676597} {\emph
  {\bibinfo {title} {{Search for chargino-neutralino production with mass
  splittings near the electroweak scale in three-lepton final states in
  $\sqrt{s}$ = 13 TeV $pp$ collisions with the ATLAS detector}}}},\ \bibinfo
  {type} {Tech. Rep.}\ \bibinfo {number} {ATLAS-CONF-2019-020}\ (\bibinfo
  {institution} {CERN},\ \bibinfo {address} {Geneva},\ \bibinfo {year}
  {2019})\BibitemShut {NoStop}%
\bibitem [{\citenamefont {Collaboration}(2020)}]{ATLAS-CONF-2020-048}%
  \BibitemOpen
  \bibfield  {author} {\bibinfo {author} {\bibfnamefont {ATLAS}\ \bibnamefont
  {Collaboration}},\ }\href {https://cds.cern.ch/record/2728058} {\emph
  {\bibinfo {title} {{Search for new phenomena in events with jets and missing
  transverse momentum in p p collisions at $\sqrt{s}$ = 13 TeV with the ATLAS
  detector}}}},\ \bibinfo {type} {Tech. Rep.}\ \bibinfo {number}
  {ATLAS-CONF-2020-048}\ (\bibinfo  {institution} {CERN},\ \bibinfo {address}
  {Geneva},\ \bibinfo {year} {2020})\BibitemShut {NoStop}%
\bibitem [{\citenamefont {Sirunyan}\ \emph
  {et~al.}(2018{\natexlab{c}})\citenamefont {Sirunyan} \emph
  {et~al.}}]{CMS:2017moi}%
  \BibitemOpen
  \bibfield  {author} {\bibinfo {author} {\bibfnamefont {A.~M.}\ \bibnamefont
  {Sirunyan}} \emph {et~al.} (\bibinfo {collaboration} {CMS}),\ }\bibfield
  {title} {\enquote {\bibinfo {title} {{Search for electroweak production of
  charginos and neutralinos in multilepton final states in proton-proton
  collisions at $\sqrt{s}=$ 13 TeV}},}\ }\href {\doibase
  10.1007/JHEP03(2018)166} {\bibfield  {journal} {\bibinfo  {journal} {JHEP}\
  }\textbf {\bibinfo {volume} {03}},\ \bibinfo {pages} {166} (\bibinfo {year}
  {2018}{\natexlab{c}})},\ \Eprint {http://arxiv.org/abs/1709.05406}
  {arXiv:1709.05406 [hep-ex]} \BibitemShut {NoStop}%
\bibitem [{\citenamefont {Akerib}\ \emph {et~al.}(2020)\citenamefont {Akerib}
  \emph {et~al.}}]{LUX-ZEPLIN:2018poe}%
  \BibitemOpen
  \bibfield  {author} {\bibinfo {author} {\bibfnamefont {D.~S.}\ \bibnamefont
  {Akerib}} \emph {et~al.} (\bibinfo {collaboration} {LUX-ZEPLIN}),\ }\bibfield
   {title} {\enquote {\bibinfo {title} {{Projected WIMP sensitivity of the
  LUX-ZEPLIN dark matter experiment}},}\ }\href {\doibase
  10.1103/PhysRevD.101.052002} {\bibfield  {journal} {\bibinfo  {journal}
  {Phys. Rev. D}\ }\textbf {\bibinfo {volume} {101}},\ \bibinfo {pages}
  {052002} (\bibinfo {year} {2020})},\ \Eprint
  {http://arxiv.org/abs/1802.06039} {arXiv:1802.06039 [astro-ph.IM]}
  \BibitemShut {NoStop}%
\bibitem [{\citenamefont {Aprile}\ \emph {et~al.}(2020)\citenamefont {Aprile}
  \emph {et~al.}}]{XENON:2020kmp}%
  \BibitemOpen
  \bibfield  {author} {\bibinfo {author} {\bibfnamefont {E.}~\bibnamefont
  {Aprile}} \emph {et~al.} (\bibinfo {collaboration} {XENON}),\ }\bibfield
  {title} {\enquote {\bibinfo {title} {{Projected WIMP sensitivity of the
  XENONnT dark matter experiment}},}\ }\href {\doibase
  10.1088/1475-7516/2020/11/031} {\bibfield  {journal} {\bibinfo  {journal}
  {JCAP}\ }\textbf {\bibinfo {volume} {11}},\ \bibinfo {pages} {031} (\bibinfo
  {year} {2020})},\ \Eprint {http://arxiv.org/abs/2007.08796} {arXiv:2007.08796
  [physics.ins-det]} \BibitemShut {NoStop}%
\bibitem [{\citenamefont {Amole}\ \emph {et~al.}(2015)\citenamefont {Amole}
  \emph {et~al.}}]{PICO:2015amc}%
  \BibitemOpen
  \bibfield  {author} {\bibinfo {author} {\bibfnamefont {C.}~\bibnamefont
  {Amole}} \emph {et~al.} (\bibinfo {collaboration} {PICO}),\ }\bibfield
  {title} {\enquote {\bibinfo {title} {{PICASSO, COUPP and PICO - Search for
  Dark Matter with Bubble Chambers}},}\ }\href {\doibase
  10.1051/epjconf/20149504020} {\bibfield  {journal} {\bibinfo  {journal} {EPJ
  Web Conf.}\ }\textbf {\bibinfo {volume} {95}},\ \bibinfo {pages} {04020}
  (\bibinfo {year} {2015})}\BibitemShut {NoStop}%
\bibitem [{\citenamefont {Bahl}\ \emph {et~al.}(2020)\citenamefont {Bahl},
  \citenamefont {Bechtle}, \citenamefont {Heinemeyer}, \citenamefont {Liebler},
  \citenamefont {Stefaniak},\ and\ \citenamefont {Weiglein}}]{Bahl:2020kwe}%
  \BibitemOpen
  \bibfield  {author} {\bibinfo {author} {\bibfnamefont {H.}~\bibnamefont
  {Bahl}}, \bibinfo {author} {\bibfnamefont {P.}~\bibnamefont {Bechtle}},
  \bibinfo {author} {\bibfnamefont {S.}~\bibnamefont {Heinemeyer}}, \bibinfo
  {author} {\bibfnamefont {S.}~\bibnamefont {Liebler}}, \bibinfo {author}
  {\bibfnamefont {T.}~\bibnamefont {Stefaniak}}, \ and\ \bibinfo {author}
  {\bibfnamefont {G.}~\bibnamefont {Weiglein}},\ }\bibfield  {title} {\enquote
  {\bibinfo {title} {{HL-LHC and ILC sensitivities in the hunt for heavy Higgs
  bosons}},}\ }\href {\doibase 10.1140/epjc/s10052-020-08472-z} {\bibfield
  {journal} {\bibinfo  {journal} {Eur. Phys. J. C}\ }\textbf {\bibinfo {volume}
  {80}},\ \bibinfo {pages} {916} (\bibinfo {year} {2020})},\ \Eprint
  {http://arxiv.org/abs/2005.14536} {arXiv:2005.14536 [hep-ph]} \BibitemShut
  {NoStop}%
\bibitem [{\citenamefont
  {Collaboration}(2018{\natexlab{b}})}]{ATL-PHYS-PUB-2018-048}%
  \BibitemOpen
  \bibfield  {author} {\bibinfo {author} {\bibfnamefont {ATLAS}\ \bibnamefont
  {Collaboration}},\ }\href {https://cds.cern.ch/record/2651927} {\emph
  {\bibinfo {title} {{Prospects for searches for staus, charginos and
  neutralinos at the high luminosity LHC with the ATLAS Detector}}}},\ \bibinfo
  {type} {Tech. Rep.}\ \bibinfo {number} {ATL-PHYS-PUB-2018-048}\ (\bibinfo
  {institution} {CERN},\ \bibinfo {address} {Geneva},\ \bibinfo {year}
  {2018})\BibitemShut {NoStop}%
\bibitem [{\citenamefont
  {Collaboration}(2018{\natexlab{c}})}]{CMS-PAS-FTR-18-001}%
  \BibitemOpen
  \bibfield  {author} {\bibinfo {author} {\bibfnamefont {CMS}\ \bibnamefont
  {Collaboration}},\ }\href {https://cds.cern.ch/record/2648538} {\emph
  {\bibinfo {title} {{Searches for light higgsino-like charginos and
  neutralinos at the HL-LHC with the Phase-2 CMS detector}}}},\ \bibinfo {type}
  {Tech. Rep.}\ \bibinfo {number} {CMS-PAS-FTR-18-001}\ (\bibinfo
  {institution} {CERN},\ \bibinfo {address} {Geneva},\ \bibinfo {year}
  {2018})\BibitemShut {NoStop}%
\bibitem [{\citenamefont {Aad}\ \emph {et~al.}(2021{\natexlab{h}})\citenamefont
  {Aad} \emph {et~al.}}]{ATLAS:2021yqv}%
  \BibitemOpen
  \bibfield  {author} {\bibinfo {author} {\bibfnamefont {Georges}\ \bibnamefont
  {Aad}} \emph {et~al.} (\bibinfo {collaboration} {ATLAS}),\ }\bibfield
  {title} {\enquote {\bibinfo {title} {{Search for charginos and neutralinos in
  final states with two boosted hadronically decaying bosons and missing
  transverse momentum in $pp$ collisions at $\sqrt{s}=13$ TeV with the ATLAS
  detector}},}\ }\href@noop {} {\  (\bibinfo {year} {2021}{\natexlab{h}})},\
  \Eprint {http://arxiv.org/abs/2108.07586} {arXiv:2108.07586 [hep-ex]}
  \BibitemShut {NoStop}%
\bibitem [{\citenamefont {Freitas}\ \emph {et~al.}(2004)\citenamefont
  {Freitas}, \citenamefont {von Manteuffel},\ and\ \citenamefont
  {Zerwas}}]{Freitas:2003yp}%
  \BibitemOpen
  \bibfield  {author} {\bibinfo {author} {\bibfnamefont {A.}~\bibnamefont
  {Freitas}}, \bibinfo {author} {\bibfnamefont {A.}~\bibnamefont {von
  Manteuffel}}, \ and\ \bibinfo {author} {\bibfnamefont {P.~M.}\ \bibnamefont
  {Zerwas}},\ }\bibfield  {title} {\enquote {\bibinfo {title} {{Slepton
  production at e+ e- and e- e- linear colliders}},}\ }\href {\doibase
  10.1140/epjc/s2004-01744-2} {\bibfield  {journal} {\bibinfo  {journal} {Eur.
  Phys. J. C}\ }\textbf {\bibinfo {volume} {34}},\ \bibinfo {pages} {487--512}
  (\bibinfo {year} {2004})},\ \Eprint {http://arxiv.org/abs/hep-ph/0310182}
  {arXiv:hep-ph/0310182} \BibitemShut {NoStop}%
\bibitem [{\citenamefont {Berggren}(2013)}]{Berggren:2013vna}%
  \BibitemOpen
  \bibfield  {author} {\bibinfo {author} {\bibfnamefont {Mikael}\ \bibnamefont
  {Berggren}},\ }\bibfield  {title} {\enquote {\bibinfo {title} {{Simplified
  SUSY at the ILC}},}\ }in\ \href@noop {} {\emph {\bibinfo {booktitle}
  {{Community Summer Study 2013}: {Snowmass on the Mississippi}}}}\ (\bibinfo
  {year} {2013})\ \Eprint {http://arxiv.org/abs/1308.1461} {arXiv:1308.1461
  [hep-ph]} \BibitemShut {NoStop}%
\bibitem [{\citenamefont {Fujii}\ \emph {et~al.}(2015)\citenamefont {Fujii}
  \emph {et~al.}}]{Fujii:2015jha}%
  \BibitemOpen
  \bibfield  {author} {\bibinfo {author} {\bibfnamefont {Keisuke}\ \bibnamefont
  {Fujii}} \emph {et~al.},\ }\bibfield  {title} {\enquote {\bibinfo {title}
  {{Physics Case for the International Linear Collider}},}\ }\href@noop {} {\
  (\bibinfo {year} {2015})},\ \Eprint {http://arxiv.org/abs/1506.05992}
  {arXiv:1506.05992 [hep-ex]} \BibitemShut {NoStop}%
\bibitem [{\citenamefont
  {Collaboration}(2018{\natexlab{d}})}]{ATL-PHYS-PUB-2018-031}%
  \BibitemOpen
  \bibfield  {author} {\bibinfo {author} {\bibfnamefont {ATLAS}\ \bibnamefont
  {Collaboration}},\ }\href {https://cds.cern.ch/record/2647294} {\emph
  {\bibinfo {title} {{ATLAS sensitivity to winos and higgsinos with a highly
  compressed mass spectrum at the HL-LHC}}}},\ \bibinfo {type} {Tech. Rep.}\
  \bibinfo {number} {ATL-PHYS-PUB-2018-031}\ (\bibinfo  {institution} {CERN},\
  \bibinfo {address} {Geneva},\ \bibinfo {year} {2018})\BibitemShut {NoStop}%
\bibitem [{\citenamefont {Dong}\ \emph {et~al.}(2018)\citenamefont {Dong} \emph
  {et~al.}}]{CEPCStudyGroup:2018ghi}%
  \BibitemOpen
  \bibfield  {author} {\bibinfo {author} {\bibfnamefont {Mingyi}\ \bibnamefont
  {Dong}} \emph {et~al.} (\bibinfo {collaboration} {CEPC Study Group}),\
  }\bibfield  {title} {\enquote {\bibinfo {title} {{CEPC Conceptual Design
  Report: Volume 2 - Physics \& Detector}},}\ }\href@noop {} {\  (\bibinfo
  {year} {2018})},\ \Eprint {http://arxiv.org/abs/1811.10545} {arXiv:1811.10545
  [hep-ex]} \BibitemShut {NoStop}%
\bibitem [{\citenamefont {Franceschini}\ \emph {et~al.}(2018)\citenamefont
  {Franceschini} \emph {et~al.}}]{deBlas:2018mhx}%
  \BibitemOpen
  \bibfield  {author} {\bibinfo {author} {\bibfnamefont {R.}~\bibnamefont
  {Franceschini}} \emph {et~al.},\ }\bibfield  {title} {\enquote {\bibinfo
  {title} {{The CLIC Potential for New Physics}},}\ }\href {\doibase
  10.23731/CYRM-2018-003} {\ \textbf {\bibinfo {volume} {3/2018}} (\bibinfo
  {year} {2018}),\ 10.23731/CYRM-2018-003},\ \Eprint
  {http://arxiv.org/abs/1812.02093} {arXiv:1812.02093 [hep-ph]} \BibitemShut
  {NoStop}%
\bibitem [{\citenamefont {Cid~Vidal}\ \emph {et~al.}(2019)\citenamefont
  {Cid~Vidal} \emph {et~al.}}]{CidVidal:2018eel}%
  \BibitemOpen
  \bibfield  {author} {\bibinfo {author} {\bibfnamefont {Xabier}\ \bibnamefont
  {Cid~Vidal}} \emph {et~al.},\ }\bibfield  {title} {\enquote {\bibinfo {title}
  {{Report from Working Group 3}: {Beyond the Standard Model physics at the
  HL-LHC and HE-LHC}},}\ }\href {\doibase 10.23731/CYRM-2019-007.585}
  {\bibfield  {journal} {\bibinfo  {journal} {CERN Yellow Rep. Monogr.}\
  }\textbf {\bibinfo {volume} {7}},\ \bibinfo {pages} {585--865} (\bibinfo
  {year} {2019})},\ \Eprint {http://arxiv.org/abs/1812.07831} {arXiv:1812.07831
  [hep-ph]} \BibitemShut {NoStop}%
\bibitem [{\citenamefont {Ellis}\ \emph {et~al.}(2019)\citenamefont {Ellis}
  \emph {et~al.}}]{Strategy:2019vxc}%
  \BibitemOpen
  \bibfield  {author} {\bibinfo {author} {\bibfnamefont {Richard~Keith}\
  \bibnamefont {Ellis}} \emph {et~al.},\ }\bibfield  {title} {\enquote
  {\bibinfo {title} {{Physics Briefing Book}: {Input for the European Strategy
  for Particle Physics Update 2020}},}\ }\href@noop {} {\  (\bibinfo {year}
  {2019})},\ \Eprint {http://arxiv.org/abs/1910.11775} {arXiv:1910.11775
  [hep-ex]} \BibitemShut {NoStop}%
\bibitem [{\citenamefont {Baer}\ \emph {et~al.}(2020)\citenamefont {Baer},
  \citenamefont {Berggren}, \citenamefont {Fujii}, \citenamefont {List},
  \citenamefont {Lehtinen}, \citenamefont {Tanabe},\ and\ \citenamefont
  {Yan}}]{Baer:2019gvu}%
  \BibitemOpen
  \bibfield  {author} {\bibinfo {author} {\bibfnamefont {Howard}\ \bibnamefont
  {Baer}}, \bibinfo {author} {\bibfnamefont {Mikael}\ \bibnamefont {Berggren}},
  \bibinfo {author} {\bibfnamefont {Keisuke}\ \bibnamefont {Fujii}}, \bibinfo
  {author} {\bibfnamefont {Jenny}\ \bibnamefont {List}}, \bibinfo {author}
  {\bibfnamefont {Suvi-Leena}\ \bibnamefont {Lehtinen}}, \bibinfo {author}
  {\bibfnamefont {Tomohiko}\ \bibnamefont {Tanabe}}, \ and\ \bibinfo {author}
  {\bibfnamefont {Jacqueline}\ \bibnamefont {Yan}},\ }\bibfield  {title}
  {\enquote {\bibinfo {title} {{ILC as a natural SUSY discovery machine and
  precision microscope: From light Higgsinos to tests of unification}},}\
  }\href {\doibase 10.1103/PhysRevD.101.095026} {\bibfield  {journal} {\bibinfo
   {journal} {Phys. Rev. D}\ }\textbf {\bibinfo {volume} {101}},\ \bibinfo
  {pages} {095026} (\bibinfo {year} {2020})},\ \Eprint
  {http://arxiv.org/abs/1912.06643} {arXiv:1912.06643 [hep-ex]} \BibitemShut
  {NoStop}%
\bibitem [{\citenamefont {Habermehl}\ \emph {et~al.}(2020)\citenamefont
  {Habermehl}, \citenamefont {Berggren},\ and\ \citenamefont
  {List}}]{Habermehl:2020njb}%
  \BibitemOpen
  \bibfield  {author} {\bibinfo {author} {\bibfnamefont {Moritz}\ \bibnamefont
  {Habermehl}}, \bibinfo {author} {\bibfnamefont {Mikael}\ \bibnamefont
  {Berggren}}, \ and\ \bibinfo {author} {\bibfnamefont {Jenny}\ \bibnamefont
  {List}},\ }\bibfield  {title} {\enquote {\bibinfo {title} {{WIMP Dark Matter
  at the International Linear Collider}},}\ }\href {\doibase
  10.1103/PhysRevD.101.075053} {\bibfield  {journal} {\bibinfo  {journal}
  {Phys. Rev. D}\ }\textbf {\bibinfo {volume} {101}},\ \bibinfo {pages}
  {075053} (\bibinfo {year} {2020})},\ \Eprint
  {http://arxiv.org/abs/2001.03011} {arXiv:2001.03011 [hep-ex]} \BibitemShut
  {NoStop}%
\bibitem [{\citenamefont {Berggren}(2020)}]{Berggren:2020tle}%
  \BibitemOpen
  \bibfield  {author} {\bibinfo {author} {\bibfnamefont {Mikael}\ \bibnamefont
  {Berggren}},\ }\bibfield  {title} {\enquote {\bibinfo {title} {{What pp SUSY
  limits mean for future e$^+$e$^-$ colliders}},}\ }in\ \href@noop {} {\emph
  {\bibinfo {booktitle} {{International Workshop on Future Linear
  Colliders}}}}\ (\bibinfo {year} {2020})\ \Eprint
  {http://arxiv.org/abs/2003.12391} {arXiv:2003.12391 [hep-ph]} \BibitemShut
  {NoStop}%
\bibitem [{\citenamefont {Baum}\ \emph {et~al.}(2020)\citenamefont {Baum},
  \citenamefont {Sandick},\ and\ \citenamefont {Stengel}}]{Baum:2020gjj}%
  \BibitemOpen
  \bibfield  {author} {\bibinfo {author} {\bibfnamefont {Sebastian}\
  \bibnamefont {Baum}}, \bibinfo {author} {\bibfnamefont {Pearl}\ \bibnamefont
  {Sandick}}, \ and\ \bibinfo {author} {\bibfnamefont {Patrick}\ \bibnamefont
  {Stengel}},\ }\bibfield  {title} {\enquote {\bibinfo {title} {{Hunting for
  scalar lepton partners at future electron colliders}},}\ }\href {\doibase
  10.1103/PhysRevD.102.015026} {\bibfield  {journal} {\bibinfo  {journal}
  {Phys. Rev. D}\ }\textbf {\bibinfo {volume} {102}},\ \bibinfo {pages}
  {015026} (\bibinfo {year} {2020})},\ \Eprint
  {http://arxiv.org/abs/2004.02834} {arXiv:2004.02834 [hep-ph]} \BibitemShut
  {NoStop}%
\bibitem [{\citenamefont {Natalia}\ \emph {et~al.}(2021)\citenamefont
  {Natalia}, \citenamefont {Andr\'es}, \citenamefont {Alfredo}, \citenamefont
  {Will}, \citenamefont {Paul},\ and\ \citenamefont {cheng}}]{Natalia:2021ssb}%
  \BibitemOpen
  \bibfield  {author} {\bibinfo {author} {\bibfnamefont {Cardona}\ \bibnamefont
  {Natalia}}, \bibinfo {author} {\bibfnamefont {Fl\'orez}\ \bibnamefont
  {Andr\'es}}, \bibinfo {author} {\bibfnamefont {Gurrola}\ \bibnamefont
  {Alfredo}}, \bibinfo {author} {\bibfnamefont {Johns}\ \bibnamefont {Will}},
  \bibinfo {author} {\bibfnamefont {Sheldon}\ \bibnamefont {Paul}}, \ and\
  \bibinfo {author} {\bibfnamefont {Tao}\ \bibnamefont {cheng}},\ }\bibfield
  {title} {\enquote {\bibinfo {title} {{Long-term LHC Discovery Reach for
  Compressed Higgsino-like Models using VBF Processes}},}\ }\href@noop {} {\
  (\bibinfo {year} {2021})},\ \Eprint {http://arxiv.org/abs/2102.10194}
  {arXiv:2102.10194 [hep-ph]} \BibitemShut {NoStop}%
\bibitem [{\citenamefont {Tumasyan}\ \emph {et~al.}(2021)\citenamefont
  {Tumasyan} \emph {et~al.}}]{CMS:2021cox}%
  \BibitemOpen
  \bibfield  {author} {\bibinfo {author} {\bibfnamefont {Armen}\ \bibnamefont
  {Tumasyan}} \emph {et~al.} (\bibinfo {collaboration} {CMS}),\ }\bibfield
  {title} {\enquote {\bibinfo {title} {{Search for electroweak production of
  charginos and neutralinos in proton-proton collisions at $\sqrt{s} = $ 13
  TeV}},}\ }\href@noop {} {\  (\bibinfo {year} {2021})},\ \Eprint
  {http://arxiv.org/abs/2106.14246} {arXiv:2106.14246 [hep-ex]} \BibitemShut
  {NoStop}%
\bibitem [{\citenamefont {Collaboration}(2021)}]{CMS-PAS-SUS-21-002}%
  \BibitemOpen
  \bibfield  {author} {\bibinfo {author} {\bibfnamefont {CMS}\ \bibnamefont
  {Collaboration}},\ }\href {https://cds.cern.ch/record/2779116} {\emph
  {\bibinfo {title} {{Search for electroweak production of supersymmetric
  particles in final states containing hadronic decays of WW, WZ, or WH and
  missing transverse momentum}}}},\ \bibinfo {type} {Tech. Rep.}\ \bibinfo
  {number} {CMS-PAS-SUS-21-002}\ (\bibinfo  {institution} {CERN},\ \bibinfo
  {address} {Geneva},\ \bibinfo {year} {2021})\BibitemShut {NoStop}%
\end{thebibliography}%

\end{document}